\documentclass[sigconf,natbib=false,nonacm=False]{aamas}

\usepackage{algorithm}
\usepackage[noend]{algorithmic}

\usepackage{balance} 

\usepackage[square,numbers,compress]{natbib}
\usepackage[utf8]{inputenc} 
\usepackage[T1]{fontenc}    
\usepackage{url}            
\usepackage{booktabs}       
\usepackage{amsfonts}       
\usepackage{nicefrac}       
\usepackage{microtype}      
\usepackage{mathtools}      

\usepackage{color}
\usepackage{amsmath}
\usepackage{amsthm}
\usepackage{wrapfig}
\usepackage{tikz}
\usetikzlibrary{positioning}
\usepackage{algorithm}
\usepackage{graphicx}
\usepackage{hyperref}

\hypersetup{
    colorlinks=true,
    linkcolor=blue,
    filecolor=magenta,
    urlcolor=cyan,
    citecolor=cyan,
}

\usepackage{siunitx}
\newcommand{\JL}[1]{}

\DeclareMathOperator*{\argmin}{arg\,min}
\newcommand{\argminSide}{\textnormal{arg\,min}}
\newcommand{\defeq}{\vcentcolon=}

\newcommand{\loss}{\mathcal{L}}  

\newcommand{\outSymbol}{A}
\newcommand{\inSymbol}{B}

\newcommand{\paramSymbol}{\boldsymbol{\theta}}
\newcommand{\outParam}{\paramSymbol_{\!\outSymbol}}
\newcommand{\inParam}{\paramSymbol_{\!\inSymbol}}
\newcommand{\bothParam}{\boldsymbol{\omega}}
\newcommand{\outLoss}{\loss_{\!\outSymbol}}
\newcommand{\inLoss}{\loss_{\!\inSymbol}}
\newcommand{\gradSymbol}{\boldsymbol{g}}
\newcommand{\bothGrad}{\hat{\gradSymbol}}
\newcommand{\hess}{\mathcal{H}}
\newcommand{\bothHess}{\hat{\hess}}
\newcommand{\outGrad}{\gradSymbol_{\!\outSymbol}}
\newcommand{\inGrad}{\gradSymbol_{\!\inSymbol}}
\newcommand{\lr}{\alpha}
\newcommand{\spectralRadius}{\rho}
\DeclareMathOperator*{\spectrum}{Sp}

\newcommand{\identity}{\boldsymbol{I}}
\newcommand{\fixedPointOp}{\boldsymbol{F}}
\newcommand{\jacFixedPointOp}{\boldsymbol{J}}
\newcommand{\dimSymbol}{d}
\newcommand{\inDim}{\dimSymbol_{\inSymbol}}
\newcommand{\outDim}{\dimSymbol_{\outSymbol}}
\newcommand{\bothDim}{\dimSymbol}
\newcommand{\eigval}{\lambda}

\newcommand*\conj[1]{\bar{#1}}

\newcommand{\transpose}{\top}

\newcommand{\currentTitle}{Lyapunov Exponents for Diversity in Differentiable Games}  

\setcopyright{ifaamas}
\acmConference[AAMAS '22]{Proc.\@ of the 21st International Conference
on Autonomous Agents and Multiagent Systems (AAMAS 2022)}{May 9--13, 2022}
{Online}{P.~Faliszewski, V.~Mascardi, C.~Pelachaud,
M.E.~Taylor (eds.)}
\copyrightyear{2022}
\acmYear{2022}
\acmDOI{}
\acmPrice{}
\acmISBN{}

\acmSubmissionID{678}

\title[\currentTitle]{\vspace{-0.065\textheight} \currentTitle}

\newcommand{\authorTextSize}{\normalsize}

\author{\authorTextSize \vspace{-0.045\textheight}Jonathan Lorraine}
\affiliation{
  \institution{\authorTextSize University of Toronto, Vector Institute}
  \country{}
}

\author{\authorTextSize \vspace{-0.045\textheight}Paul Vicol}
\affiliation{
  \institution{\authorTextSize University of Toronto, Vector Institute}
  \country{}
}

\author{\authorTextSize \vspace{-0.045\textheight}Jack Parker-Holder}
\affiliation{
  \institution{\authorTextSize University of Oxford}
  \country{}
}

\author{\authorTextSize \vspace{-0.01\textheight}Tal Kachman}
\affiliation{
  \institution{\authorTextSize Radboud University}
  \country{}
}

\author{\authorTextSize \vspace{-0.01\textheight}Luke Metz}
\affiliation{
  \institution{\authorTextSize Google Research, Brain Team}
  \country{}
}

\author{\authorTextSize \vspace{-0.01\textheight}Jakob Foerster}
\affiliation{
  \institution{\authorTextSize University of Oxford}
  \country{}
}

\begin{abstract}
    Ridge Rider (RR) is an algorithm for finding diverse solutions to optimization problems by following eigenvectors of the Hessian (``ridges'').
    RR is designed for conservative gradient systems (i.e., settings involving a single loss function), where it branches at saddles --- easy-to-find bifurcation points.
    We generalize this idea to non-conservative, multi-agent gradient systems by proposing a method -- denoted Generalized Ridge Rider (GRR) -- for finding arbitrary bifurcation points.
    We give theoretical motivation for our method by leveraging machinery from the field of dynamical systems.
    We construct novel toy problems where we can visualize new phenomena while giving insight into high-dimensional problems of interest.
    Finally, we empirically evaluate our method by finding diverse solutions in the iterated prisoners' dilemma and relevant machine learning problems including generative adversarial networks.
\end{abstract}

\newcommand{\BibTeX}{\rm B\kern-.05em{\sc i\kern-.025em b}\kern-.08em\TeX}

\begin{document}
    \pagestyle{fancy}
    \fancyhead{}
    
    \maketitle 
    
    \section{Introduction}\label{sec:introduction}
        In machine learning it is often useful to select particular solutions with desirable properties that an arbitrary (global or local) minimum might not have.
        For example, finding solutions in image classification using shapes which generalize more effectively than textures.
        Important instances of this in single-objective minimization are seeking solutions that generalize to unseen data in supervised learning~\citep{geirhos2018imagenet, geirhos2020shortcut}, in policy optimization~\citep{cully2015robots}, and generative models \citep{song2020score}.
        Many real-world systems are not so simple and instead involve multiple agents each of which uses a different subset of parameters to minimize their own objective.
        Some examples are generative adversarial networks (GANs)~\citep{goodfellow2014generative,pfau2016connecting},
        actor-critic models~\citep{pfau2016connecting},
        curriculum learning~\citep{bengio2009curriculum, baker2019emergent, balduzzi2019open, sukhbaatar2017intrinsic},
        hyperparameter optimization~\citep{domke2012generic,maclaurin2015gradient,lorraine2018stochastic, mackay2019self, raghu2020teaching, lorraine2020optimizing, raghu2021meta}, 
        adversarial examples~\citep{bose2020adversarial, yuan2019adversarial}, 
        learning models~\citep{rajeswaran2020game, baconlagrangian, nikishin2021control},
        domain adversarial adaptation~\citep{acuna2021fdomainadversarial},
        neural architecture search~\citep{zoph2016neural,real2019regularized,liu2018darts,grathwohl2018gradient, adam2019understanding},
        multi-agent settings~\citep{foerster2018learning}
        and meta-learning~\citep{ren2018meta, rajeswaran2019meta,ren2020flexible}.
        In these settings, the aim is to find one equilibrium (of potentially many equilibria) where agents exhibit some desired behavior.
        
        For example, in the iterated prisoners' dilemma (Sec. \ref{subsec:new_problems}), solutions favoring reciprocity over unconditional defection result in higher returns for all agents.
        In GANs, solutions often generate a subset of the modes from the target distribution~\citep{arjovsky2017wasserstein}, and in Hanabi, some solutions coordinate far better with humans~\citep{hu2020other}.
        Existing methods often find solutions in small subspaces -- even after many random restarts, as shown in Table~\ref{tab:results}.
        By finding a diverse set of equilibria in these games, we may be able to (a) find solutions with a better joint outcome, (b) develop stronger generative models in adversarial learning, or (c) find solutions that coordinate better with humans.
         
        Recently, Ridge Rider (RR)~\cite{parker2020ridge} proposed a general method for finding diverse solutions in \emph{single-objective} optimization.
        RR is a branching tree search, which starts at a stationary point and then follows different \emph{eigenvectors of the Hessian} (``ridges'') with negative eigenvalues, moving downhill from saddle point to saddle point.
        In settings where multiple agents each have their own objective (i.e., games), the relevant generalization of the Hessian --- the \emph{game Hessian}~\citep{balduzzi2018mechanics} in Eq.~\ref{eq:game_hess} --- is not symmetric. 
        Thus, in general, the game Hessian has complex eigenvalues (EVals) with associated complex eigenvectors (EVecs), making RR not directly applicable.
        
        In this paper, we generalize RR to multi-agent settings by leveraging machinery from \emph{dynamical systems}.
        We connect RR with methods for finding \emph{bifurcation} points, i.e. points where small changes in the initial parameters lead to very different learning dynamics and optimization outcomes.
        We propose novel metrics, inspired by \emph{Lyapunov exponents}~\citep{katok1997introduction} that measure how quickly learning trajectories separate.
        These metrics generalize the branching criterion from RR, allowing us to locate a broad class of potential bifurcations, and enabling us to find more diverse behavior.
        
        Starting points with rapid trajectory separation can be found via gradient-based optimization by differentiating through the exponent calculation, which is implemented efficiently for large models with Jacobian-vector products.
        
        \vspace{0.3em}
        
        Our contributions include:
        \vspace{-0.2em}
        \begin{itemize}
            \item Connections between finding diverse solutions and Lyapunov exponents, allowing us to leverage the broad body of work in dynamical systems.
            \item Proposing a method, Generalized Ridge-Rider (GRR), that scales to high-dimensional differentiable games.
            \item Presenting novel diagnostic problem settings based on high dimensional games like the iterated prisoners dilemma (IPD) and GANs to study a range of different bifurcation types.
            \item Compared to existing methods, GRR finds diverse solutions in the IPD, spanning cooperation, defection and reciprocity.
            \item Compared to RR, our GRR method also provides a more efficient estimator of the negative EVecs.
            \item Lastly, we present larger-scale experiments on GANs --- a model class of high interest to the ML community.
        \end{itemize}

    \vspace{-0.0125\textheight}
    \section{Background}\label{sec:background}
        App. Table~\ref{tab:TableOfNotation} summarizes our notation.
        Consider the single-objective optimization problem:\vspace{-0.0125\textheight}
        \begin{equation}\label{eq:single_objective_opt}
            \smash{
                \paramSymbol^{*} \defeq \argminSide_{\paramSymbol} \loss(\paramSymbol) 
            }
        \end{equation}
        We denote the gradient of the loss at parameters $\paramSymbol^{j}$ by $\gradSymbol^j \defeq\ \gradSymbol(\paramSymbol^j) \defeq \left. \nabla_{\paramSymbol} \loss(\paramSymbol) \right|_{\paramSymbol^{j}}$.
        We can locally minimize the loss $\loss$ using gradient descent with step size $\lr$:\vspace{-0.005\textheight}
        \begin{equation}\label{eq:single_objective_gd}
            \smash{\paramSymbol^{j+1} = \paramSymbol^{j} - \alpha \gradSymbol^j}
        \end{equation}
        Due to the potential non-convexity of the $\loss$, multiple stationary points can exist, and gradient descent will only find a particular solution based on the initialization $\paramSymbol^{0}$.
        \vspace{-0.01\textheight}
        \subsection{Ridge Rider}
            Ridge Rider (RR)~\citep{parker2020ridge} finds diverse solutions in single-objective minimization problems.
            The method first finds a saddle point, either analytically, e.g. in tabular reinforcement learning, or by minimizing the gradient norm.
            
            Then, RR branches the optimization procedure following different directions (or ``ridges'') given by the EVecs of the Hessian $\hess = \nabla_{\paramSymbol} \gradSymbol = \nabla_{\paramSymbol} (\nabla_{\paramSymbol} \mathcal{L})$. 
            Full computation of the eigendecomposition of $\hess$, i.e. its EVecs and EVals, is often prohibitively expensive; however, we can efficiently access a subset of the eigenspaces via Hessian-vector products $\hess \boldsymbol{v} = \nabla_{\paramSymbol} ((\nabla_{\paramSymbol} \mathcal{L}) \boldsymbol{v})$ ~\citep{pearlmutter1994fast, tensorflow2015-whitepaper, paszke2017automatic, jax2018github}.
        
        \vspace{-0.01\textheight}
        \subsection{Optimization in Games}
            Instead of simply optimizing a single loss, optimization in games involves multiple agents, each with a loss function that can depend on other agents.
            For simplicity, we look at $\num{2}$-player games with players (denoted by $A$ and $B$) who want to minimize their loss -- $\loss_A(\paramSymbol_A, \paramSymbol_B)$ or  $\loss_B(\paramSymbol_A, \paramSymbol_B)$ --  with their parameters -- $\paramSymbol_A$ or $\paramSymbol_B$.
            \begin{align}\label{eq:nash_equilibrium}
                \smash{\outParam^{*} \!\in\! \argminSide_{\outParam}\! \outLoss(\outParam,\! \inParam^{*}),
                \inParam^{*} \!\in\! \argminSide_{\inParam}\! \inLoss(\outParam^{*},\! \inParam)}
            \end{align}
            If $\inLoss$ and $\outLoss$ are differentiable in $\inParam$ and $\outParam$, we say the game is differentiable.
            One of the simplest optimization methods is to find local solutions by simply following the players' gradients, but this is unstable when the game Hessian has complex EVals~\citep{bailey2020finite}.
            Here, \smash{$\outGrad^j \defeq \outGrad(\outParam^j, \inParam^j)$} is an estimator for $ \nabla_{\outParam} \outLoss |_{\smash{\outParam^j, \inParam^j}}$, and the simultaneous gradient update is:\vspace{0.001\textheight}
            \begin{align}\label{eq:multi_objective_gd_long}\tag{SimSGD}
                \smash{
                \outParam^{j+1} = \outParam^{j} - \lr \outGrad^j,\quad\quad
                \inParam^{j+1} = \inParam^{j} - \lr \inGrad^j
                }
            \end{align}
            We simplify notation with the concatenation of all players' parameters (or joint-parameters) $\smash{\bothParam \!\defeq\! [\outParam, \inParam] \!\in\! \mathbb{R}^{\bothDim}}$ and the joint-gradient vector field $\bothGrad: \mathbb{R}^{\bothDim} \to \mathbb{R}^{\bothDim}$, denoted at the $j^{th}$ iteration by:\vspace{-0.005\textheight}
            \begin{equation}
                \bothGrad^j \defeq \bothGrad(\bothParam^j) \defeq [\outGrad(\bothParam^j), \inGrad(\bothParam^j)] = [\outGrad^j, \inGrad^j]
            \end{equation}
            We write the next iterate in (\ref{eq:multi_objective_gd_long}) with fixed-point operator $\fixedPointOp$:\vspace{-0.005\textheight}
            \begin{equation}\label{eq:fixed_point_operator}
                \smash{
        	    \bothParam^{j+1} \!=\! \fixedPointOp_{SGD} (\bothParam^{j}) \!=\! \bothParam^{j} - \lr \bothGrad^{j}
        	    }
        	\end{equation}
        	The Jacobian of the fixed point operator $\fixedPointOp$ -- denoted $\jacFixedPointOp$ -- is useful for analysis, including bounding convergence rates near fixed points~\citep{bertsekas2008nonlinear} and finding points where local changes to parameters may cause convergence to qualitatively different solutions~\citep{hale2012dynamics}.
        	The fixed point operator's Jacobian crucially depends on the Jacobian of the joint-gradient $\bothGrad$, which is called the \emph{game Hessian}~\citep{letcher2019differentiable} because it generalizes the Hessian:\vspace{-0.01\textheight}
        	\begin{equation}\label{eq:game_hess}
        	    \bothHess \defeq \nabla_{\bothParam} \bothGrad = \begin{bmatrix}
        	        \nabla_{\outParam}^2 \outLoss & \nabla_{\outParam}\! \nabla_{\inParam} \outLoss \\
        	        \nabla_{\inParam} \!\nabla_{\outParam} \inLoss^\top & \nabla_{\inParam}^2 \inLoss
        	    \end{bmatrix}
        	\end{equation}\vspace{-0.01\textheight}
        	\begin{equation}\label{eq:jac)fixed_point_operator}
                \smash{
        	    \jacFixedPointOp_{SGD} \defeq \nabla_{\bothParam} \fixedPointOp_{SGD} (\bothParam) \!=\! \identity - \lr \bothHess
        	    }
        	\end{equation}
        	In Fig.~\ref{fig:fig_new_problems_baseline} we show a game with a solution that we can only converge to by using an appropriate optimizer.
            Thus, we need to incorporate information about the optimizer when generalizing RR, which we do by working with the (largest) EVals/EVecs of $\jacFixedPointOp$ instead of (most negative) EVals/EVecs of $\bothHess$.
            
            We can understand the difference between optimization with a single and multiple objectives as follows:
            In single-objective optimization following the gradient forms trajectories from a conservative vector field because $\bothHess = \hess$ is the Hessian of the loss which is symmetric and has real EVals.
        	However, in games with multiple objectives, $\bothHess$ can be non-symmetric and have complex EVals, resulting in a non-conservative vector field from optimization, opening the door to many new phenomena.

    \section{Methods for Generalizing RR}\label{sec:theory}
        Here, we present two key contributions for generalizing RR to games.
        We first connect diversity in optimization to the general concept of bifurcations, places where a small change to the parameters causes a large change to the optimization trajectories.
        Second, we introduce Lyapunov exponents~\citep{katok1997introduction} and easy-to-optimize variants as a tool for finding these aforementioned bifurcations. 
        
        \vspace{-0.01\textheight}
        \subsection{Connecting Diversity and Bifurcations}
            In dynamical systems, \emph{bifurcations} are regions of the parameter space where small perturbations result in very different optimization trajectories, and in general a dynamical system can contain a variety of different \emph{types of bifurcations}.
            In contrast, in conservative gradient vector fields, saddle points are the only relevant class of bifurcations. 
            As a consequence, their EVecs play a key role in the shape of the phase portraits, which are geometric representations of the underlying dynamics.
            In particular, the negative EVecs are orthogonal to separatrices~\citep{tabor1989chaos}, boundaries between regions in our system with different dynamical behavior, thus providing directions to move in for finding different solutions.
            This perspective provides a novel view on RR.
            RR branches at saddle points, the only relevant class of bifurcation points in single loss optimization. 
            
            However, in the dynamical systems literature, many bifurcation types have been studied~\citep{katok1997introduction}.
            This inspires generalizing RR to non-conservative gradient fields (e.g. multi-agent settings) where a broad variety of bifurcations occur. 
            See Fig.~\ref{fig:fig_new_problems_baseline} for a \emph{Hopf bifurcation}~\citep{hale2012dynamics} or Fig.~\ref{fig:complicated_toy} for various others.
            
        \newcommand{\lyap}{\hat{\eigval}}
        \newcommand{\displacement}{\boldsymbol{d}}
        \vspace{-0.01\textheight}
        \subsection{Lyapunov Exponents for Bifurcations}\label{sec:lyap_for_bifurcation}
\begin{figure}[t!]
                \vspace{-0.06\textheight}
                \centering
                \begin{tikzpicture}
                    \centering
                    \node (img){\includegraphics[trim={.0cm .0cm .0cm .0cm},clip, width=.97\linewidth]{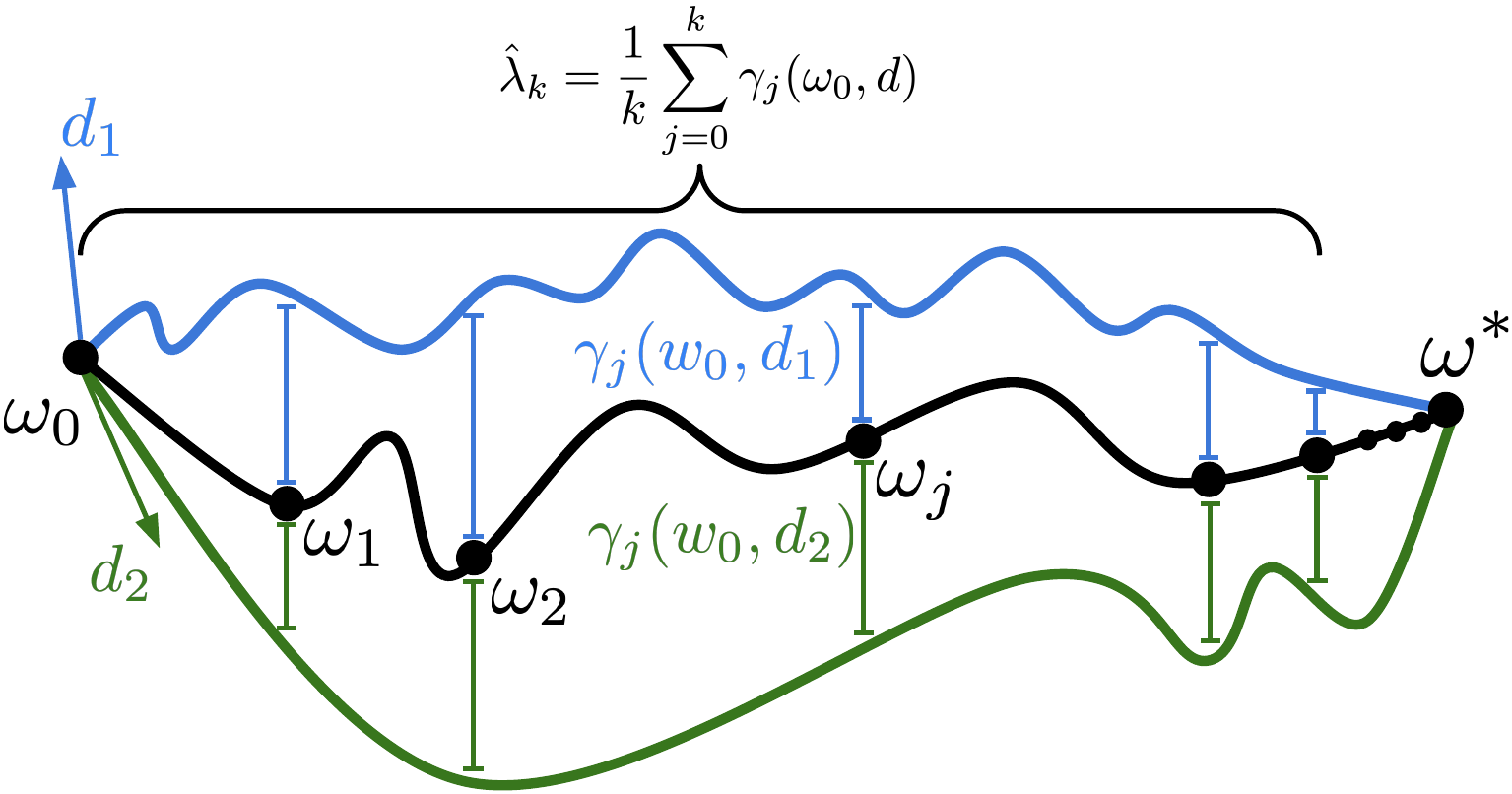}};
                \end{tikzpicture}
                \vspace{-0.035\textheight}
                \caption{
                    Visualization of the components to a Lyapunov exponent $\lyap_k(\bothParam_0, \displacement)$ described in Eqs.~\ref{eq:lyap_calc2},~\ref{eq:lyap_calc}, which measures how quickly trajectories separate starting at a point $\bothParam_0$ in direction $\displacement$.
                    Here, the optimization trajectory iterates $\bothParam_j$ accumulate at a fixed point $\bothParam^{*}$.
                    We show two displacements -- $\color{blue}\displacement_1$ and ${\color{green}\displacement_2}$ -- resulting in separate "perturbed" trajectories shown in $\color{blue}blue$ and $\color{green}green$.
                    We measure the separation rate between the true and perturbed trajectories at the $j^{th}$ optimizer iteration with the Lyapunov term $\gamma_j(\bothParam_0, \displacement)$.
                    The exponent $\lyap_k(\bothParam_0, \displacement)$ is the average of the first $k$ terms.
                    See Fig.~\ref{fig:fig_mix} for actual trajectories on a toy problem used in an exponent calculation.
                }\label{fig:lyap_calc}
                \vspace{-0.01\textheight}
            \end{figure}

\begin{figure}
            \vspace{-0.05\textheight}
            \centering
            \begin{tikzpicture}
                \centering
                \node (img){\includegraphics[trim={.25cm .25cm 1.05cm .25cm},clip, width=.89\linewidth]{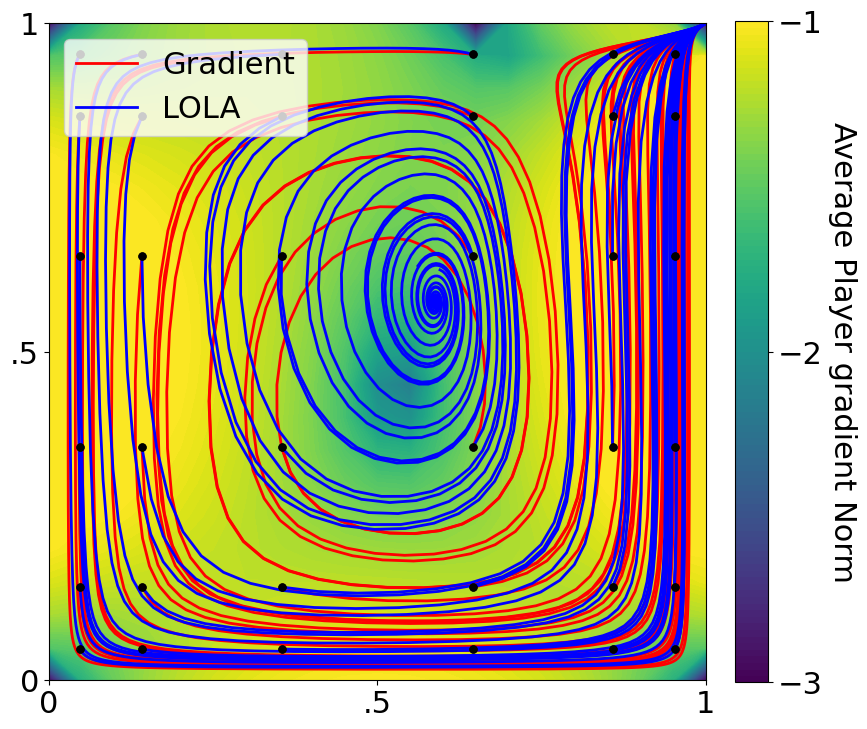}};
                \node[left=of img, node distance=0cm, rotate=90, xshift=1.5cm, yshift=-.9cm, font=\color{black}] {Player 1 Strategy};
                \node[below=of img, node distance=0cm, xshift=-.0cm, yshift=1.25cm,font=\color{black}] {Player 2 Strategy};
                \node[right=of img, node distance=0cm, rotate=270, xshift=-2.75cm, yshift=-.9cm, font=\color{black}] {Joint-player grad. log-norm $\log(\|\bothGrad\|)$};
            \end{tikzpicture}
            \vspace{-0.015\textheight}
            \caption{
                The phase portrait for two standard optimization algorithms on the mixed small IPD and Matching pennies problem.
                We show trajectories following the gradient with simSGD in {\color{red}red} and LOLA~\citep{foerster2018learning}  -- a method for learning in games -- in {\color{blue}blue}.
                All initializations of SimSGD only find the solution in the top right, because the center solution has imaginary EVals, while LOLA finds all solutions.
                For comparisons over more test problems see App. Fig.~\ref{fig:fig_new_problems_baseline_full}.
            }\label{fig:fig_new_problems_baseline}
            \vspace{-0.02\textheight}
        \end{figure}

            Using tools from dynamical systems research we look at \emph{how} to find general bifurcation points. 
            Our objectives are inspired by the Lyapunov exponent, which measures asymptotic separation rates of optimization trajectories for small perturbations.
            We propose a similar quantity, but for finite length trajectories.
            Given a $k$-step trajectory generated by our fixed point operator $\fixedPointOp$ -- i.e., optimizer -- with initialization $\bothParam_0$ and Jacobian at iteration $j$ of $\jacFixedPointOp^{j}$, we measure the separation rate for an initial, normalized displacement $\displacement$ with:\vspace{-0.005\textheight}
            \begin{align}\label{eq:lyap_calc2}
                \lyap_k(\bothParam_0, \displacement) &= \frac{1}{k}\smash{\sum_{j = 0}^{k}} \gamma_j(\bothParam_0, \displacement),\\
                \textnormal{where }\gamma_j(\bothParam_0, \displacement) &\defeq \log(\displacement^\transpose (\jacFixedPointOp^{j}(\bothParam_0))^\transpose \jacFixedPointOp^{j}(\bothParam_0) \displacement)\label{eq:lyap_calc}
            \end{align}
            We call $\gamma_j$ the \emph{$j^{th}$ Lyapunov term}.
            When $k=0$, the $\lyap$ are called the \emph{local Lyapunov exponents}, while as $k \to \infty$ these are called the \emph{(global) Lyapunov exponents}~\citep{wolff1992local}.
            For a variable $k$, we denote this as the \emph{$k$-step or truncated Lyapunov exponent}.
            Fig.~\ref{fig:lyap_calc} visualizes the exponents' calculation providing additional intuition.
            For notational simplicity, we suppress the dependency of $\jacFixedPointOp^{j}$ on $\bothParam_0$ going forward.
            
            Within a basin of attraction to a given fixed point the global Lyapunov exponent is constant~\citep{katok1997introduction}.
            Intuitively, this is because an arbitrarily high number of Lyapunov terms near the fixed point dominate the average defining the exponent in Eq.~\ref{eq:lyap_calc2}.
            This property prevents us from optimizing the global exponent using gradients, making it a poor objective for bifurcations.
            As such, our interest in the truncated exponent is motivated from multiple directions:
            \begin{enumerate}
                \item Non-zero gradient signals for finding bifurcations
                \item Computationally tractability
                \item A better separation rate description for the finite trajectories used in practice
            \end{enumerate}
            However, unlike the global exponent, the truncated version lacks theoretical results.
            
            In more than one dimension, the $k$-step Lyapunov exponent is a function of the specific direction of the perturbation $\displacement$~\citep{tabor1989chaos}.
            We look at using the direction for maximal separation --- i.e., the \emph{max $k$-step Lyapunov exponent}:\vspace{-0.005\textheight}
            \begin{equation}\label{eq:lyap_max}
                \lyap_k^{max}(\bothParam_0) = \max_{\displacement, \|\displacement\| = 1} \lyap_k(\bothParam_0, \displacement)
            \end{equation}
            For dynamical systems with basins of attraction, common in optimization, the max exponent is largest on the boundary between basins, which motivates maximizing the max exponent to find bifurcations.
            The max exponent can be evaluated by finding the largest EVal of an average of the Jacobians over the optimization steps~\citep{katok1997introduction}:\vspace{-0.005\textheight}
            \begin{align}\label{eq:jac_sum}
                 \jacFixedPointOp^{\dagger} \defeq \frac{1}{k} \sum_{j = 0}^{k} (\jacFixedPointOp^j)^{\transpose}(\jacFixedPointOp^j),\\
                 \smash{\lyap_k^{max}(\bothParam_0)} = \max_{\lambda \in \spectrum(\jacFixedPointOp^{\dagger})} |\lambda|
            \end{align}
            Importantly, in higher dimensions one eigenvalue dominates the spectrum of $\jacFixedPointOp$ after a large number of steps \citep{loreto1996characterization,kachman2017numerical} and is thus a point of maximal exploration in our solution space.
            
            We note some practical points for computing these exponents:
            when $k=0$ the max exponent is the max eigenvalue of $\jacFixedPointOp^0$.
            As $k \to \infty$ and our fixed point operator converges to a fixed point $\bothParam^*$, the max exponent is the max EVal of $\jacFixedPointOp$ at $\bothParam^*$.
            Calculating $\lyap_k^{max}$ is easiest when $k=0$ or $k \! \to \! \infty,$ e.g., by power iteration on the relevant $\jacFixedPointOp$.
            For intermediary $k$, directly using leading EVals of $\jacFixedPointOp^{\dagger}$ involves re-evaluating the entire optimization trajectory many times.
            Instead, it is often easier to work with bounds.
            A simple lower bound is formed by using the leading EVec at any single step, or an upper bound by using the leading EVec at each step, which are tight as $k \to \infty$~\citep{loreto1996characterization}.
            We investigate these strategies in App. Fig.~\ref{fig:lyap_dir_mix}.
            
            Commonly, our goal is to obtain many qualitatively different solutions from a single starting point, which motivates simultaneously optimizing the exponents corresponding to multiple different directions.
            Relatedly, the sum of positive global Lyapunov exponents gives an estimate of the Kolmogorov–Sinai or metric entropy by Pesin's theorem~\citep{pesin1977characteristic}.
            We use this to motivate different performance metrics in Section~\ref{sec:grr}.
            
    \section{Proposed Algorithms}\label{sec:algs}
        Having given an overview of the key mathematical concepts, we now present our overall algorithm.
        First, we introduce a general branching-tree search framework for finding diverse solutions in differentiable games.
        Next, we present our method -- Generalized Ridge Rider (GRR) -- which implements this framework using truncated Lyapunov exponents (Eq.~\ref{eq:lyap_max}) as the branching criterion.
        Lastly, we highlight the differences between GRR and RR.
        
        \subsection{Branching Optimization Tree Searches}
\begin{figure}[t!]
                \vspace{-0.04\textheight}
                \centering
                \begin{tikzpicture}
                    \centering
                    \node (img){\includegraphics[trim={.5cm 2.75cm 4.5cm 1.9cm},clip, width=.99\linewidth]{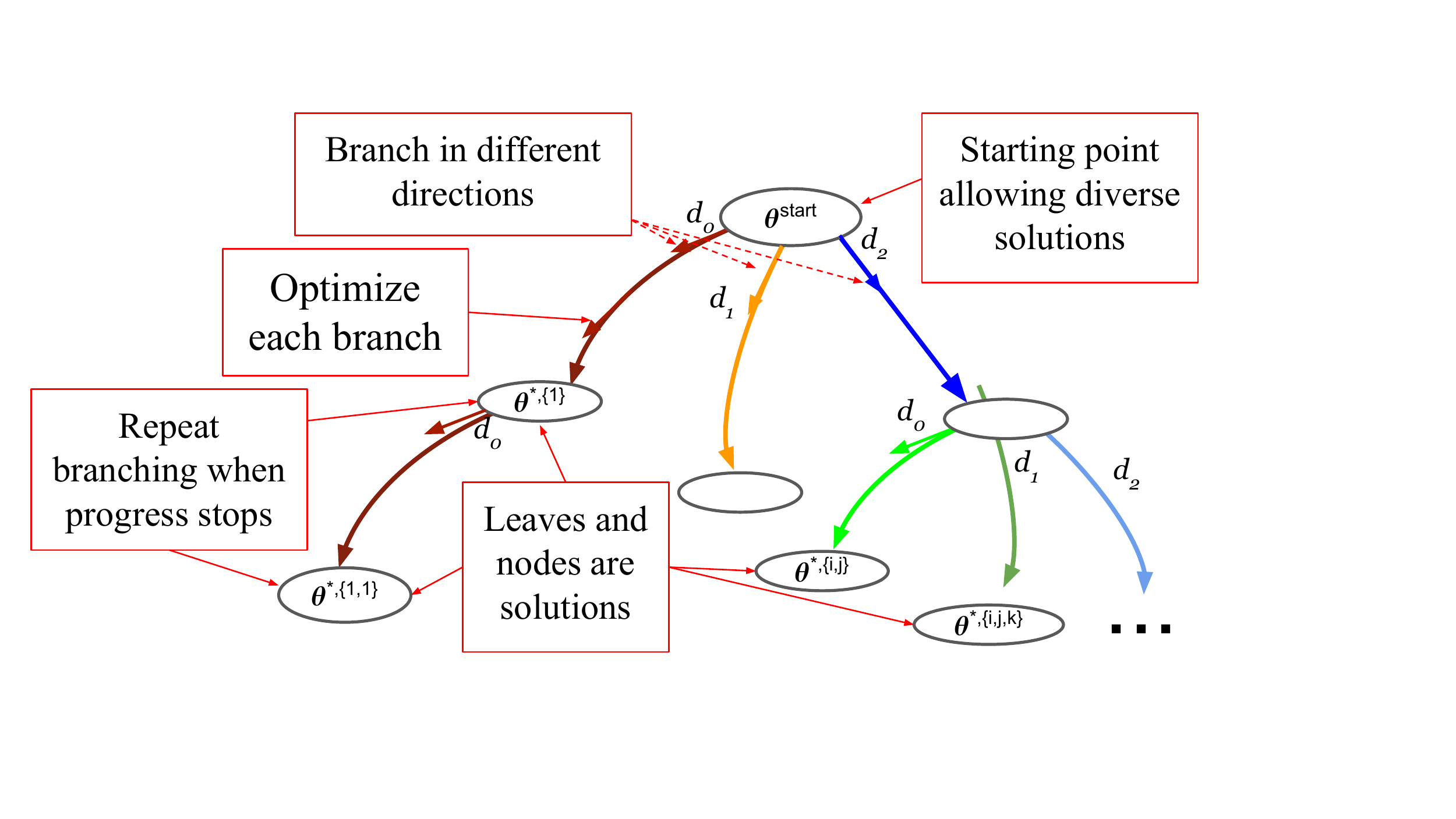}};
                \end{tikzpicture}
                \vspace{-0.015\textheight}
                \caption{
                    Visualization of branching optimization tree search.
                    The key components are: (1) selecting the starting point, (2) creating different branches, (3) optimizing each branch, and (4) choosing when to re-branch.
                }\label{fig:branching}
                \vspace{-0.015\textheight}
            \end{figure}

\begin{figure}[t!]
                \vspace{-.05\textheight}
                \centering
                \begin{tikzpicture}
                    \centering
                    \node (img){\includegraphics[trim={.25cm .25cm 1.05cm .25cm},clip, width=.81\linewidth]{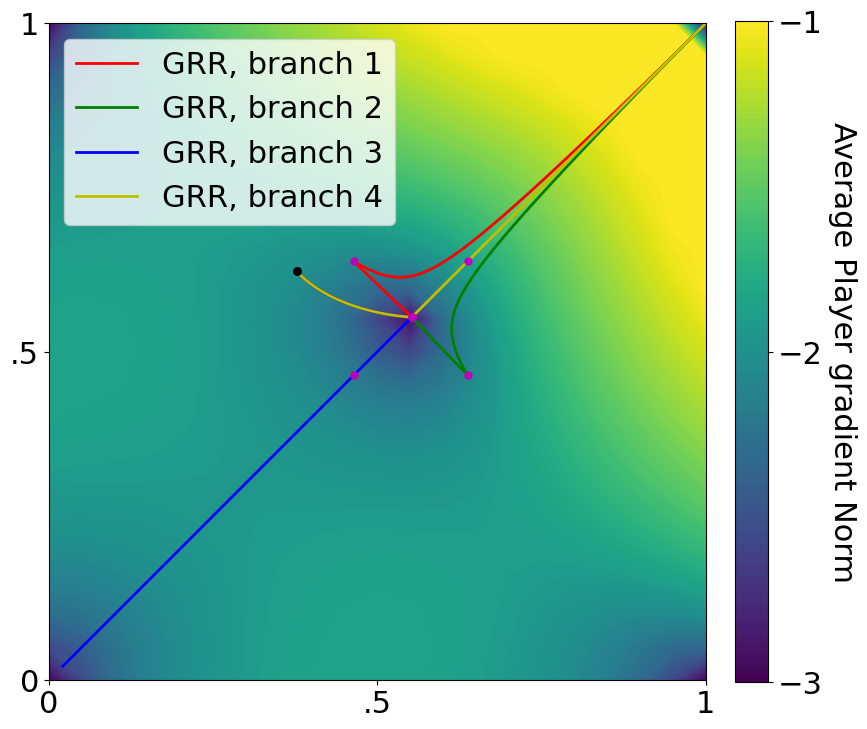}};
                    \node[left=of img, node distance=0cm, rotate=90, xshift=1.5cm, yshift=-.9cm, font=\color{black}] {Player 1 Strategy};
                    \node[right=of img, node distance=0cm, rotate=270, xshift=0cm, yshift=-.9cm, font=\color{black}] {Joint-player grad. log-norm $\log(\|\bothGrad\|)$};
                    \node[above=of img, node distance=0cm, yshift=-1.1cm,font=\color{black}] {GRR -- Our Method -- at a Saddle Bifurcation};
                    
                    \node [below=of img, yshift=1.25cm] (img2){\includegraphics[trim={.25cm .25cm 1.05cm .25cm},clip, width=.81\linewidth]{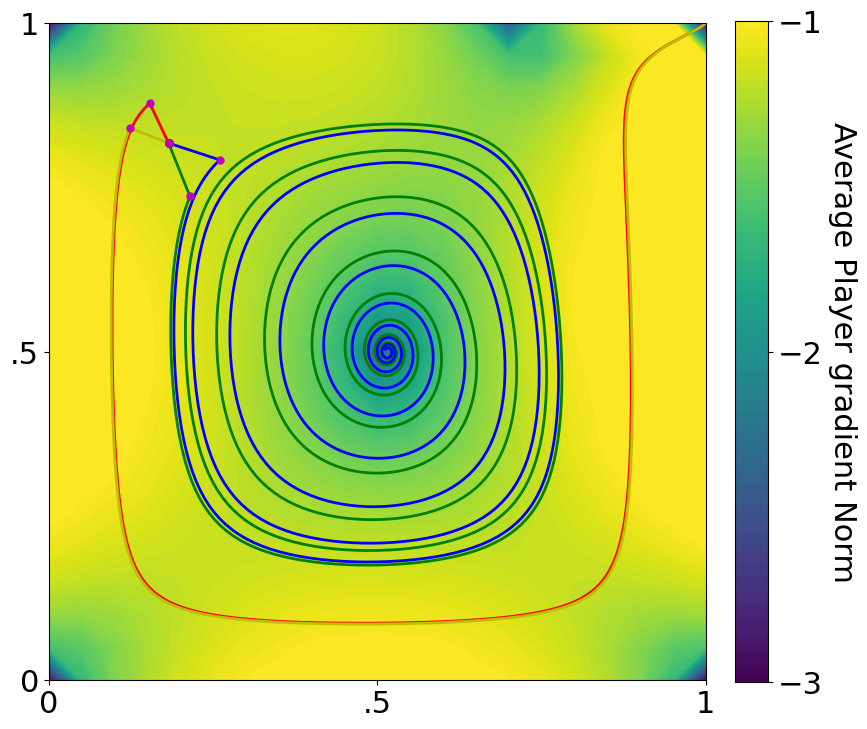}};
                    \node[left=of img2, node distance=0cm, rotate=90, xshift=1.5cm, yshift=-.9cm, font=\color{black}] {Player 1 Strategy};
                    \node[below=of img2, node distance=0cm, xshift=-.25cm, yshift=1.15cm,font=\color{black}] {Player 2 Strategy};
                    \node[below=of img2, node distance=0cm, xshift=-.1cm,yshift=.6cm,font=\color{black}] {GRR at a Hopf Bifurcation};
                \end{tikzpicture}
                \vspace{-.02\textheight}
                \caption{
                    We show branching at different types of bifurcations, obtained by optimizing a Lyapunov exponent as shown in Fig.~\ref{fig:fig_mix}.
                    In each setup, we have two EVecs and branch in opposite directions, giving four paths, displayed in different colors.
                    Steps with the eigenvector have {\color{magenta}magenta} circles marking boundaries.
                    \emph{Top}: In the small IPD, finding, then branching at a saddle -- where the joint-player grad. log-norm $\log(\|\bothGrad\|)$ is $0$ -- allows us to find defect-defect and tit-for-tat solutions.
                    \emph{Bottom:} In the Mixed Problem of Small IPD and Matching Pennies, branching at the Hopf bifurcation allows us to find both solutions.
                    Here, there are no saddle points near the bifurcation, so RR's starting point does not allow branching to find both solutions.
                }\label{fig:branch_toy}
                \vspace{-.025\textheight}
            \end{figure}

            Our framework is a generalized version of RR and contains the following components:
            \begin{enumerate}
                \item A method for finding a suitable \emph{starting point} for our branching process - see Fig.~\ref{fig:fig_mix}.
                \item A process for selecting \emph{branching directions} (or perturbations) from a given branching point - see Fig.~\ref{fig:branch_toy}.
                \item A prescription for how to \emph{continue the optimization process} along a given branch after the initial perturbation.
                \item A re-branching decision rule, i.e., when to go back to step (2).
                This was important in RR because optimizers in high-dimensional non-convex ML problems often finish at saddle points~\citep{yao2020pyhessian}.
                \item Lastly, a metric to rank the different solutions.
            \end{enumerate}
            We visualize this process in Fig.~\ref{fig:branching}.
            RR is an instance of this general process, where each component is suitable for single-objective optimization.
            In the next section, we present another instance of this method, designed for optimization in games.
            We include a more detailed description of branching tree searches in App. Alg.~\ref{algorithm:branching_tree}, highlighting the important changes compared to RR.
            
        \subsection{Generalized Ridge Rider (GRR)}\label{sec:grr}
            \textbf{Starting point:}\label{sec:method_start_obj}
                Motivated by Section~\ref{sec:lyap_for_bifurcation}, we look at optimizing the maximal $k$-step Lyapunov exponent from Eq.~\ref{eq:lyap_max} to obtain our starting point:
                \begin{equation}\label{eq:1d_obj}
                    \loss(\bothParam_0) = -\lyap_k^{max}(\bothParam_0) = -\max_{\displacement, \|\displacement\| = 1} \lyap_k(\bothParam_0, \displacement)
                \end{equation}
                However, using a single exponent only guarantees trajectory separation in a single direction.
                If we want to branch across multiple bifurcations in different directions, we need an objective using exponents in multiple, different directions.
                We look at the simple objective choice summing over exponents:
                \begin{align}
                    &\loss_{n}^{\textnormal{sum}}(\bothParam_0) = -\max_{\displacement_1, \dots, \displacement_n} \sum_{l=1}^{n} \lyap_k(\bothParam_0, \displacement_l),\\
                    \textnormal{such that } \|\displacement_l\| &= 1, \displacement_l^\transpose \displacement_m = 0 \textnormal{ for all $l,m \in {1, \dots, n}$, $l \neq m$}
                    \label{eq:constraint}
                \end{align}
                Intuitively, the constraint guarantees we have different directions to separate in, by making them orthogonal.
                It is straightforward to evaluate this objective, by evaluating the top-$n$ EVals of the matrix from Eq.~\ref{eq:jac_sum}.
                More generally, convex functions of the $k$-step exponents in different directions form reasonable objectives that are more amenable to optimization.
                Specifically, we also look at:
                \begin{align}
                    &\loss_{n}^{\textnormal{min}}(\bothParam_0) = -\max_{\displacement_1, \dots, \displacement_n} \min_{l=1\dots n} \lyap_k(\bothParam_0, \displacement_l)\\
                    \textnormal{such that } \|\displacement_l\| &= 1, \displacement_l^\transpose \displacement_m = 0 \textnormal{ for all $l,m \in {1, \dots, n}$, $l \neq m$}
                \end{align}

            \textbf{Branching the parameter optimization:}
                We must choose what direction to branch in; our procedure for evaluating Lyapunov exponent objectives creates natural candidates.
                Specifically, evaluating the max exponent involves finding the direction maximizing trajectory separation, which we re-use for branching.
                Notably, this is the most negative EVal of the Hessian if we start at a saddle point and use SGD when calculating the trajectories, generalizing the choice from RR.
                For each direction, we can move in both a positive and negative direction, giving two branches.
                
                Also, we must choose how far to move in the directions.
                If we move too far, we may leap into entirely different parts of the parameter space -- e.g., missing interesting regions and recovering similar solution modes.
                If we are not exactly at a bifurcation -- only near it -- then we may need to move some minimum distance to cross the separatrix and find a new solution.
                We look at two simple strategies to move sufficiently far.
                First, we try taking a single step with the normalized exponent direction scaled according to the exponent.
                Second, we look at taking small steps in the exponent direction until the alignment with the joint-gradient flips, which generalizes RR's ``riding a ridge'' (following an EVec of the Hessian) while it is a descent direction.
            
            \textbf{Optimizing each branch:}
                For optimization in games, the stability properties of solutions can crucially depend on optimizer choice~\citep{gidel2018negative}.
                One should choose an optimizer suited to the problem.
                In our experiments, we use Learning with Opponent Learning Awareness (LOLA)~\citep{foerster2018learning} which can converge to periodic solutions and is attracted to high-welfare solutions in the IPD.
                In Section~\ref{sec:toy_opt} we contrast finding diverse solutions using LOLA with simultaneous SGD (simSGD) -- a method that works well for single-objective optimization, but cannot find periodic solutions.
                App. ~\ref{sec:related-work} summarizes other optimizer choices.

            \textbf{Re-branching:}
                In single-objective optimization in ML, our optimizer often finishes at a high-dimensional saddle~\citep{yao2020pyhessian}, which makes re-branching important.
                Specifically, we can re-branch at the saddle in negative EVec directions to try to find critical points with less negative EVecs.
                In our setup, we are interested in re-branching if our optimizer finishes at a point where EVals of the Jacobian of the fixed point operator $\jacFixedPointOp$ are greater than 1.
                These are directions where our optimizer will continue moving the parameters.
                \citep{lorraine2021complex} observed EVals of $\jacFixedPointOp$ larger than $1$ at the end of GAN training, indicating that we may want to re-branch for games in machine learning.
        %
        
        \subsection{Comparing GRR and RR}
            RR is a branching optimization search specifically for single-objective optimization -- which is less general than optimization in games -- so it can outperform GRR.
            E.g., non-conservative systems have more bifurcation types than conservative ones.
            If we are only concerned with saddle bifurcations, we can just find a saddle stationary point by minimizing the gradient norm.
            We know that this (relatively) easy-to-find stationary point lies on the separatrix.
            However, Hopf bifurcations are not necessarily near stationary points.
            Thus optimizing gradient norms does not work in general, while optimizing a Lyapunov exponent does (Fig.~\ref{fig:fig_new_problems_baseline}).
            
            While it might be overkill to find a separatrix with Lyapunov exponents in a single-objective setting, we take some lessons from GRR back to RR.
            It is useful to view RR as a method for finding bifurcations and branching across them.
            This motivates ways to sort between different stationary points to start at -- an open problem from RR.
            For example, using the point with the largest Kolmogorov-Sinai entropy~\citep{pesin1977characteristic}.
            At stationary points, this is simply the (negative) sum of negative EVals.
            Another limitation of RR is effectively estimating the most negative EVals of the Hessian.
            It is often simpler -- in computation and implementation -- to estimate the leading EVals of the Jacobian of the fixed point operator $\jacFixedPointOp$ instead of the most negative eigenvalues of the Hessian $\bothHess$.
            In Section~\ref{subsec:exp_coop} we show that our method saves Hessian-vector product evaluations when estimating EVecs in setups from RR.
            
    \begin{figure*}[t!]
            \vspace{-0.04\textheight}
            \centering
            \begin{tikzpicture}
                \centering
                \node (img){\includegraphics[trim={1.0cm 1.0cm 55.0cm 1.15cm},clip, width=.3\linewidth]{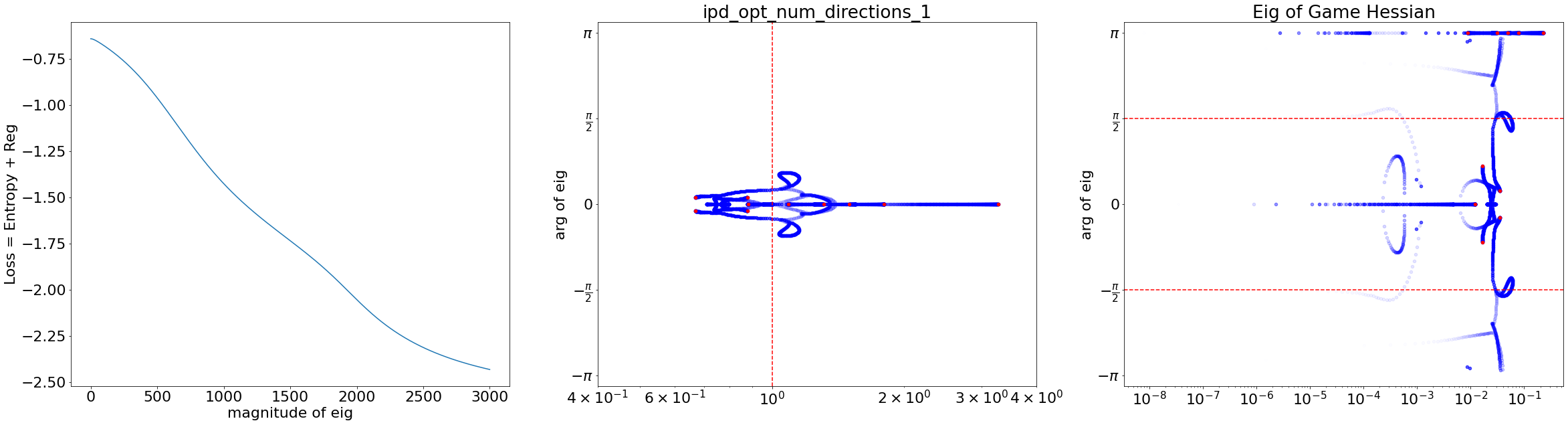}};
                \node[left=of img, node distance=0cm, rotate=90, xshift=1.8cm, yshift=-.7cm, font=\color{black}] {loss $\loss(\bothParam_0) = -\lyap_k^{max}(\bothParam_0)$};
                \node[below=of img, node distance=0cm, xshift=-.0cm, yshift=1.0cm,font=\color{black}] {optimization iteration};
                
                \node [right=of img, xshift=-.5cm](img2){\includegraphics[trim={30.0cm 1.0cm 27.5cm 1.15cm},clip, width=.29\linewidth]{images/experiments/ipd/tune_lyap_ipd_main.png}};
                \node[left=of img2, node distance=0cm, rotate=90, xshift=1.6cm, yshift=-.75cm, font=\color{black}] {argument of EVal $\arg(\eigval)$};
                \node[below=of img2, node distance=0cm, xshift=3.0cm, yshift=1.0cm,font=\color{black}] {log-norm of EVal $\log(|\eigval|)$};
                \node[above=of img2, node distance=0cm, xshift=-.0cm, yshift=-1.15cm,font=\color{black}] {EVals of Jac. of fixed point operator $\spectrum(\jacFixedPointOp)$};
                
                \node [right=of img2, xshift=-.75cm](img3){\includegraphics[trim={57.75cm 1.0cm 0.5cm 1.15cm},clip, width=.29\linewidth]{images/experiments/ipd/tune_lyap_ipd_main.png}};
                \node[above=of img3, node distance=0cm, xshift=-.0cm, yshift=-1.15cm,font=\color{black}] {EVals of game Hessian $\spectrum(\bothHess)$};
            \end{tikzpicture}
            \vspace{-0.02\textheight}
            \caption{
                We display gradient descent optimization on the $1$-step max Lyapunov exponent objective (Eq.~\ref{eq:1d_obj}) on the IPD.
                \textbf{\emph{Takeaway:}} We effectively reduce our loss and correspondingly raise the max EVal of $\jacFixedPointOp$.
                \emph{Left:} We display our loss -- i.e., the negative Lyapunov exponent objective-- as optimization progresses.
                \emph{Middle:} We visualize the spectrum of the Jacobian of our fixed point operator in log-polar coordinates as optimization progresses.
                The spectrum is shown with a scatter-plot in {\color{blue}blue}, with a progressively larger alpha at each iteration.
                The final spectrum is shown in {\color{red}red}.
                A vertical {\color{red}red} line is shown where the EVal norm equals 1, signifying the cutoff between (locally) convergent and divergent eigenspaces.
                We effectively maximize the norm of the largest EVal.
                \emph{Right:} We display the spectrum of the game Hessian.
                A horizontal {\color{red}red} line is shown where the real part of the EVal transitions from negative to positive, signifying the cutoff between (locally) convergent and divergent eigenspaces under gradient flow.
                Log-polar coordinates are required to see structure in the spectrum.
            }\label{fig:tune_lyap_ipd}
            \vspace{-0.01\textheight}
        \end{figure*}
    \newcommand{\numTrain}{\num{40000}}
    \newcommand{\numValid}{\num{10000}}
    \newcommand{\numTest}{\num{10000}}
    \newcommand{\batchSize}{\num{64}}
    \newcommand{\numEpochs}{\num{200}}
    \newcommand{\codeURL}{https://colab.research.google.com/drive/1ayW3g_wI6a_6C_deKwGV-MM1pyMxTewb?usp=sharing}
    \section{Experimental Setting}\label{subsec:new_problems}
        We experimentally investigate GRR on a variety of problems summarized in this section and described in detail in  App. ~\ref{sec:app_test_problems}.
        We chose these as they cover different types of dynamics and contain different kinds of bifurcations.
        Some are standard benchmarks, while others -- i.e., Random Subspaces -- are novel to this work.
        We also summarize our gradient computation for these problems.
        
        \textbf{Matching Pennies} is a simplified 2-parameter version of rock-paper-scissors.
            This problem's game Hessian has purely imaginary EVals unlike the small IPD, but only a single solution.
            Thus, by itself, is a poor fit for evaluating methods for a diversity of solutions, but nevertheless a useful test when probing GRR behavior.
        
        The \textbf{Iterated Prisoners' Dilemma (IPD)} is the discounted, infinitely iterated Prisoner's Dilemma~\citep{poundstone1993prisoner}.  
            Each agent's policy conditions on the actions in the prior time step, so there are 5 parameters for each agent -- the probability of cooperating initially and those given both agents' preceding actions.
            There are several different relevant equilibria in the IPD, including \emph{unconditional} defection (DD), leading to the worst-case joint outcome, and \emph{tit-for-tat} (TT), where agents initially cooperate, then copy the opponents' action (giving a higher reward).
            We turn the IPD into a differentiable game by calculating the analytical expected return as a function of the joint policy of the two agents.
        
        The \textbf{Small IPD} is
            a 2-parameter simplification of IPD, which has both DD and TT Nash equilibria, allowing us to visualize optimization difficulties from the full-scale IPD.
            However, the game Hessian has strictly real EVals, unlike the full-scale IPD.
        
        \textbf{Mixing Small IPD and Matching Pennies} interpolates between the Small IPD and Matching pennies games with an interpolation factor $\tau \in [0, 1]$.
            This problem has two solutions -- one where both players cooperate and one where both players select actions uniformly, with a Hopf bifurcation separating these.
        
        \textbf{Generative Adversarial Networks (GANs):}
            We use a setup from \citep{metz2016unrolled, balduzzi2018mechanics, letcher2018stable}, where the task is to learn a Gaussian mixture distribution using GANs.
            The data is sampled from a multimodal distribution to investigate the tendency to collapse on a subset of modes during training -- see App. Fig.~\ref{fig:gan_2d_samples} for the ground truth.
        
        \textbf{Random Subspace IPD/GAN:}
            To see how robustly we can find bifurcations with the exponents, we construct more complicated toy problems by taking higher-dimensional problems and optimizing in a random subspace.
            For each player, we select a random direction to optimize in, by sampling a vector $\mathbf{v}$ with entries from $U[0, 1]$ and normalizing it.
            Additionally, we select a random offset $\mathbf{b}$ from whatever an appropriate initialization is for the higher-dimensional problem.
            So, the first player controls $x$-coordinate and optimizes the loss $\outLoss(\mathbf{v}_{\outSymbol}x + \mathbf{b}_{\outSymbol}, \mathbf{v}_{\inSymbol}y + \mathbf{b}_{\inSymbol})$, while the second player controls the $y$-coordinate and optimizes $\inLoss(\mathbf{v}_{\outSymbol}x + \mathbf{b}_{\outSymbol}, \mathbf{v}_{\inSymbol}y + \mathbf{b}_{\inSymbol})$.
            
        \textbf{Single-objective problems:}
            We apply our method to find bifurcations on single-objective optimization problems.
            There are various relevant problems in machine learning, but we focus on comparisons with RRs EVec estimation in MNIST classification.
            
        \textbf{Optimizing the starting point objective:}
            To optimize these objectives, we use automatic differentiation libraries (like Jax~\citep{jax2018github} or PyTorch~\citep{paszke2017automatic}) to compute gradients through methods that calculate our Lyapunov exponent-inspired objectives.
            The scalability of this approach depends on the implementation of our exponent calculation, which can depend on estimating the top eigenvalues of the positive semi-definite (PSD) symmetric matrix in Eq.~\ref{eq:jac_sum}.
            In simple settings we can differentiate through the full spectrum calculation via \texttt{jax.linalg.numpy.eigh}; we investigate this on the toy experiment in Sec.~\ref{sec:toy_find_start} and the IPD in Sec.~\ref{sec:opt_lyap_ipd}.
            However, in ML, the matrix from Eq.~\ref{eq:jac_sum} is typically too large for the full spectrum.
            Directly estimating the top EVals with an iterative method allows us to (automatically) differentiate through them.
            We differentiate through \texttt{jax.numpy.linalg.eigh} in Fig.\ref{fig:fig_mix} and investigate using power iteration with Hessian-vector products in App. Fig.~\ref{fig:lyap_dir_mix}.
            
    \section{Experimental Results}
        First, in Sec.~\ref{sec:diagnostic} we use the diagnostic problems to demonstrate and ablate the key parts of our algorithms -- i.e. optimizer choice (Sec.~\ref{sec:toy_opt}), starting point selection (Sec.~\ref{sec:toy_find_start}), and branching (Sec. \ref{sec:toy_branch}).
        Next, in Sec.~\ref{sec:scaling} we scale GRR to \emph{large-scale problem settings} by (a) demonstrating that we improve RR's EVec estimation for neural network classifiers in Sec.~\ref{subsec:exp_coop}, and (b) calculating Lyapunov exponents for GANs in Sec.~\ref{subsec:exp_gan}.
        \vspace{-0.02\textheight}
        \subsection{Diagnostic Experiments}\label{sec:diagnostic}
        \subsubsection{Optimizer Choice}\label{sec:toy_opt}
            Here, we give a system with complex EVecs showing (a) the importance of selecting a convergent optimizer in GRR, and (b) an example task where RR cannot be applied.
            Fig.~\ref{fig:fig_new_problems_baseline} shows the phase portrait for baseline methods on our Mixed Problem.
            LOLA (and other game optimizers) can find both solutions, while na\"ively following the gradient always finds a single solution.
        
        \vspace{-0.3cm}
        \subsubsection{Starting Point Selection}\label{sec:toy_find_start}
\begin{figure}
                \vspace{-.05\textheight}
                \centering
                \begin{tikzpicture}
                    \centering
                    \node (img){\includegraphics[trim={.25cm .25cm 1.05cm 1.25cm},clip, width=.84\linewidth]{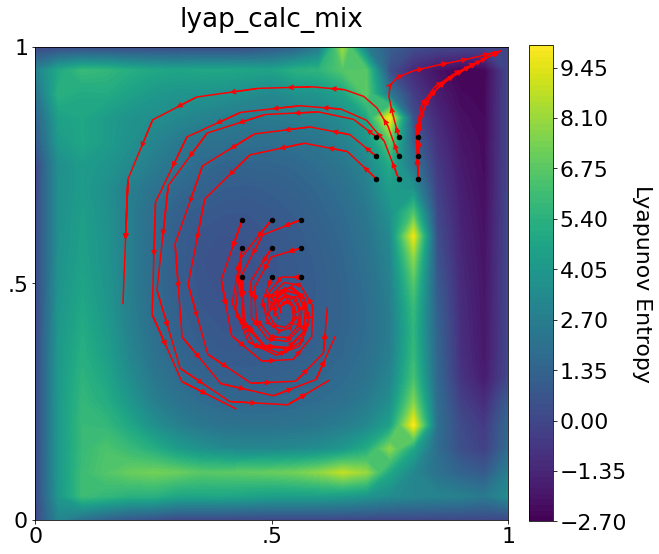}};
                    \node[left=of img, node distance=0cm, rotate=90, xshift=1.5cm, yshift=-.9cm, font=\color{black}] {Player 1 Strategy};
                    \node[right=of img, node distance=0cm, rotate=270, xshift=0cm, yshift=-.9cm, font=\color{black}] {Max $10$-step Lyapunov Exponent};
                    \node[above=of img, node distance=0cm, xshift=-.35cm, yshift=-1.3cm,font=\color{black}] {Calculating the exponent};
                    
                    \node [below=of img, yshift=1.25cm] (img2){\includegraphics[trim={.25cm .25cm 1.05cm 1.25cm},clip, width=.84\linewidth]{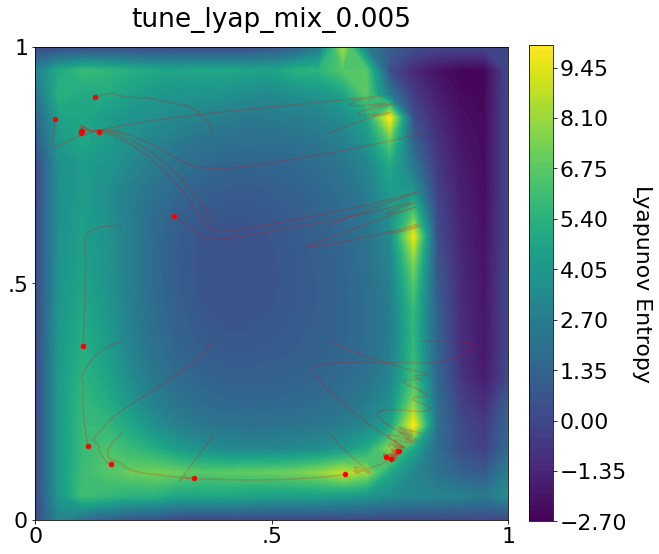}};
                    \node[left=of img2, node distance=0cm, rotate=90, xshift=1.5cm, yshift=-.9cm, font=\color{black}] {Player 1 Strategy};
                    \node[below=of img2, node distance=0cm, xshift=-.25cm, yshift=1.25cm,font=\color{black}] {Player 2 Strategy};
                    \node[below=of img2, node distance=0cm, xshift=-.25cm, yshift=.75cm,font=\color{black}] {Optimizing the exponent};
                \end{tikzpicture}
                \vspace{-.01\textheight}
                \caption{
                    Calculation and optimization of a max $10$-step Lyapunov exponent from Eq.~\ref{eq:lyap_max} on the mixed small IPD and Matching Pennies problem.
                    Gradient-based optimization on this objective effectively finds the bifurcation. 
                    \emph{Top:} We show a heatmap of the exponent, and visualize the calculation of each exponent in two regions.
                    This involves simulating $10$-step trajectories shown in red starting at the black points, then finding a direction that maximizes trajectory separation.
                    We use this exponent to find bifurcations -- in this case between the solution in the top right and the center.
                    \emph{Bottom:} We show optimization trajectories for gradient ascent on the exponent for a grid of initializations.
                    For a variety of starting points, the optimization procedure finds large exponent locations (final iterate shown with red circles).
                }\label{fig:fig_mix}
                \vspace{-0.4cm}
            \end{figure}
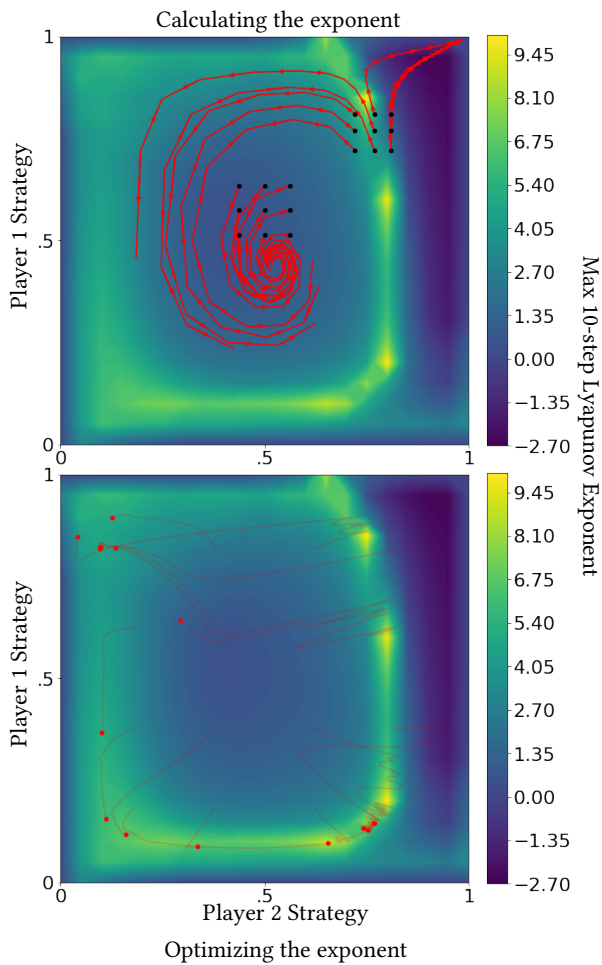

            Fig.~\ref{fig:fig_mix} shows the effect of optimizing the starting point for the max $10$-step Lyapunov exponent on the Mixed Problem.
            We find gradient-based optimization can find bifurcations.
            Next, Fig.~\ref{fig:lyap_dir_mix} contrasts different direction choices for the exponent calculation.
            We find that re-estimating the top EVecs at each iteration performs best, though the simple methods also work. 
            App. Fig.~\ref{fig:lyap_step_mix} shows the max $k$-step exponent for multiple numbers of steps $k$, showing that a moderate number of steps -- e.g., $10$ -- allows us to find bifurcations.
            App. Fig.~\ref{fig:entropy_mix} shows different Lyapunov exponent objectives, trying to guarantee trajectory divergence in multiple directions.
            We can find bifurcations while guaranteeing trajectory separation in every direction.
            
            \emph{Impact of inner optimizer choices on bifurcation structure:}
                App. Fig.~\ref{fig:lyap_opt_mix} contrasts the exponents for LOLA and simSGD, showing that we find optimizer-dependent bifurcations.
                App. Fig.~\ref{fig:lyap_opt_param_mix} investigates the impact of optimization-algorithm parameter choices on bifurcation structure.
                This shows that if the step size is too large, the optimizer does not converge, resulting in bifurcations between complicated limit cycle trajectories~\citep{strogatz2018nonlinear}, and making GRR difficult to apply.
            
            \emph{Starting points on single-objective problems:}
                App. Fig.~\ref{fig:lyap_coop_toy} investigates our algorithm in single-objective problems, showing that our method finds bifurcations in the same setup as RR.
                App. Fig.~\ref{fig:lyap_oneDim_toy} shows our method on the logistic map, giving intuition for our method on a canonical example for bifurcations.

            \vspace{-0.2cm}
            
        \subsubsection{Branching at Bifurcations}\label{sec:toy_branch}
            In Fig.~\ref{fig:branch_toy} we demonstrate branching at bifurcations to find multiple solutions to toy problems.
            This shows the branching process, and an explicit example where RR's starting point does not work, but GRR's does.
            The small IPD has a saddle bifurcation, while the Mixed Problem has a Hopf bifurcation.
        
        \vspace{-0.2cm}
        \subsubsection{A Range of Complicated Toy Problems}
\begin{figure}
                \vspace{-.05\textheight}
                \centering
                \begin{tikzpicture}
                    \centering
                    \node (img){\includegraphics[trim={.25cm .25cm 1.05cm 1.25cm},clip, width=.81\linewidth]{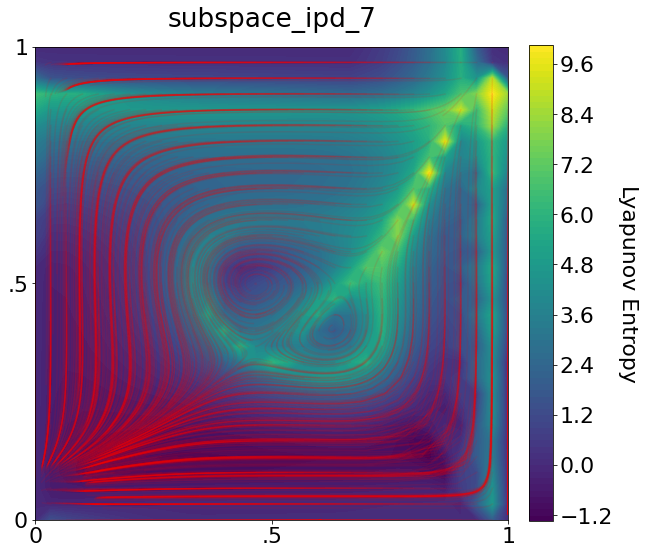}};
                    \node[left=of img, node distance=0cm, rotate=90, xshift=1.5cm, yshift=-.9cm, font=\color{black}] {Player 1 Strategy};
                    \node[right=of img, node distance=0cm, rotate=270, xshift=0cm, yshift=-.9cm, font=\color{black}] {Max $10$-step Lyapunov Exponent};
                    \node[below=of img, node distance=0cm, xshift=-.1cm,yshift=1.25cm,font=\color{black}] {Player 2 Strategy};
                    \node[above=of img, node distance=0cm, xshift=-.25cm, yshift=-1.2cm,font=\color{black}] {Random subspace test problems};
                    
                    \node [below=of img, xshift=.15cm, yshift=0.9cm] (img2){\includegraphics[trim={2.75cm 1.0cm 1.05cm 1.25cm},clip, width=.78\linewidth]{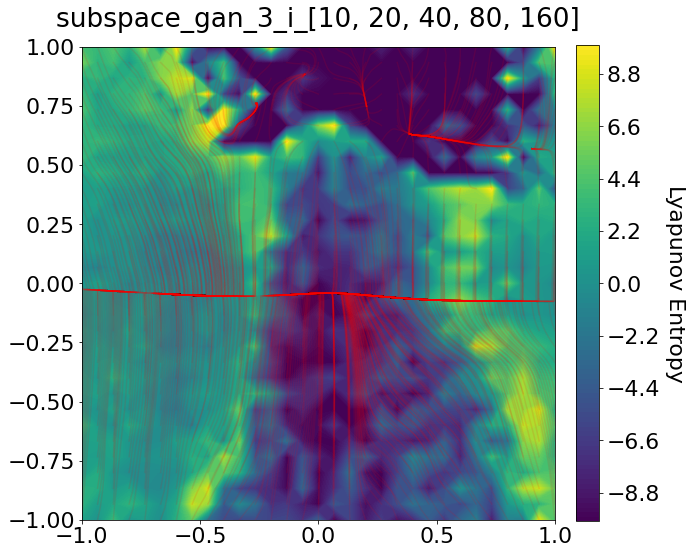}};
                    \node[left=of img2, node distance=0cm, rotate=90, xshift=1.15cm, yshift=-.75cm, font=\color{black}] {Discriminator};
                    \node[below=of img2, node distance=0cm, xshift=-.35cm, yshift=1.1cm,font=\color{black}] {Generator};
                \end{tikzpicture}
                \vspace{-.01\textheight}
                \caption{
                    We show the Lyapunov exponent heatmap (as in Fig.~\ref{fig:fig_mix}) on more complicated toy problems to see how robustly we can find different bifurcations.
                    The exponent peaks near where trajectories (shown in {\color{red}red}) separate, showing that we find various bifurcations.
                    See App. Fig.~\ref{fig:complicated_toy_moreSeeds} for other sampled subspaces.
                    Sec.~\ref{subsec:new_problems} describes how we construct these examples by taking higher-dimensional problems and optimizing them in a random subspace.
                    \emph{Top:} An IPD subspace with multiple Hopf bifurcations.
                    \emph{Bottom:} A GAN subspace with various bifurcations.
                }\label{fig:complicated_toy}
                \vspace{-0.5cm}
            \end{figure}
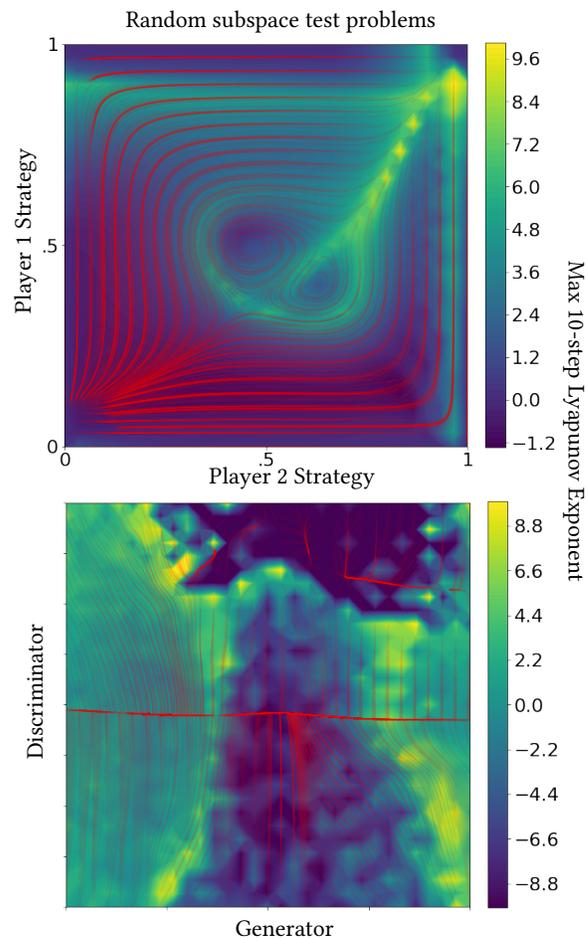

            In Fig.~\ref{fig:complicated_toy} we look at calculating Lyapunov exponents on toy problems with more complicated bifurcation structures.
            We create a variety of more complex toy problems by taking the high-dimensional IPD and GAN problems and selecting a random subspace to optimize in.
            We can effectively highlight bifurcations in this setup.

            \vspace{-0.2cm}
        \subsection{Scaling the Results}\label{sec:scaling}
        \subsubsection{Optimizing Lyapunov Exponents on IPD}\label{sec:opt_lyap_ipd}
            Here, we investigate our ability to use gradient-based optimizers on Lyapunov exponents in the IPD.
            Fig.~\ref{fig:tune_lyap_ipd} shows the feasibility of using gradient descent to tune the $1$-step max Lyapunov exponent.
            App. Fig.~\ref{fig:tune_lyap_ipd_moreDirections} optimizes an objective using multiple exponents, showing that we effectively optimize multiple exponents, which gives trajectory separation in multiple directions.
            App. Fig.~\ref{fig:tune_lyap_ipd_difObjectives} compares objectives using multiple exponents, showing that using the minimum of the top $n$ exponents gives trajectory separation in all $n$ directions, unlike the na\"ive choice of optimizing their sum.
            The sum of exponents finds solutions separating extremely fast in the top directions, while (slowly) converging in the bottom directions.
            In contrast, the min of the exponents does not allow convergence in the bottom directions.
            App. Fig.~\ref{fig:tune_lyap_ipd_difNumLyap} compares optimizing the $k$-step max Lyapunov exponent for variable $k$, showing that we effectively minimize multi-step exponents in higher-dimensional problems if required. 

        \subsubsection{GRR Applied to the IPD}\label{subsec:exp_ipd}
            Here, we use our method on the IPD, where existing methods have difficulty finding diverse solutions.
            There are two solution modes: ones where both agents end up defecting and cooperating respectively.
            Table~\ref{tab:results} compares our method to baselines of following gradients and LOLA, each run with random initializations.
            Our method finds both solutions modes, unlike existing approaches.
            We found that it was sufficient to use the max Lyapunov exponent as our objective, which only guarantees separation in 1 direction.
            Similarly, we found that it was sufficient to use a $1$-step or local Lyapunov exponent objective, though we may require more steps to find bifurcations in other problems.
            \begin{table*}[t!]
                \vspace{0.05\textheight}
            	\centering
            	\begin{tabular}{lllllll}
            		                                & Player 1 Loss & \multicolumn{5}{c}{Player 1 Strategy Distribution, [min, max]} \\
            		                                        \cmidrule(lr){3-7}
            		Search Strategy                  & {$\loss$ [min, max]} & {$p(C_0)$}           & {$p(C | CC)$}      & {$p(C | CD)$}      & {$p(C | DC)$}      & {$p(C | DD)$} \\
            		\midrule
            		GRR: tune max Lyap + top EVec branch + simSGD           & $[1.000, 2.000]$ & $[.003,.999]$  & $[.032,.999]$  & $[.004, .884]$  & $[.001, .912]$  & $[.000, .013]$\\
            		GRR: tune max Lyap + top EVec branch + LOLA           & $[1.000, 2.000]$ & $[.002,.999]$  & $[.063,.993]$  & $[.001, .910]$  & $[.000, .922]$  & $[.005, .103]$\\
            		20 Random init + SimSGD           & $[1.997, 1.998]$ & $[.043, .194]$  & $[.142, .480]$  & $[.041, .143]$  & $[.055, .134]$  & $[.001, .001]$\\
            		20 Random init + LOLA           & $[1.000,1.396]$ & $[.000, 1.00]$  & $[.093, 1.00]$  & $[.000, .966]$  & $[.057, 1.00]$  & $[.000, .947]$\\
            		1 Random init + top EVec branch      & $[2.000, 2.000]$ & $[.001, .003]$  & $[.027, .030]$  & $[.003, .007]$  & $[.008, .009]$  & $[.000, .000]$\\
            	\end{tabular}
            	\vspace{0.01\textheight}
            	\caption{
            	    We show strategies for finding diverse solutions in the iterated prisoner's dilemma (IPD).
            	    \textbf{\emph{Takeaway:}} Our method finds solutions at both loss modes, while existing approaches of using random initializations, then following the gradient or using LOLA do not find diverse solutions.
            	    The IPD has two solution modes -- i.e., solutions where both agents end up defecting with a loss of $2$ and where both agents end up cooperating with a loss of $1$ (like tit-for-tat).
            	    We assess which modes were found by showing (P)layer 1's strategy, which is the chance of (C)ooperating given both players' last action -- ex., $p(C|DC)$ is the chance if previously P1 defected and P2 cooperated.
            	    We compare GRR flavors with just following gradients via \ref{eq:multi_objective_gd_long} and LOLA~\citep{foerster2018learning} from random (init)ializations.
            	    We compare with 20 random inits because GRR follows at most 20 branches, and because we have 10 EVecs in either direction (+/-).
            	    GRR only branches in directions where EVals of the Jacobian of the fixed point operator are greater than 1 (i.e., trajectories locally diverge) as visualized in Fig.~\ref{fig:tune_lyap_ipd} (middle).
            	    We look at the impact of starting at an approximate bifurcation in GRR, by branching on the EVecs at a random init.
            	    If the max Lyapunov exponent is not tuned, then each branch finds the same solution.
            	}
            	\label{tab:results}
            	\vspace{-0.01\textheight}
            \end{table*}

        \subsubsection{Improving RR's EVec Estimation}\label{subsec:exp_coop}
            We investigate efficiently finding the most negative EVecs in RR by estimating the largest EVecs of the Jacobian of our fixed point operator.
            We measure efficiency by comparing the number of Hessian-vector product (HVP) evaluations because HVP evaluations dominate the cost of EVec estimation here.
            Table~\ref{tab:hvp_evals} shows how many HVP evaluations we require to reach different MNIST classifier accuracies by following EVecs.
            Our method can more efficiently use HVP evaluations than the RR method because we do not need to repeatedly re-estimate the most negative EVal.
            
            We stress that this problem is not designed to train a single, strong classifier; it is easy to simply train our network by following the gradient to $100\%$ train accuracy.
            This problem was selected from RR's experiments because it requires us to accurately and efficiently estimate negative EVecs many times.
            A downstream use of this is training an ensemble of classifiers for generalization. 
            
            \begin{table}
            	\centering
            	\begin{tabular}{lll}
            		    &\multicolumn{2}{c}{MNIST Accuracy} \\
            		\cmidrule(lr){2-3}
            		\# HVP Evaluations & Our method & Method from RR\\
            		\midrule
            		\num{10000} & $\num{19}\%({\color{green}+8\%})$ & $\num{11}\%$ \\
            		\num{100000} & $\num{89}\%({\color{green}+6\%})$ & $\num{83}\%$ \\
            		\num{1000000} & $\num{93}\%({\color{green}+2\%})$ & $\num{91}\%$ \\
            	\end{tabular}
            	\vspace{0.01\textheight}
            	\caption{
            	    We show how many HVP evaluations we require to reach different MNIST classifier accuracies by following EVecs, repeating the exp. in RR's Fig.~4.
            	    This experiment is not designed to train a single strong classifier, but to test our ability to efficiently follow negative EVecs -- see Sec.~\ref{subsec:exp_coop}.
            	 }
            	\label{tab:hvp_evals}
            	\vspace{-0.01\textheight}
            \end{table}
        
        \subsubsection{Calculating Lyapunov Exponents for GANs}\label{subsec:exp_gan}
            Here, we investigate scaling our exponent calculations to machine learning models where the (game) Hessian is so large we cannot materialize it and can only use Hessian-vector products.
            Specifically, we use the GAN described in Section~\ref{subsec:new_problems}.
            We look at calculating our exponent for various hyperparameters and random re-starts.
            We evaluate the quality of using our exponent to find diverse solutions, by calculating the log-probability of samples from an ensemble of GANs from the top 5 optimization branches.
            Table~\ref{tab:gan_lyap_calc} shows the mean and standard deviation (over 10 random restarts) of the max $10$-step Lyapunov exponent and the resulting ensemble's log-probability.
            Each GAN was trained for $\num{10000}$ updates, so evaluating each ensemble cost approximately $\num{50000}$ evaluations of both players' gradients.
            In contrast, each exponent cost less than $\num{1000}$ evaluations of both gradients to compute.
            
            This shows that we effectively scale our exponent calculation to larger models of interest from machine learning, and find that a large (mean) exponent aligns with regions where we can branch to train the strongest ensemble of GANs.
            
            \begin{table}
                \vspace{-0.03\textheight}
            	\centering
            	\begin{tabular}{lll}
            		Init scale, step size & Max Lyap Coeff & Ensemble log-prob\\
            		\midrule
            		\num{0.001}, 1.0 & $\num{0.952} \pm  \num{0.834}$& $\num{-16342} \pm \num{817}$ \\
            		\num{0.1}, 1.0 & $\num{6.485} \pm  \num{1.155}$& $\num{-13691} \pm \num{1317}$ \\
            		\num{10.0}, 1.0 & $\num{0.053} \pm  \num{0.128}$ & $\num{-46659} \pm \num{26793}$ \\
            		\num{0.001}, 0.1 & $\num{0.849} \pm  \num{0.765}$ & $\num{-12321} \pm \num{126}$ \\
            		\num{0.1}, 0.1 & $\num{6.571} \pm  \num{0.953}$ & $\num{-10846} \pm \num{256}$\\
            		\num{10.0}, 0.1 & $\num{-0.012} \pm  \num{0.014}$ & $\num{-23459} \pm \num{12693}$ \\
            	\end{tabular}
            	\vspace{0.01\textheight}
            	\caption{
            	    We display the mean and standard deviation (over 10 random restarts) of the max $10$-step Lyapunov exponent and the log-probability of an ensemble of $5$ GANs obtained by branching in the top $5$ directions at the initialization.
            	    We show that the better performing ensembles also have higher Lyapunov coefficients as well as demonstrating that our exponent calculation is scalable to larger problems.
            	    The best GANs log-prob. from the best ensemble was $\num{-12861} \pm \num{356}$, which is worse than the ensemble's performance of $\num{-10846} \pm \num{256}$.
            	    This indicates that each GAN may be learning a different part of the data distribution (samples in App. Fig.~\ref{fig:gan_2d_samples}).
            	 }
            	\label{tab:gan_lyap_calc}
            	\vspace{-0.03\textheight}
            \end{table}
            
\vspace{-0.3cm}
    \section{Conclusion}\label{sec:conclusion}
        In this paper we introduced Generalized Ridge Rider, an extension of the Ridge Rider algorithm to settings with multiple losses.
        We showed that, in these settings, a broader class of bifurcation points needs to be considered, and that GRR indeed discovers them in a variety of problems.
        Experimentally, we isolate each component of GRR demonstrating their effectiveness, and show that -- in contrast to baseline methods -- GRR obtains a diversity of qualitatively different solutions in multi-agent settings such as the iterated prisoner's dilemma.
        We also provide empirical justification for our method by using tools from the dynamical systems literature, allowing us to find arbitrary bifurcations.
        This hints at a multitude of approaches and tools from dynamical systems, that can be used for understanding game dynamics and learning diversity.

    \newpage

\begin{acks}
    Resources used in preparing this research were provided, in part, by the Province of Ontario, the Government of Canada through CIFAR, and companies sponsoring the Vector Institute.
    Work, in part, was done while Jonathan Lorraine and Jack Parker-Holder were on internship at FAIR.
    We would also like to thank C. Daniel Freeman, H\'erve J\'egou, Noam Brown, and David Acuna for feedback on this work and acknowledge the Python community ~\citep{van1995python, oliphant2007python} for developing the tools that enabled this work, including numpy 
    ~\citep{oliphant2006guide, van2011numpy, harris2020array}, Matplotlib~\citep{hunter2007matplotlib} and SciPy~\citep{jones2001scipy}.
\end{acks}

\typeout{}
{\small
\vspace{-0.0\textheight}
\bibliography{references}
}
\newpage

\appendix

\begin{table*}[htbp]\caption{Notation}
        \begin{center}
            \begin{tabular}{c c}
                \toprule
                RR & Ridge Rider~\citep{parker2020ridge}\\
                IPD & Iterated Prisoners' Dilemma\\
                GAN & Generative Adversarial Network~\citep{goodfellow2014generative}\\
                LOLA & Learning with opponent learning awareness~\citep{foerster2018learning}\\
                EVec, EVal & Shorthand for Eigenvector or Eigenvalue\\
                SGD & Stochastic Gradient Descent\\
                SimSGD & Simultaneous SGD\\
                $\defeq$ & Defined to be equal to\\
                $x, y, z, \dots \in \mathbb{C}$ & Scalars\\
                $\boldsymbol{x}, \boldsymbol{y}, \boldsymbol{z}, \dots \in \mathbb{C}^{n}$ & Vectors\\
                $\boldsymbol{X}, \boldsymbol{Y}, \boldsymbol{Z}, \dots \in \mathbb{C}^{n \times n}$ & Matrices\\
                $\boldsymbol{X}^\transpose$ & The transpose of matrix $\boldsymbol{X}$\\
                $\identity$ & The identity matrix\\
                $\Re(z), \Im(z)$ & The real or imaginary component of $z \in \mathbb{C}$\\
                $i$ & The imaginary unit. $z \in \mathbb{C} \implies z = \Re(z) + i \Im(z)$\\
                $\conj{z}$ & The complex conjugate of $z \in \mathbb{C}$\\
                $|z| \defeq \sqrt{z \conj{z}}$ & The magnitude or modulus of $z \in \mathbb{C}$ \\
                $\arg(z)$ & The argument or phase of $z \in \mathbb{C} \implies z = |z| \exp(i\arg(z))$ \\
                $\outSymbol, \inSymbol$ & A symbol for the outer/inner players\\
                $\outDim, \inDim \in \mathbb{N}$ & The number of weights for the outer/inner players\\
                $\paramSymbol$ & A symbol for the parameters or weights of a player\\
                $\outParam \in \mathbb{R}^{\outDim}, \inParam \in \mathbb{R}^{\inDim}$ & The outer/inner parameters or weights\\
                $\loss: \mathbb{R}^{n} \to \mathbb{R}$ & A symbol for a loss\\
                $\outLoss(\outParam, \inParam), \inLoss(\outParam, \inParam)$ & The outer/inner losses -- $\mathbb{R}^{\outDim + \inDim} \mapsto \mathbb{R}$ \\
                $\outGrad(\outParam, \inParam), \inGrad(\outParam, \inParam)$ & Gradient of outer/inner losses w.r.t. their weights in $\mathbb{R}^{\outDim/\inDim}$ \\
                $\inParam^*\!(\outParam\!) \!\defeq\! \argmin\limits_{\inParam} \!\!\inLoss\!(\!\outParam \!, \!\inParam\!)$&The best-response of the inner player to the outer player\\
                $\outLoss^*(\outParam) \!\defeq\! \outLoss\!(\!\outParam \!, \!\inParam^*(\outParam)\!)$ & The outer loss with a best-responding inner player\\
                $\outParam^* \!\defeq\! \argmin\limits_{\outParam} \outLoss^*(\outParam\!) $ & Outer optimal weights with a best-responding inner player\\
                $\bothDim \defeq \outDim + \inDim$ & The combined number of weights for both players\\
                $\bothParam \defeq [\outParam, \inParam] \in \mathbb{R}^{\bothDim}$ & A concatenation of the outer/inner weights\\
                $\bothGrad(\bothParam) \!\defeq\! [\outGrad(\bothParam), \inGrad(\bothParam)] \in \mathbb{R}^{\bothDim}$ & A concatenation of the outer/inner gradients\\
                $\bothParam^{\num{0}} = [\outParam^{\num{0}}, \inParam^{\num{0}}] \in \mathbb{R}^{\bothDim}$ & The initial parameter values\\
                $j$ & An iteration number\\
                $\bothGrad^j \defeq \bothGrad(\bothParam^j) \in \mathbb{R}^{\bothDim}$ & The joint-gradient vector field at weights $\bothParam^j$\\
                $\nabla_{\bothParam} \bothGrad^j \defeq \nabla_{\bothParam}\bothGrad |_{\bothParam^j} \in \mathbb{R}^{\bothDim \times \bothDim}$ & The Jacobian of the joint-gradient $\bothGrad$ at weights $\bothParam^j$\\
                $\bothHess$ & The game Hessian \\
                $\lr \in \mathbb{C}$ & The step size or learning rate\\
                $\eigval \in \mathbb{C}, e$ & Notation for an arbitrary Eval \\
                $\spectrum(\boldsymbol{M}) \in \mathbb{C}^{n}$ & The spectrum -- or set of eigenvalues -- of $\boldsymbol{M} \in \mathbb{R}^{n \times n}$ \\
                $\spectralRadius(\boldsymbol{M}) \!\defeq\! \max_{z \in \spectrum(\boldsymbol{M})} |z|$ & The spectral radius in $\mathbb{R}^{+}$ of $\boldsymbol{M} \in \mathbb{R}^{n \times n}$ \\
                $\fixedPointOp(\bothParam)$ & Fixed point operator for our optimization \\
                $\jacFixedPointOp$ & The Jacobian of the fixed point operator \\
                $\displacement$ & A displacement for a Lyapunov exponent \\
                $\gamma_j(\bothParam_0, \displacement) \defeq \log(\displacement^\transpose (\jacFixedPointOp^{j}(\bothParam_0))^\transpose \jacFixedPointOp^{j}(\bothParam_0) \displacement)$ & A Lyapunov term for a Lyapunov exponent \\
                $\lyap_k(\bothParam_0, \displacement) = \frac{1}{k}\smash{\sum_{j = 0}^{k}} \gamma_j(\bothParam_0, \displacement)$ & A $k$-step Lyapunov exponent \\
                $\lyap_k^{max}(\bothParam_0) = \max_{\displacement, \|\displacement\| = 1} \lyap_k(\bothParam_0, \displacement)$ & The max $k$-step Lyapunov exponent \\
                $\loss(\bothParam_0)$ & A starting point loss using Lyapunov exponents \\
                $\loss_{n}^{\textnormal{sum}}(\bothParam_0)$ & A loss using the sum of top $n$ exponents \\
                $\loss_{n}^{\textnormal{min}}(\bothParam_0)$ & A loss using the min of top $n$ exponents \\
                \bottomrule
            \end{tabular}
        \end{center}
        \label{tab:TableOfNotation}
    \end{table*}
    
    \section{Related Work}\label{sec:related-work}
        \textbf{Diversity in machine learning. }
            Finding diverse solutions is often desirable in machine learning, for example improving performance for model ensembles \cite{1990_ensembles}, with canonical approaches directly optimizing for negative correlation amongst model predictions \cite{NCL}.
            In recent times these ideas have begun to re-emerge, improving ensemble performance \cite{dibs, lit2020, marSra16}, robustness \cite{pang2019improving, cully2015robots} and boosting exploration in reinforcement learning \cite{Lehman08exploitingopen, eysenbach2018diversity, parkerholder2020effective, qdnature}.
        
            Many of these approaches seek to find diverse solutions by following gradients of an altered, typically multi-objective loss function.
            By contrast, the recent \emph{Ridge Rider} (RR, \cite{parker2020ridge}) algorithm searches for diverse solutions by following EVecs of the Hessian with respect to the original loss function, producing orthogonal (loss reducing) search directions.
        
        \textbf{Finding solutions in games. }
            There are first-order methods for finding solutions in games including extragradient~\cite{korpelevich1976extragradient,azizian2020tight}, optimistic gradient~\cite{rakhlin2013optimization,daskalakis2017training}, negative momentum~\citep{gidel2018negative}, complex momentum~\citep{lorraine2021complex}, and iterate averaging~\citep{gidel2018variational}.
            There are also higher-order methods like consensus optimization~\cite{mescheder2017numerics}, symplectic gradient adjustment (SGA)~\cite{letcher2019differentiable}, local symplectic surgery (LSS)~\cite{mazumdar2019finding}, competitive gradient descent (CGD)~\cite{schafer2019competitive}.
            Recently, \citep{vicol2020implicit} looked at the effects of overparameterization while doing online learning in games.
        
        \textbf{Diversity in multi-agent RL. }
            In recent times a series of works have explore the benefit of diversity in both competitive \cite{balduzzi2019open, garnelo2021pick, vinyals2019grandmaster} and cooperative \cite{yang2020multi, lupu2021trajectory} multi-agent RL.
            However, once again these approaches all consider augmented loss functions. Instead, we take inspiration from RR and extend it to the multi-agent setting.
        
        \textbf{Tree searches in MDP. }
            Tree searches are a classic technique for planning and control in discrete Markov Decision Processes (MDP).
            In these settings, the locations to branch from, and the potential branch choices are already given, and all that remains is choosing which to explore.
            Much interest has been developed in performing these searches ranging from depth/breadth first search, A* \citep{Hart1968}, to more sophisticated methods such as Monte Carlo Tree Search \citep{abramson2014expected} potentially with learned value functions \citep{silver2016mastering}.
            Searching in continuous action spaces has been explored in \citep{moerland2018a0c, kim2020monte, mao2020poly}.
            All of these methods search in the action space of the agents. In contrast, we seek to employ these search methods to find the parameters of the agents.
        
            Rapidly-exploring random tree is an algorithm that builds a space-filling tree to cover a given continuous optimization space \citep{lavalle1998rapidly, lavalle2001randomized}. Unlike RR-based methods, the branching points are not determined by the underlying properties of the loss landscape (e.g. only at saddles, or bifurcation). This technique is typically used in robotic motion planning \citep{rodriguez2006obstacle}.
        
        \textbf{Branching in evolutionary optimization. }
            Designing optimization algorithms in a multi-modal loss landscape has been the focus of the  evolutionary optimization community \citep{singh2006comparison, wong2015evolutionary}.
            These methods implicitly build a tree over candidate solutions.
            Evolutionary algorithms specifically to encourage diversity have been explored in the context of multi player games \citep{balduzzi2019open, vinyals2019grandmaster, arulkumaran2019alphastar} by encouraging populations of agents to learn different strategies which perform well against other agents resulting in a growing collection of strategies.
        
        \textbf{Bifurcations. }
            This work builds on the prior workshop submission of ~\citep{lorraine2021using}.
            The works of \citep{zeeman1980population, yang2018bifurcation, chotibut2020route, piliouras2020catastrophe, leonardos2020exploration, bielawski2021follow} leverage the bifurcations for learning in games.
            \citep{yang2018bifurcation} introduces two mechanisms -- hysteresis and optimal control mechanisms - to control equilibrium selection, with the goal of improving social welfare.
            We do not focus on maximizing social welfare and instead focus just on finding multiple solutions.
            \citep{chotibut2020route} studies the properties of bifurcations in routing games, showing various analyses for the social cost.
            \citep{piliouras2020catastrophe} combines catastrophe theory with mechanism design to destabilize inefficient solutions.
            \citep{leonardos2020exploration} studies how changing an exploration parameter in multi-agent learning can affect equilibrium selection.
            Our work focuses on connecting bifurcations with finding diverse solutions in optimization with Ridge Rider, and finding those bifurcations - potentially with gradient-based methods.
        
        \textbf{Lyapunov exponents. }
            The works of \citep{cheung2019vortices, cheung2020chaosb, sato2002chaos} use Lyapunov exponents when learning in games.
            \citep{cheung2019vortices} shows that some learning algorithms are Lyapunov chaotic in payoff space.
            Our work focuses on using Lyapunov exponents variants to identify bifurcations (which are used to find diverse solutions in optimization), and potentially using optimization methods on the exponent.

        \textbf{Separatrices. }
            The works of ~\citep{panageas2016average, zhang2015equilibrium, nagarajan2020chaos} look at computing separatrices in games which are leveraged for various purposes.
            \citep{nagarajan2020chaos} finds the shape of separatrices using invariant functions with online learning dynamics.

    \section{Proposed Algorithms}
        \subsection{Branching Optimization Tree Searches}
            \begin{algorithm}[H]
                \caption{Branching Optimization Tree Search-- important changes from RR to GRR in {\color{red}red} components}
                \begin{algorithmic}[1]
                    \STATE \textbf{Input:}  ${\color{red}\mathrm{FindStartingPoint}}$, $\mathrm{ChooseBranch}$, $\color{red}\mathrm{SplitBranch}$,
                    \STATE \qquad \quad $\color{red}\mathrm{EndRide}$, ${\color{red}\mathrm{Optimize}}, \mathrm{VerifySolution}$
                    \STATE \qquad \quad $\mathrm{ContinueBranching}$
                    \STATE Select optimization parameters $\alpha$
                    \STATE Find starting parameters $\bothParam^{start} = {\color{red}\mathrm{FindStartingPoint}(\alpha)}$
                    \STATE Initialize a branch $\psi^{init} = \mathrm{InitBranch}(\bothParam^{start}, \alpha)$
                    \STATE Initialize the set of branches $\mathcal{B} = {\color{red}\mathrm{SplitBranch}}(\psi^{init})$
                    \STATE Initialize the set of solutions $\mathcal{S} = \emptyset$
                    \WHILE{$\textnormal{Branches } \mathcal{B} \textnormal{ non-empty}$}
                      \STATE $\psi, \mathcal{B}  = \mathrm{ChooseBranch}(\mathcal{B})$
                      \STATE $\bothParam^* = {\color{red}\mathrm{Optimize}}(\psi.\bothParam, \psi.\alpha)$ \# Optimize our parameters
                      \IF{$\mathrm{VerifySolution}(\bothParam^{*})$} 
                           \STATE $\mathcal{S} = \mathcal{S} \cup \{ \bothParam^{*} \}$
                      \ENDIF
                      \STATE Make new branch to split $\psi' = \mathrm{copy}(\psi)$
                      \STATE Store the optimized parameters $\psi'.parameters = \bothParam^*$
                      \IF{$\mathrm{ContinueBranching}(\psi')$} 
                           \STATE $\mathcal{B} = \mathcal{B} \cup {\color{red}\mathrm{SplitBranch}}(\psi')$
                      \ENDIF
                    \ENDWHILE
                    \STATE \textbf{return} $\mathcal{S}$
                \end{algorithmic}
                \label{algorithm:branching_tree}
            \end{algorithm}
    \section{Experiments}
        \subsection{Test Problems}\label{sec:app_test_problems}
            \textbf{Iterated Prisoners' Dilemma (IPD)}: This game is an infinite sequence of the Prisoner's Dilemma, where the future payoff is discounted by a factor $\gamma \in [0, 1)$.
            In other words, the sequences of rewards $r_j$ for each player summed via $\sum_{j=0}^{\infty} \gamma^j r_j$.
            Each agent is conditioned on the actions in the prior state ($s$).
            Thus, there are 5 parameters for each agent $i$: $P^{i}(C | s)$ is the probability of cooperating at start state $s_0 = \emptyset$ or state $s_t = (a^{1}_{t-1}, a^{2}_{t-1})$ for $t > 0$.
            There are two Nash equilibria which we are interested in: Defect-Defect (DD) where agents are selfish (resulting in poor reward), and \emph{tit-for-tat} (TT) where agents initially cooperate, then copy the opponents' action (resulting in higher reward).
            
            \textbf{Small IPD}: This is a 2-parameter 
             simplification of IPD, which allows DD and TT Nash equilibria.
            We fix the strategy if our opponent defects, to defect with high probability.
            We also constrain the probability of cooperating to only depend on if the opponent cooperates, and in the initial state we assume our opponent cooperated.
            This game allows us to visualize some of the optimization difficulties for the full-scale IPD, however, the game Hessian has strictly real EVals unlike the full-scale IPD.
            See Fig~\ref{fig:fig_new_problems_baseline} top for a visualization of the strategy space.
            
            \textbf{Matching Pennies}: This is a simplified 2-parameter version of rock-paper-scissors, where each player selects Cooperate or Defect.
            This game has a Nash equilibrium where each player selects its action with uniform probability.
            Notably, this problem's game Hessian has purely imaginary EVals, so following the gradient does not converge to solutions and we need a method for learning in games like LOLA.
            Also, this game only has a single solution; thus it is a poor fit for evaluating RR, which finds a diversity of solutions.
            See Fig~\ref{fig:fig_new_problems_baseline}, bottom for a visualization of the strategy space.
            
\begin{figure*}
                \centering
                \begin{tikzpicture}
                    \centering
                    \node (img){\includegraphics[trim={.25cm .25cm 3.5cm .25cm},clip, width=.31\linewidth]{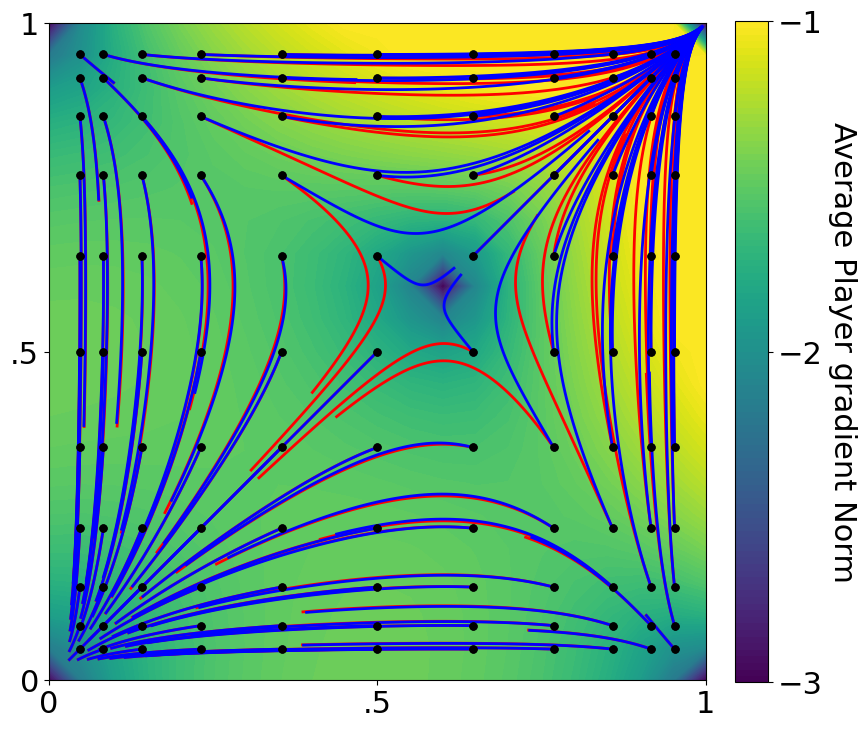}};
                    \node[left=of img, node distance=0cm, rotate=90, xshift=1.5cm, yshift=-.9cm, font=\color{black}] {Player 1 Strategy};
                    \node[below=of img, node distance=0cm, xshift=-.0cm, yshift=1.25cm,font=\color{black}] {Player 2 Strategy};
                    \node[above=of img, node distance=0cm, xshift=-.0cm, yshift=-1.25cm,font=\color{black}] {Small IPD};
                    
                    \node [right=of img, xshift=-1.25cm] (img2){\includegraphics[trim={1.0cm .25cm 3.5cm .25cm},clip, width=.295\linewidth]{images/standard_methods/small_mixture_grid_standard_mix=0.25.png}};
                    \node[below=of img2, node distance=0cm, xshift=-.0cm, yshift=1.25cm,font=\color{black}] {Player 2 Strategy};
                    \node[above=of img2, node distance=0cm, xshift=-.0cm, yshift=-1.25cm,font=\color{black}] {Mixed Problem};
                    
                    \node [right=of img2, xshift=-1.25cm] (img3){\includegraphics[trim={1.0cm .25cm 1.05cm .25cm},clip, width=.335\linewidth]{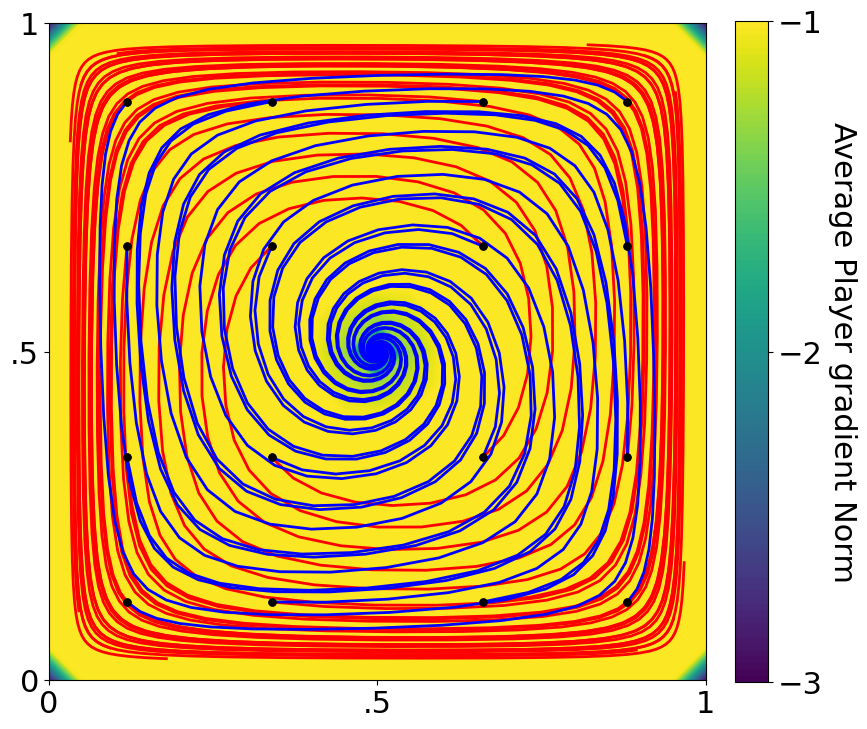}};
                    \node[right=of img3, node distance=0cm, rotate=270, xshift=-2.75cm, yshift=-.9cm, font=\color{black}] {Joint-player grad. log-norm $\log(\|\bothGrad\|)$};
                    \node[below=of img3, node distance=0cm, xshift=-.35cm, yshift=1.25cm,font=\color{black}] {Player 2 Strategy};
                    \node[above=of img3, node distance=0cm, xshift=-.35cm, yshift=-1.25cm,font=\color{black}] {Matching Pennies};
                \end{tikzpicture}
                \caption{
                    This shows the phase portrait for two standard optimization algorithms on a range of problems.
                    Following the gradient is shown in {\color{red}red}, while LOLA -- a method for learning in games -- is shown in {\color{blue}blue}.
                    \emph{Left}: The small IPD, which has solutions in the top right and bottom left.
                    \emph{Middle}: Matching Pennies, which has a single solution in the middle.
                    Following the gradient does not find this solution because it has imaginary EVals, so we must use a method like LOLA.
                    \emph{Right}: A mixture of small IPD and Matching Pennies.
                    Following the gradient only finds the solution in the top right, because the center solution has imaginary EVals;
                    LOLA can find either solution.
                    \textbf{Takeaway}: The mixture game has a range of phenomena, including an imaginary EVal solution, a real EVal solution, and a Hopf bifurcation.
                    We may want to use a method for learning in games, so we can robustly converge to different solutions.
                }\label{fig:fig_new_problems_baseline_full}
            \end{figure*}

            \textbf{Mixing Small IPD and Matching Pennies}: This game interpolates between the Small IPD and Matching Pennies games, with the loss for player $j$:
            \begin{equation}
                \mathcal{L}_{mix, P_{j}, \tau} = \tau \mathcal{L}_{smallIPD, P_{j}} + (1 - \tau)\mathcal{L}_{matchingPennies, P_{j}}
            \end{equation}.
            This problem has two solutions -- one where both players cooperate, and one where both players select actions uniformly.
            The uniform action solution has imaginary EVals, so it is only stable under a method for learning in games, while the both cooperate solution has real EVals.
            There is a Hopf bifurcation separating these solutions.
            See Fig~\ref{fig:fig_new_problems_baseline} for standard methods on this problem and Appendix Fig.~\ref{fig:fig_new_problems_baseline_full} to contrast this problem with Small IPD or Matching Pennies.
        
            \begin{figure*}
                \centering
                \begin{tikzpicture}
                    \centering
                    \node (img11){\includegraphics[trim={.25cm .25cm 5.0cm 1.25cm},clip, width=.22\linewidth]{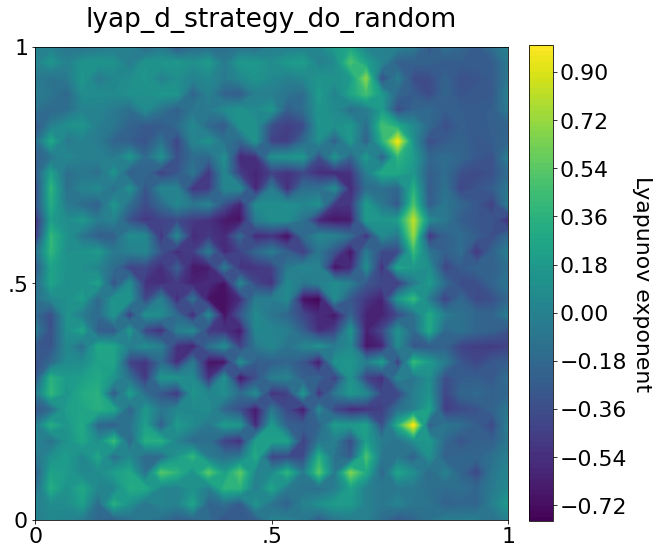}};
                        \node[left=of img11, node distance=0cm, rotate=90, xshift=1.5cm, yshift=-.9cm, font=\color{black}] {Player 1 Strategy};
                        \node[below=of img11, node distance=0cm, xshift=6.5cm,yshift=1.0cm,font=\color{black}] {Player 2 Strategy};
                        \node[above=of img11, node distance=0cm, xshift=6.5cm,yshift=-.75cm,font=\color{black}] {Different direction estimation strategies in $10$-step Lyapunov exponent calculation};
                        \node[above=of img11, node distance=0cm, xshift=-.15cm, yshift=-1.2cm,font=\color{black}] {Random Direction};
                    
                    \node [right=of img11, xshift=-1cm](img12){\includegraphics[trim={.25cm .25cm 5.2cm 1.25cm},clip, width=.22\linewidth]{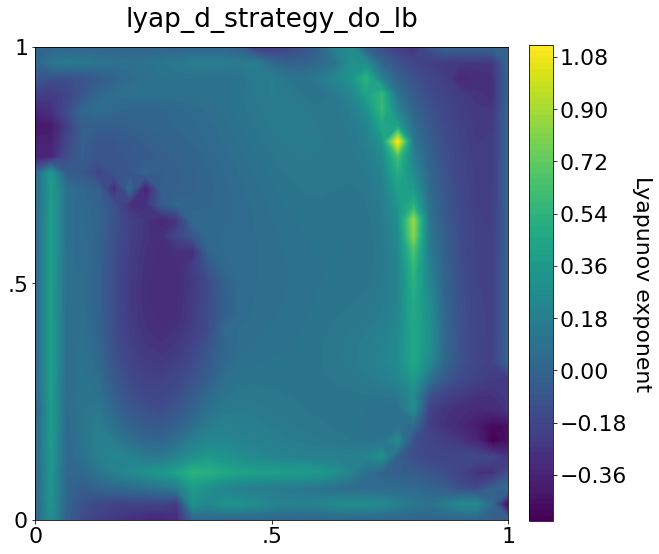}};
                        \node[above=of img12, node distance=0cm, xshift=-.1cm, yshift=-1.2cm,font=\color{black}] {Power Iteration, first step};
                    
                    \node [right=of img12, xshift=-1cm](img13){\includegraphics[trim={.25cm .25cm 5.0cm 1.25cm},clip, width=.22\linewidth]{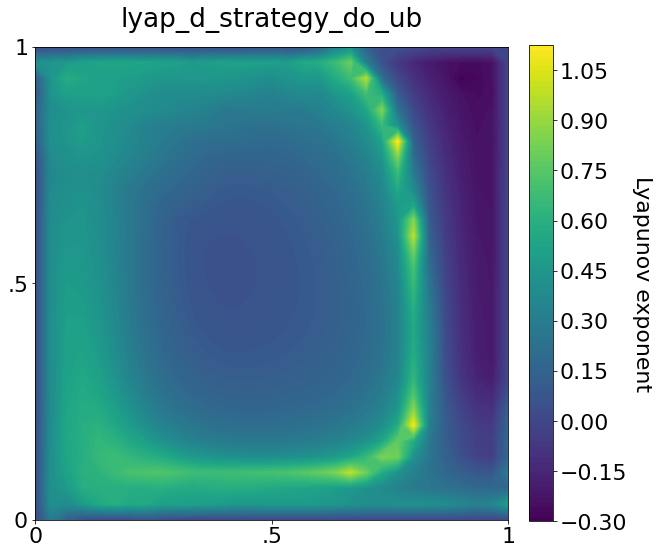}};
                        \node[above=of img13, node distance=0cm, xshift=-.05cm, yshift=-1.2cm,font=\color{black}] {Power Iteration, every step};
                    
                    \node [right=of img13, xshift=-1cm](img14){\includegraphics[trim={.25cm .25cm 1.05cm 1.25cm},clip, width=.26\linewidth]{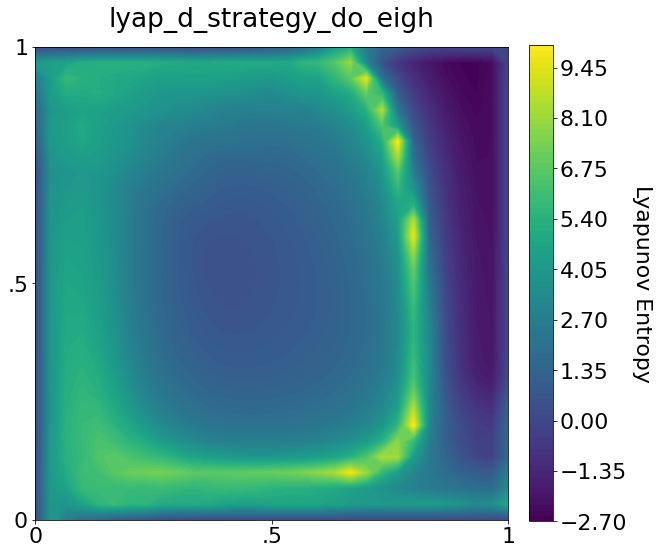}};
                        \node[above=of img14, node distance=0cm, xshift=-.25cm, yshift=-1.2cm,font=\color{black}] {\small{\texttt{jax.numpy.linalg.eigh}} every step};
                    \node[right=of img14, node distance=0cm, rotate=270, xshift=-.25cm, xshift=-2cm, yshift=-.9cm, font=\color{black}] {Max $10$-step Lyapunov Exponent};
                \end{tikzpicture}
                \vspace{-.01\textheight}
                \caption{
                    We compare different methods for choosing a direction in the max $10$-step Lyapunov exponent calculation on the Mixed Objective.
                    \textbf{\emph{Takeaway:}} Re-estimating the top EVecs at each iteration performs best, but is most expensive.
                    \emph{Left}: We sample a random normalized direction uniformly for our displacement $\displacement$ at each exponent calculation, which does not clearly show the bifurcation.
                    \emph{Middle}: We perform $10$ steps of power iteration to tune $\displacement$.
                    First, we tune the displacement at only the first iteration, which shows the bifurcation.
                    Next, we tune the displacement at every iteration in the exponent calculation, which shows the bifurcation more clearly.
                    \emph{Right}: We use \texttt{jax.numpy.linalg.eigh} to tune $\displacement$, which also clearly shows the bifurcation.
                }\label{fig:lyap_dir_mix}
            \end{figure*}
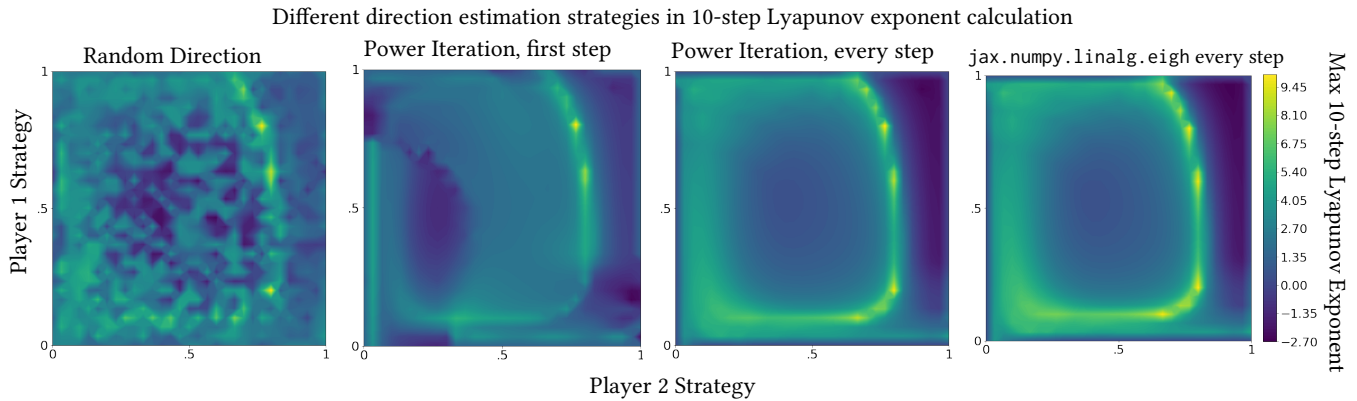
            
            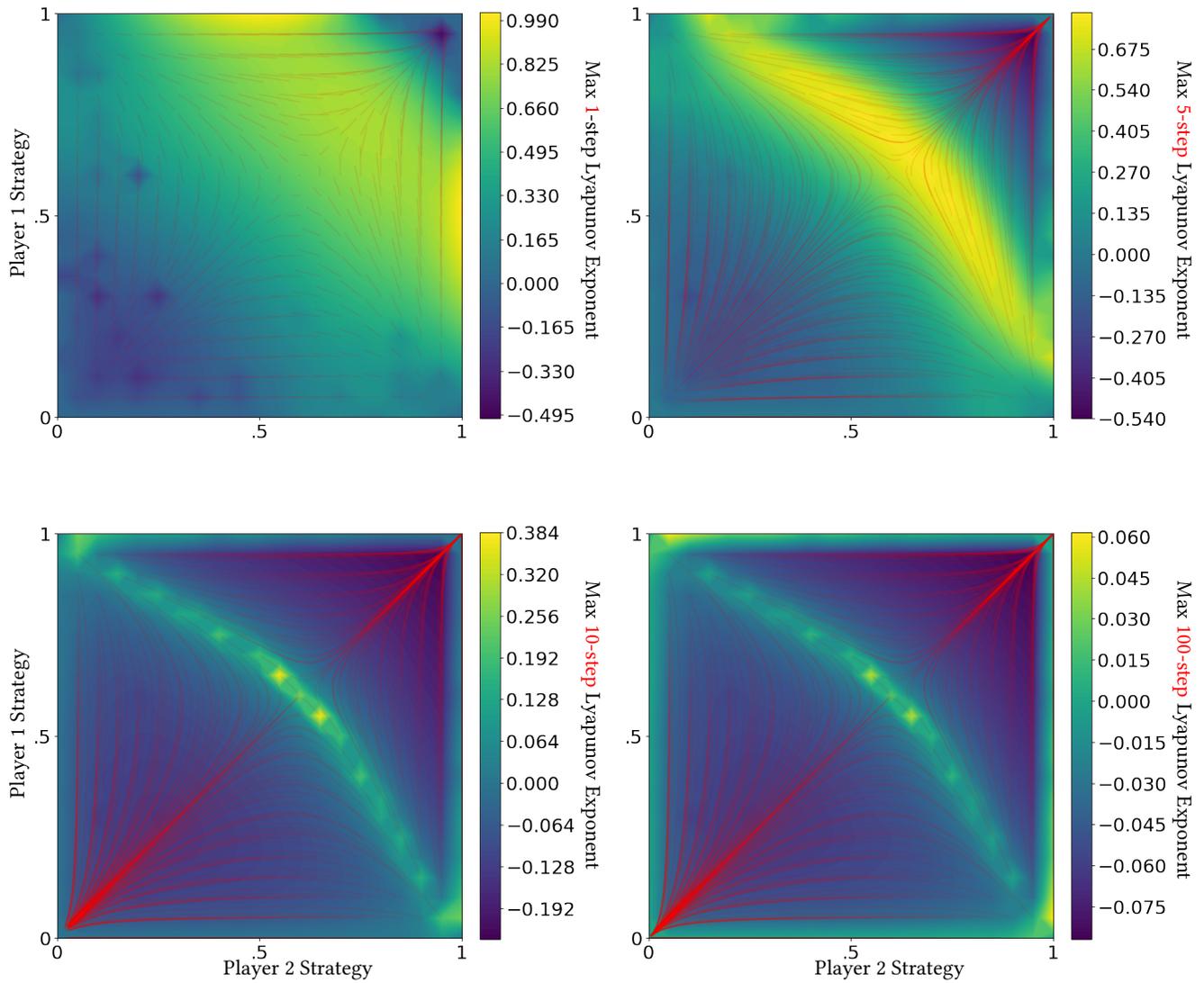
\begin{figure*}
                \centering
                \begin{tikzpicture}
                    \centering
                    \node (img11){\includegraphics[trim={.25cm .25cm 1.05cm 1.25cm},clip, width=.45\linewidth]{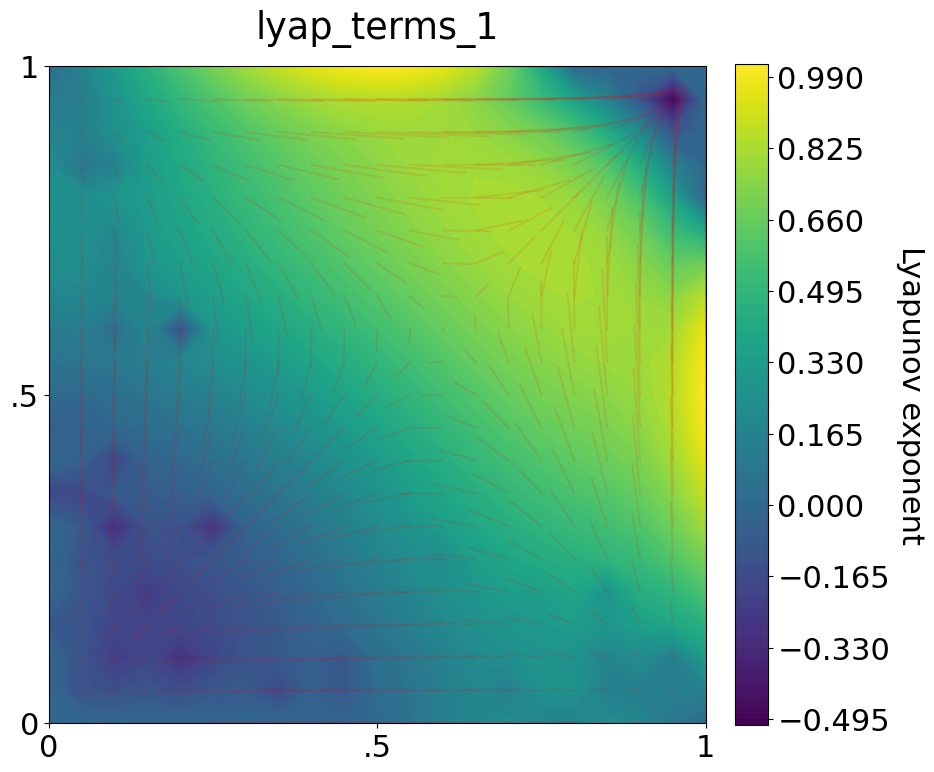}};
                        \node[left=of img11, node distance=0cm, rotate=90, xshift=1.5cm, yshift=-.9cm, font=\color{black}] {Player 1 Strategy};
                        \node[right=of img11, node distance=0cm, rotate=270, xshift=-2.5cm, yshift=-.9cm, font=\color{black}] {Max {\color{red}$1$}-step Lyapunov Exponent};
                    
                    \node [right=of img11, xshift=-.5cm](img12){\includegraphics[trim={.25cm .25cm 1.05cm 1.25cm},clip, width=.45\linewidth]{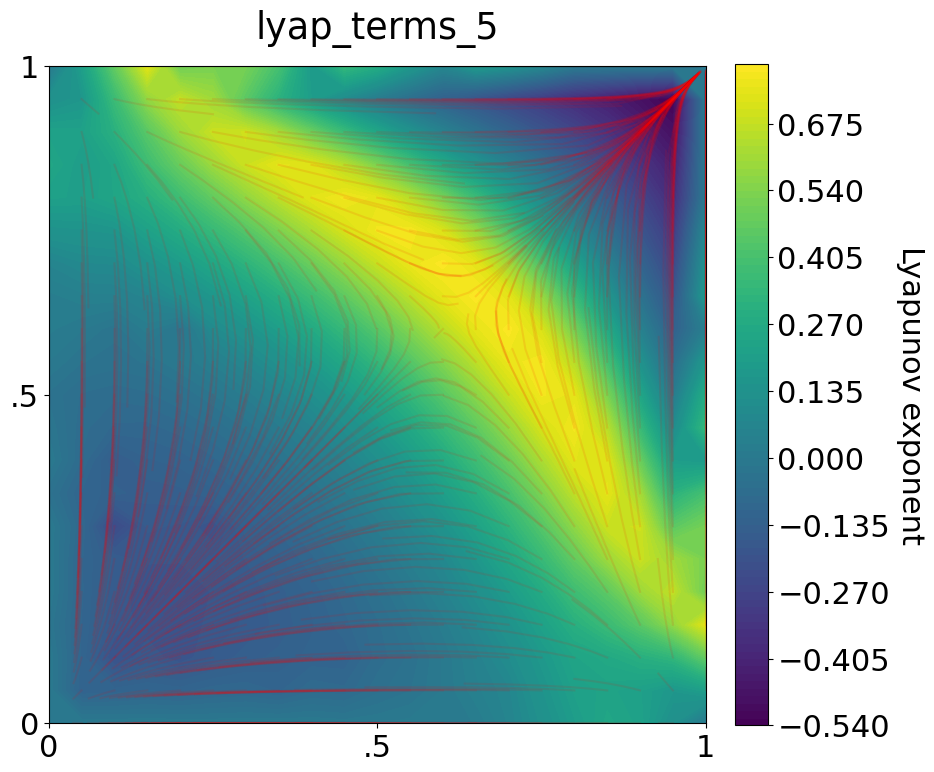}};
                        \node[right=of img12, node distance=0cm, rotate=270, xshift=-2.5cm, yshift=-.9cm, font=\color{black}] {Max {\color{red}$5$-step} Lyapunov Exponent};
                    
                    \node [below=of img11, xshift=-0cm](img21){\includegraphics[trim={.25cm .25cm 1.05cm 1.25cm},clip, width=.45\linewidth]{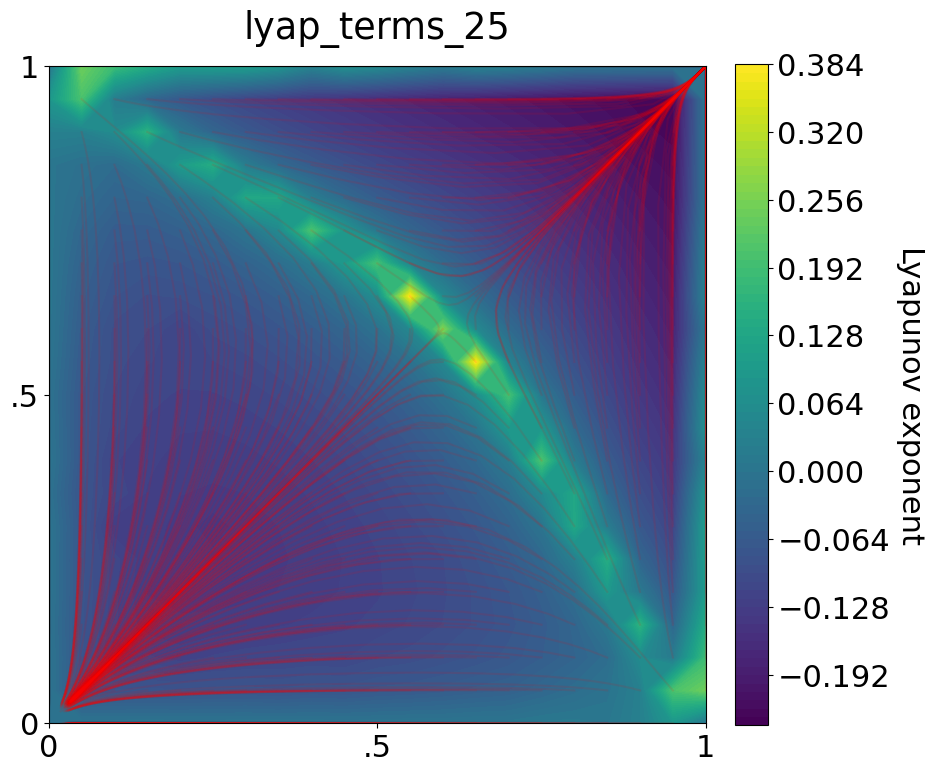}};
                        \node[right=of img21, node distance=0cm, rotate=270, xshift=-2.5cm, yshift=-.9cm, font=\color{black}] {Max {\color{red}$10$-step} Lyapunov Exponent};
                        \node[left=of img21, node distance=0cm, rotate=90, xshift=1.5cm, yshift=-.9cm, font=\color{black}] {Player 1 Strategy};
                        \node[below=of img21, node distance=0cm, xshift=-.1cm,yshift=1.25cm,font=\color{black}] {Player 2 Strategy};
                        
                    \node [right=of img21, xshift=-.5cm](img22){\includegraphics[trim={.25cm .25cm 1.05cm 1.25cm},clip, width=.45\linewidth]{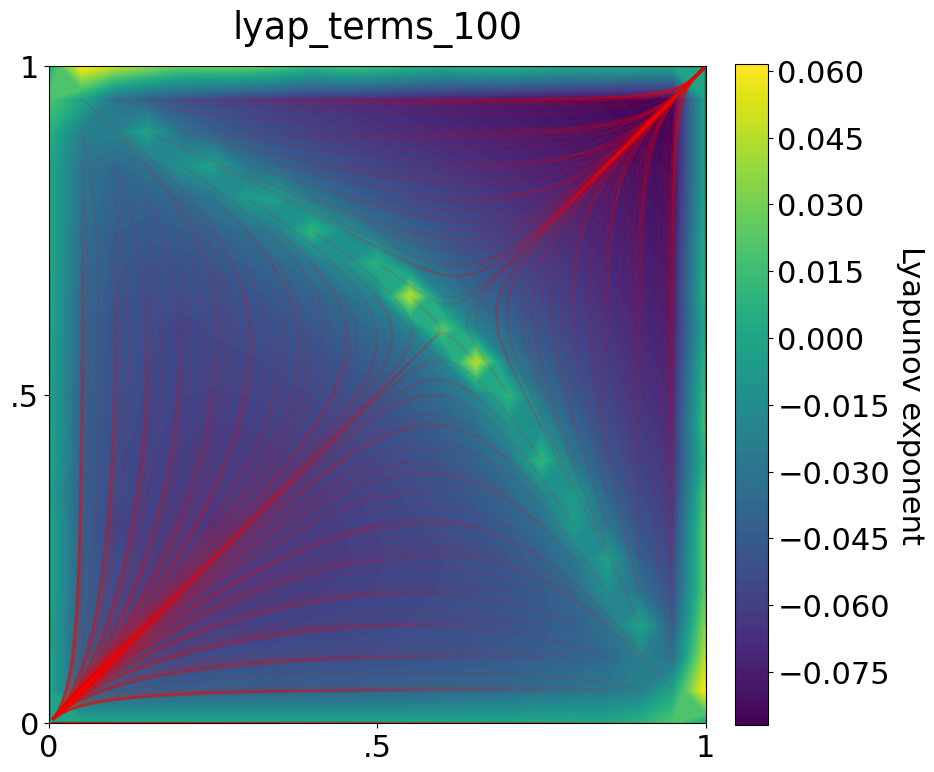}};
                        \node[right=of img22, node distance=0cm, rotate=270, xshift=-2.5cm, yshift=-.9cm, font=\color{black}] {Max {\color{red}$100$-step} Lyapunov Exponent};
                        \node[below=of img22, node distance=0cm, xshift=-.1cm,yshift=1.25cm,font=\color{black}] {Player 2 Strategy};
                \end{tikzpicture}
                \vspace{-.01\textheight}
                \caption{
                    We show the calculation for the $k$-step Lyapunov exponent for various $k$ on the small IPD.
                    The $k$-step optimization trajectories used for the exponent calculation are shown in {\color{red}red}, allowing us to see the horizon that the associated exponent measures separation over.
                    \textbf{\emph{Takeaway:}} We can effectively find bifurcations with a $k$-step exponent for various $k$.
                    Using only 1-step does not show the bifurcation, while using more does.
                    As $k$ gets larger, the scale of the gradients changes, causing difficult optimization in the limit --  note the changing color bar scale.
                }\label{fig:lyap_step_mix}
            \end{figure*}
            
            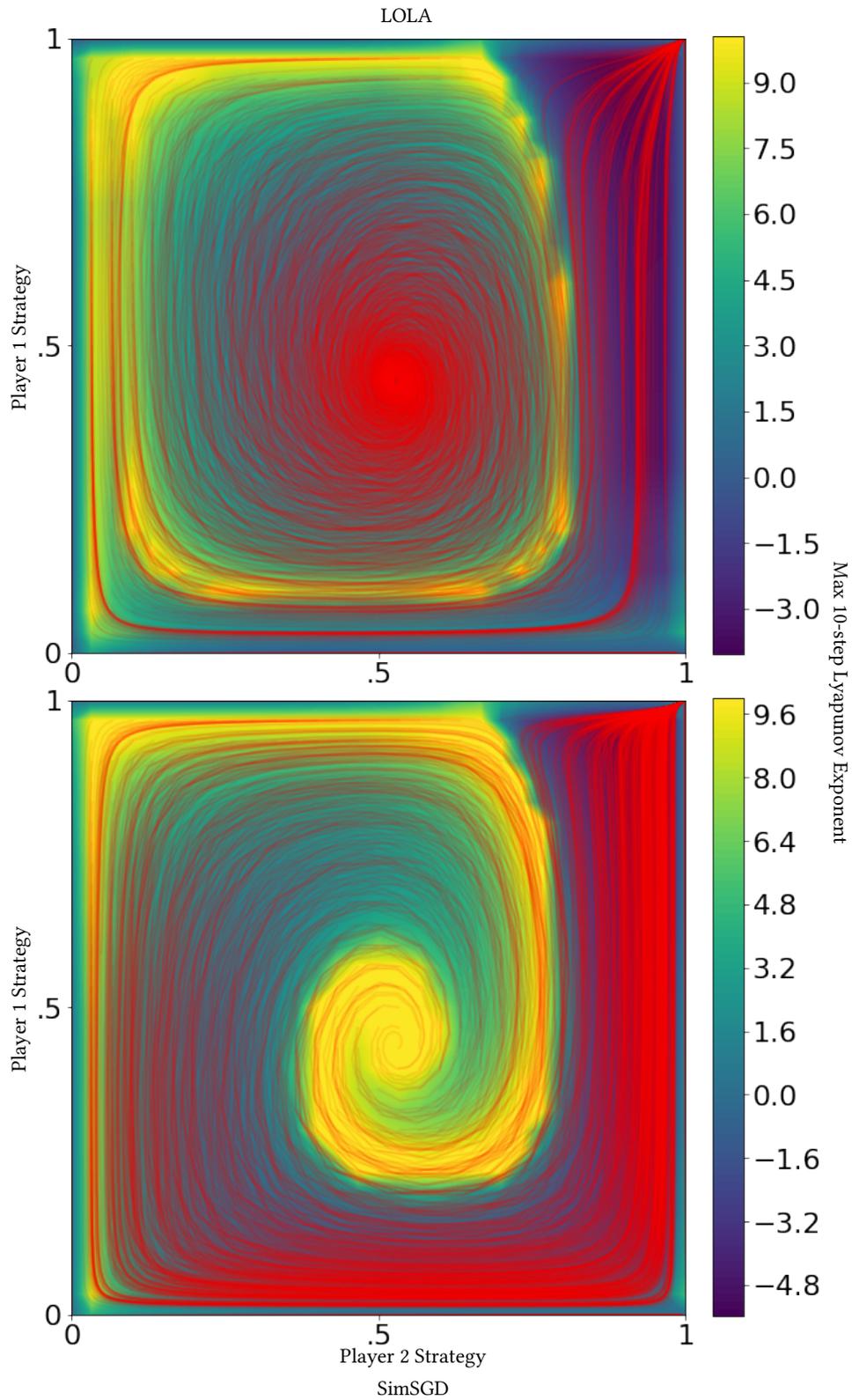
\begin{figure*}
                \vspace{-.06\textheight}
                \centering
                \begin{tikzpicture}
                    \centering
                    \node (img){\includegraphics[trim={.25cm .25cm 1.05cm 1.25cm},clip, width=.67\linewidth]{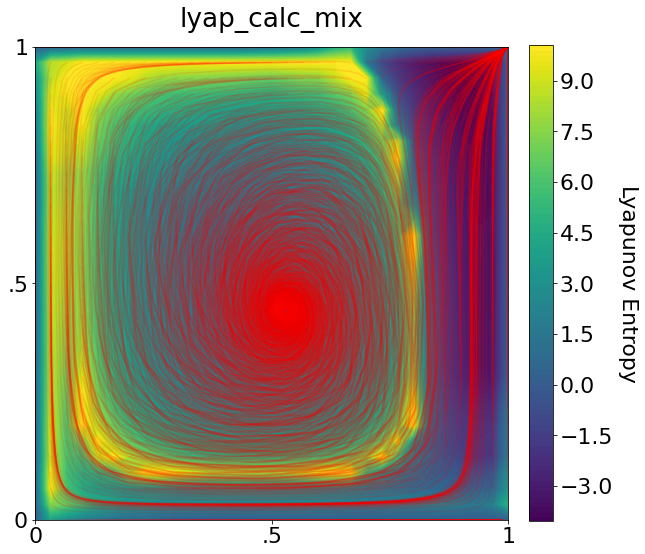}};
                    \node[left=of img, node distance=0cm, rotate=90, xshift=1.5cm, yshift=-.9cm, font=\color{black}] {Player 1 Strategy};
                    \node[right=of img, node distance=0cm, rotate=270, xshift=3cm, yshift=-.9cm, font=\color{black}] {Max $10$-step Lyapunov Exponent};
                    \node[above=of img, node distance=0cm, xshift=-.35cm, yshift=-1.2cm,font=\color{black}] {LOLA};
                    
                    \node [below=of img, yshift=1.25cm] (img2){\includegraphics[trim={.25cm .25cm 1.05cm 1.25cm},clip, width=.67\linewidth]{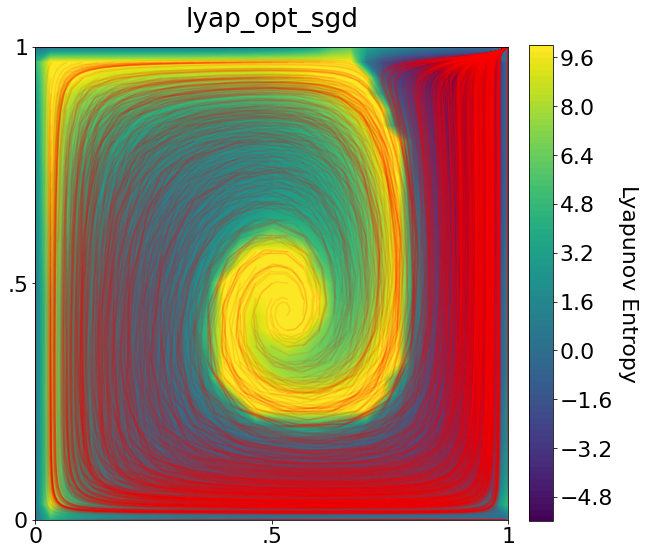}};
                    \node[left=of img2, node distance=0cm, rotate=90, xshift=1.5cm, yshift=-.9cm, font=\color{black}] {Player 1 Strategy};
                    \node[below=of img2, node distance=0cm, xshift=-.25cm, yshift=1.25cm,font=\color{black}] {Player 2 Strategy};
                    \node[below=of img2, node distance=0cm, xshift=-.25cm, yshift=.75cm,font=\color{black}] {SimSGD};
                \end{tikzpicture}
                \vspace{-.01\textheight}
                \caption{
                    We contrast the max $20$-step Lyapunov exponent calculations between LOLA and simSGD.
                    The $20$-step optimization trajectories used for the exponent calculation are shown in {\color{red}red}, allowing us to see the trajectories the associated exponent measures separation over.
                    \textbf{Takeaway:} We can find optimizer-dependant bifurcations with our method.
                    \emph{Top:} $10$ steps from optimization with LOLA at each point where we calculate an exponent for the heatmap.
                    \emph{Bottom:} The same visualization as top, except with an optimizer of simSGD.
                }\label{fig:lyap_opt_mix}
            \end{figure*}
            
            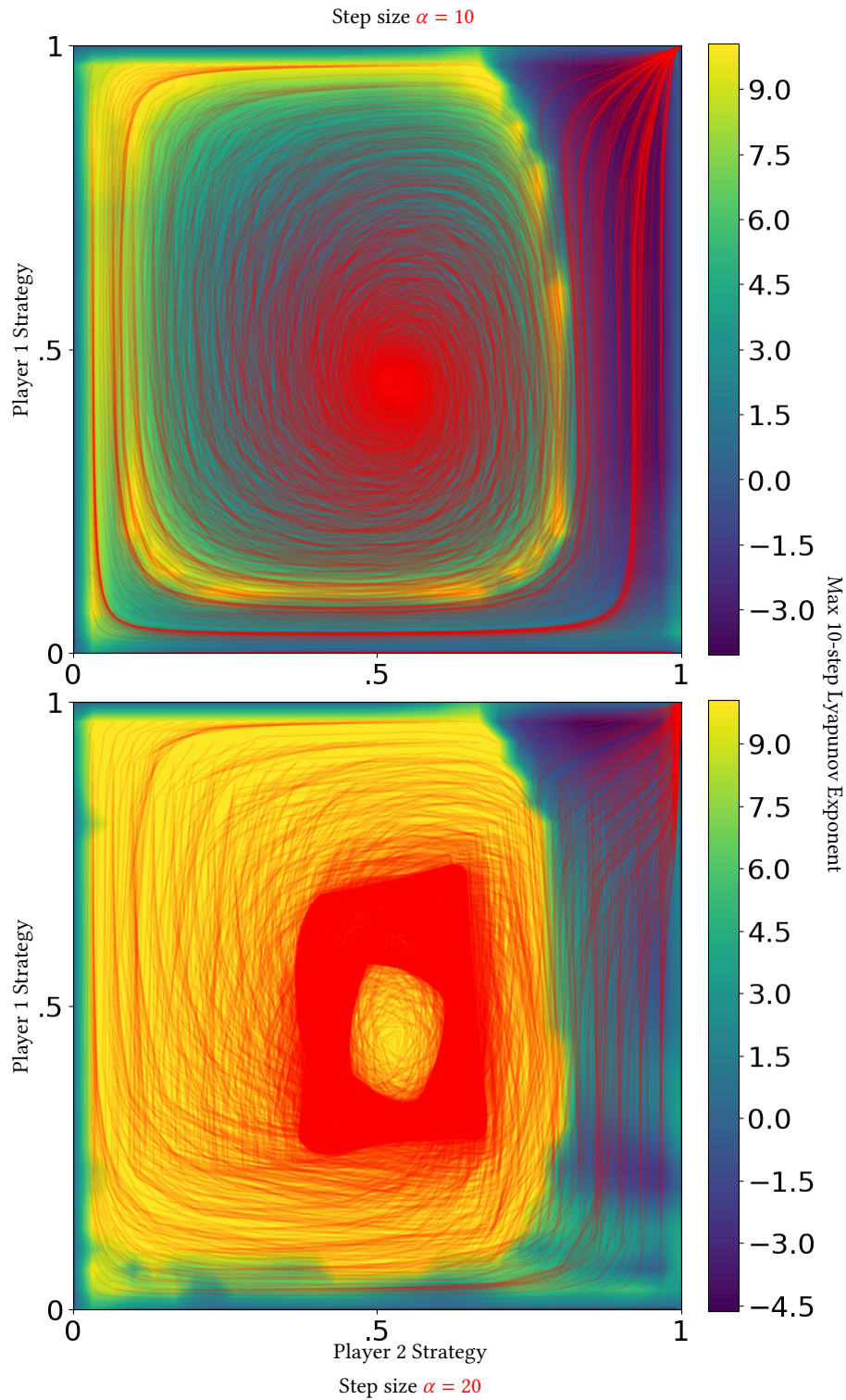
\begin{figure*}
                \vspace{-.06\textheight}
                \centering
                \begin{tikzpicture}
                    \centering
                    \node (img){\includegraphics[trim={.25cm .25cm 1.05cm 1.25cm},clip, width=.63\linewidth]{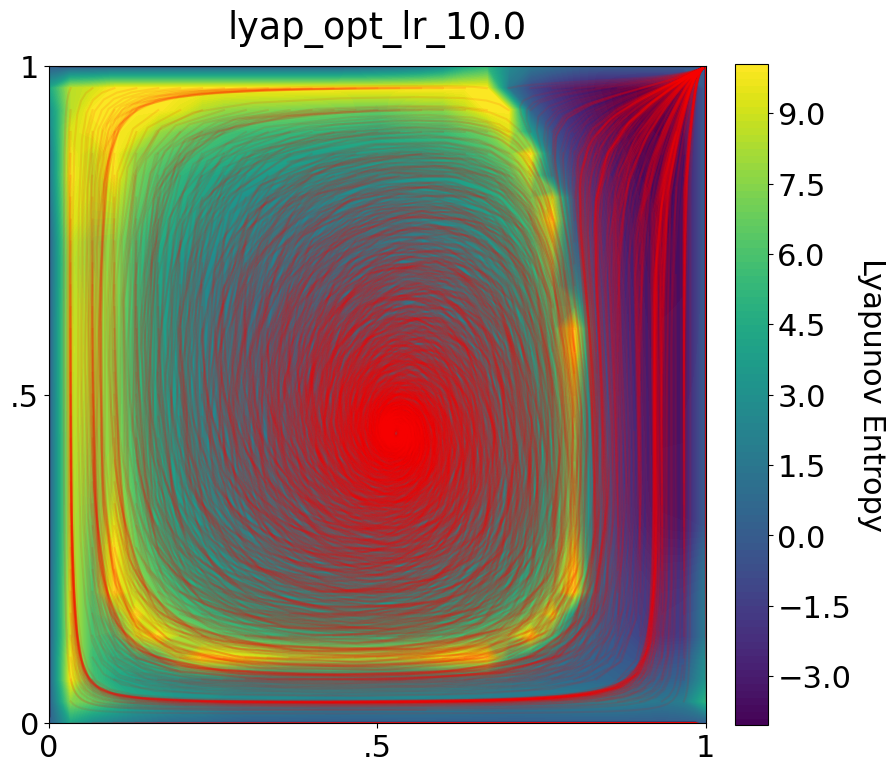}};
                    \node[left=of img, node distance=0cm, rotate=90, xshift=1.5cm, yshift=-.9cm, font=\color{black}] {Player 1 Strategy};
                    \node[right=of img, node distance=0cm, rotate=270, xshift=3cm, yshift=-.9cm, font=\color{black}] {Max $10$-step Lyapunov Exponent};
                    \node[above=of img, node distance=0cm, xshift=-.35cm, yshift=-1.2cm,font=\color{black}] {Step size {\color{red}$\alpha = 10$}};
                    
                    \node [below=of img, yshift=1.25cm] (img2){\includegraphics[trim={.25cm .25cm 1.05cm 1.25cm},clip, width=.63\linewidth]{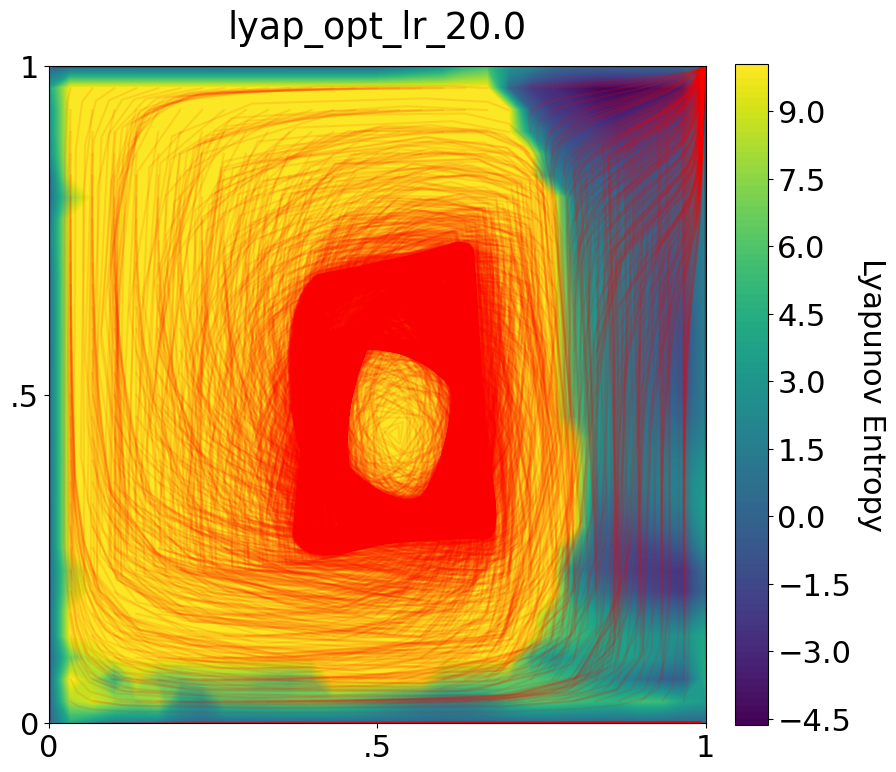}};
                    \node[left=of img2, node distance=0cm, rotate=90, xshift=1.5cm, yshift=-.9cm, font=\color{black}] {Player 1 Strategy};
                    \node[below=of img2, node distance=0cm, xshift=-.25cm, yshift=1.25cm,font=\color{black}] {Player 2 Strategy};
                    \node[below=of img2, node distance=0cm, xshift=-.25cm, yshift=.75cm,font=\color{black}] {Step size {\color{red}$\alpha = 20$}};
                \end{tikzpicture}
                \vspace{-.01\textheight}
                \caption{
                    We investigate the impact of optimization algorithm parameters on the $10$-step max Lyapunov exponent calculation with LOLA.
                    The $10$-step optimization trajectories used for the exponent calculation are shown in {\color{red}red}, allowing us to see the trajectories the associated exponent measures separation over.
                    \textbf{Takeaway:} If the step size is too large, the optimizer does not converge to solutions, resulting in complicated limit cycle bifurcations.
                    \emph{Top:} $10$ steps from the optimization with a step size of $\alpha = 10$ at each point where we calculate an exponent for the heatmap.
                    \emph{Bottom:} The same visualization as top, except with step size of $\alpha = 10$.
                    The optimization trajectories only converge at the solution in the top right.
                    In the center, the trajectories accumulate at a limit cycle, rotating around the solution.
                }\label{fig:lyap_opt_param_mix}
            \end{figure*}
            
            \begin{figure*}
                \vspace{-.05\textheight}
                \centering
                \begin{tikzpicture}
                    \centering
                    \node (img){\includegraphics[trim={.25cm .25cm 1.05cm 1.25cm},clip, width=.73\linewidth]{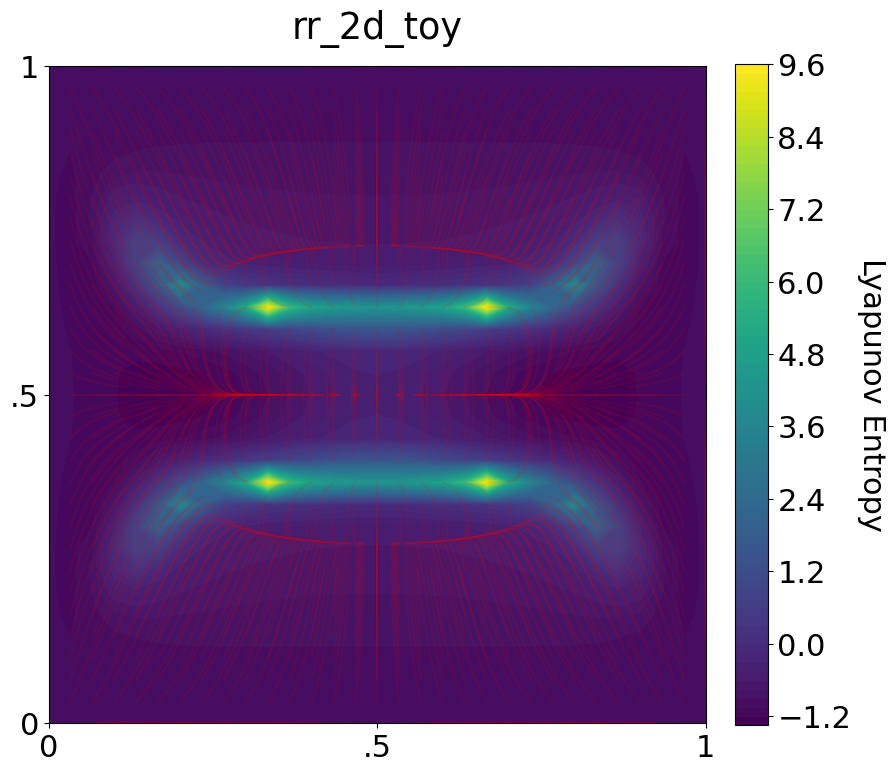}};
                    \node[left=of img, node distance=0cm, rotate=90, xshift=0cm, yshift=-.5cm, font=\color{black}] {$y$};
                    \node[right=of img, node distance=0cm, rotate=270, xshift=-1.75cm, yshift=-.9cm, font=\color{black}] {Max $10$-step Lyapunov Exponent};
                    \node[below=of img, node distance=0cm, xshift=-.5cm, yshift=.75cm,font=\color{black}] {$x$};
                \end{tikzpicture}
                \vspace{-.01\textheight}
                \caption{
                    We investigate finding bifurcations with a $10$-step max Lyapunov exponent on a single-objective optimization problem from RR.
                    The optimization trajectories (used for the exponent calculation) are shown in {\color{red}red}, allowing us to verify where bifurcations are.
                    \textbf{Takeaway:} Our method effectively highlights bifurcations in RRs example.
                }\label{fig:lyap_coop_toy}
            \end{figure*}
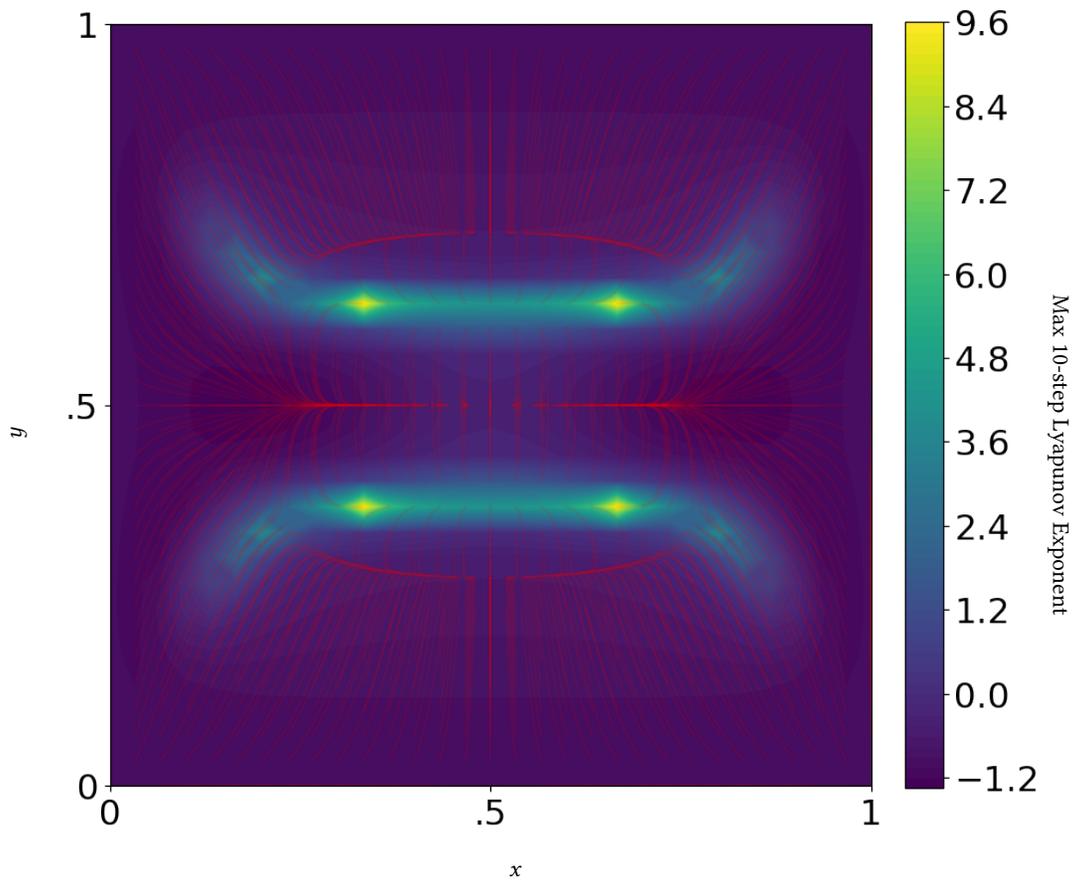
            
            \begin{figure*}
                \centering
                \begin{tikzpicture}
                    \centering
                    \node (img){\includegraphics[trim={.25cm 1.0cm 1.05cm 1.14cm},clip, width=.8\linewidth]{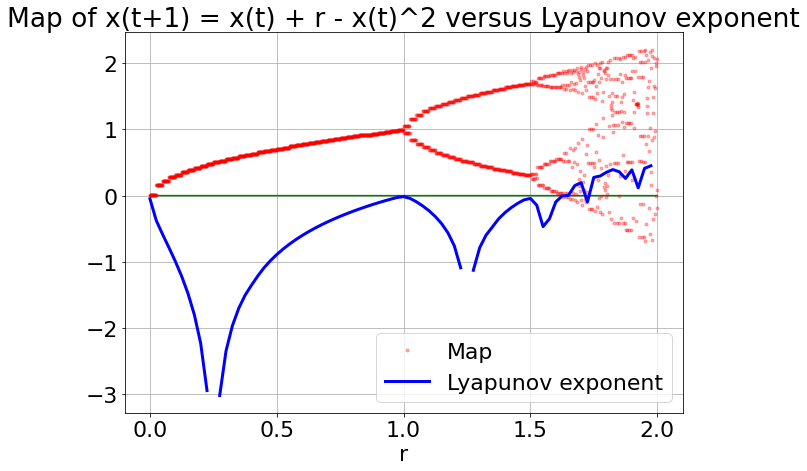}};
                    \node[left=of img, node distance=0cm, rotate=90, xshift=1.5cm, yshift=-2.0cm, font=\color{black}] {$x_0$};
                    \node[below=of img, node distance=0cm, xshift=-.35cm, yshift=1.0cm,font=\color{black}] {$r$};
                \end{tikzpicture}
                \vspace{-.01\textheight}
                \caption{
                    We display the Lyapunov exponent on the logistic map: $x(t+1) = x(t) + r + x(t)^2$.
                    \textbf{Takeaway:} Intuition for Lyapunov exponents on a canonical 1-dimensional example for bifurcations.
                }\label{fig:lyap_oneDim_toy}
            \end{figure*}
            
            \begin{figure*}
                \vspace{-.05\textheight}
                \centering
                \begin{tikzpicture}
                    \centering
                    \node (img){\includegraphics[trim={.25cm .25cm 1.05cm 1.25cm},clip, width=.69\linewidth]{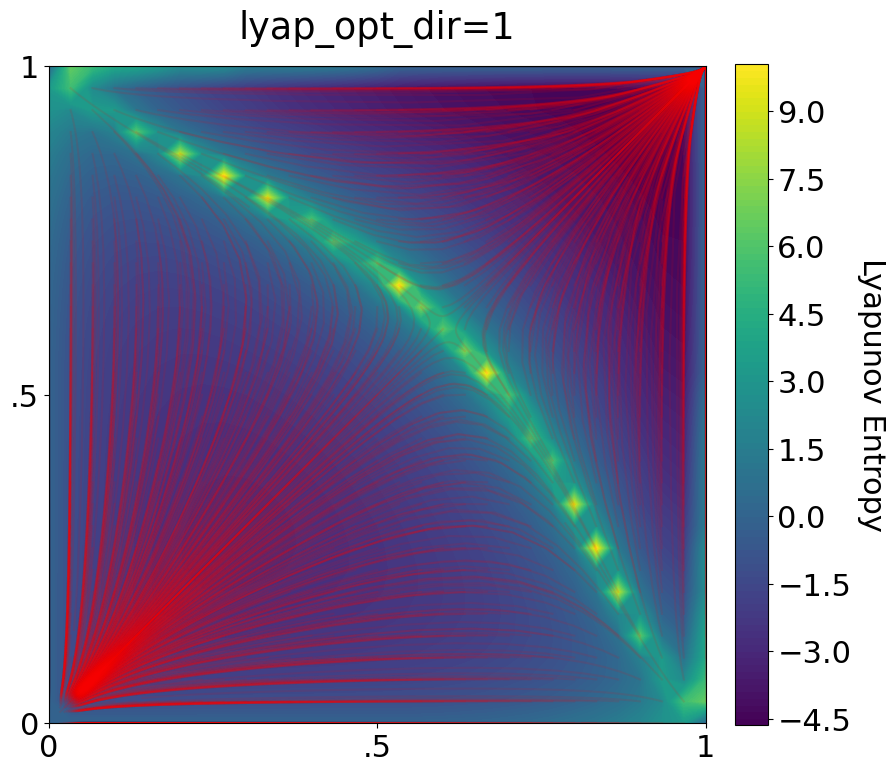}};
                        \node[left=of img, node distance=0cm, rotate=90, xshift=1.5cm, yshift=-.9cm, font=\color{black}] {Player 1 Strategy};
                        \node[right=of img, node distance=0cm, rotate=270, xshift=-2.75cm, yshift=-.9cm, font=\color{black}] {{\color{red}Max = 1-direction} Lyapunov Exponent};
                    
                    \node [below=of img, yshift=1.25cm] (img2){\includegraphics[trim={.25cm .25cm 1.05cm 1.25cm},clip, width=.69\linewidth]{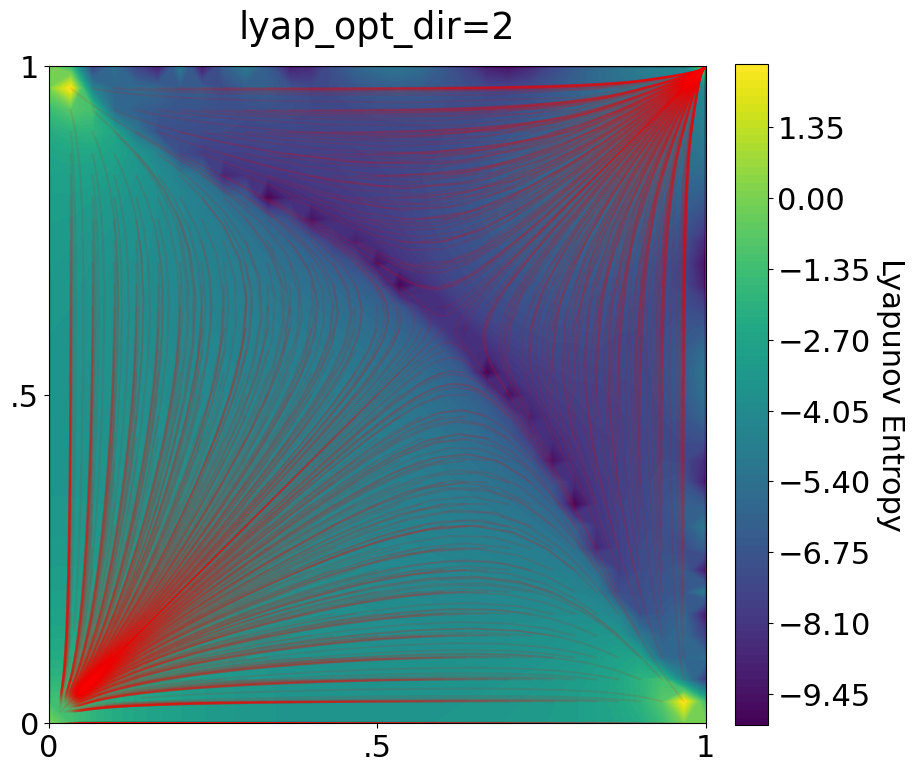}};
                        \node[left=of img2, node distance=0cm, rotate=90, xshift=1.5cm, yshift=-.9cm, font=\color{black}] {Player 1 Strategy};
                        \node[below=of img2, node distance=0cm, xshift=-.25cm, yshift=1.25cm,font=\color{black}] {Player 2 Strategy};
                        \node[right=of img2, node distance=0cm, rotate=270, xshift=-2.5cm, yshift=-.9cm, font=\color{black}] { {\color{red}Sum of top 2} Lyapunov Exponents};
                \end{tikzpicture}
                \vspace{-.01\textheight}
                \caption{
                    We compare calculating different $10$-step Lyapunov exponent objectives trying to guarantee divergence in multiple directions on the small IPD.
                    The optimization trajectories used for the exponent calculation are shown in {\color{red}red}, allowing us to see the trajectories for the associated objective.
                    \textbf{Takeaway:} We can find bifurcations, while guaranteeing trajectory separation in every direction.
                    Interestingly, local maxima -- not saddles -- allow trajectory separation in all directions here.
                    \emph{Top:} The max $10$-step Lyapunov exponent.
                    \emph{Bottom:} The sum of the top 2 $10$-step Lyapunov exponents.
                }\label{fig:entropy_mix}
            \end{figure*}
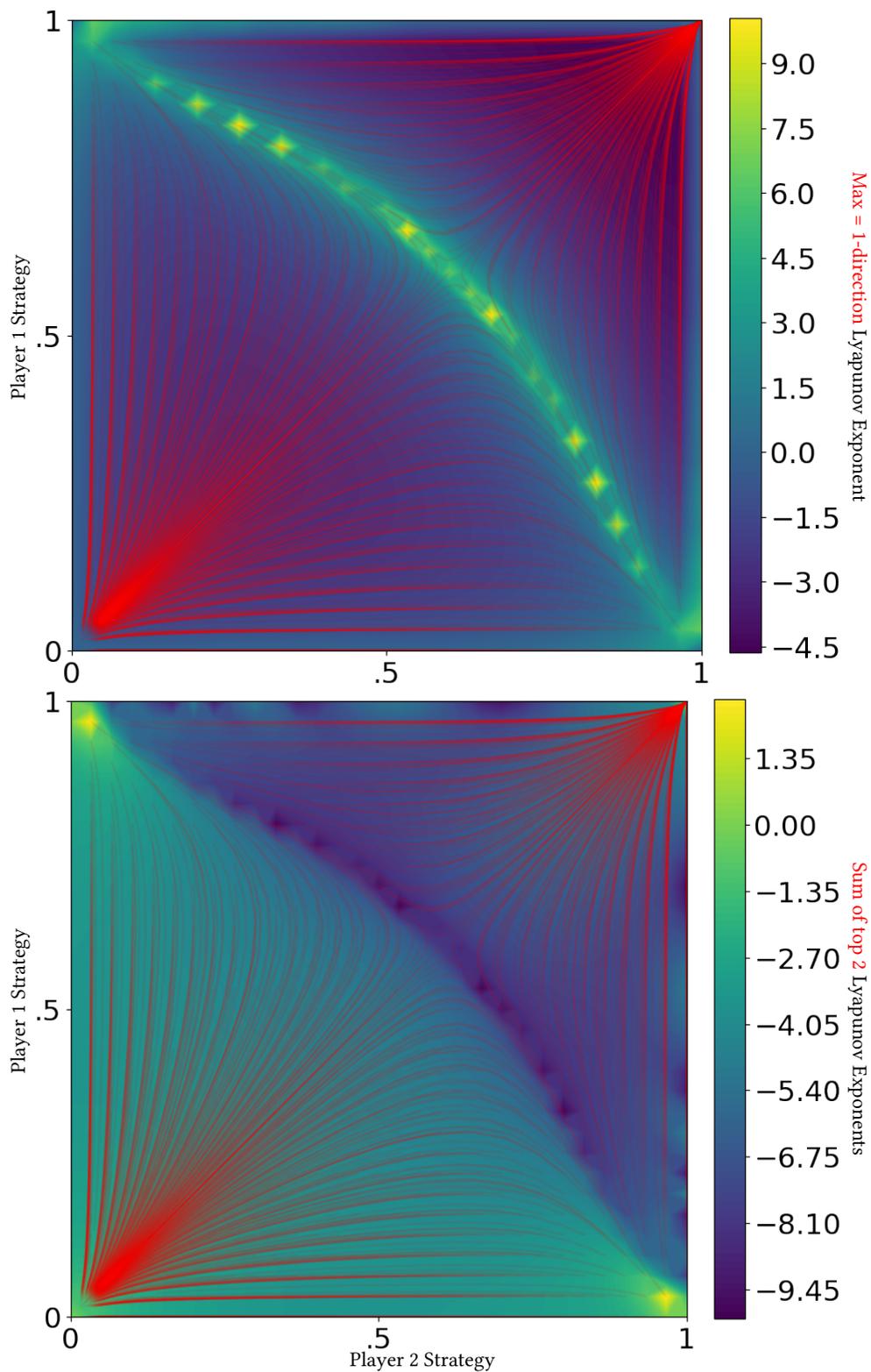
            
            \begin{figure*}
                \includegraphics[trim={2.3cm 1.75cm .0cm 1.0cm},clip,width=.3\linewidth]{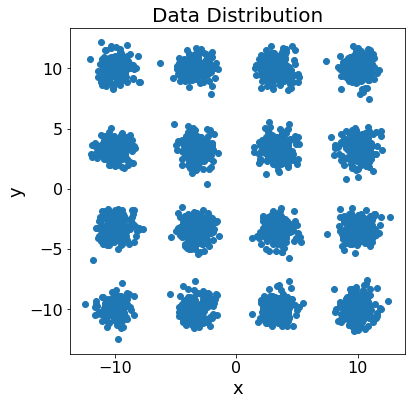}
                \caption{
                    Ground truth samples for our Mixture of Gaussian experiment.
                }
                \label{fig:gan_2d_samples}
            \end{figure*}
            
            \begin{figure*}
                \centering
                \begin{tikzpicture}
                    \centering
                    \node (img11){\includegraphics[trim={.25cm .25cm 4.5cm 1.25cm},clip, width=.22\linewidth]{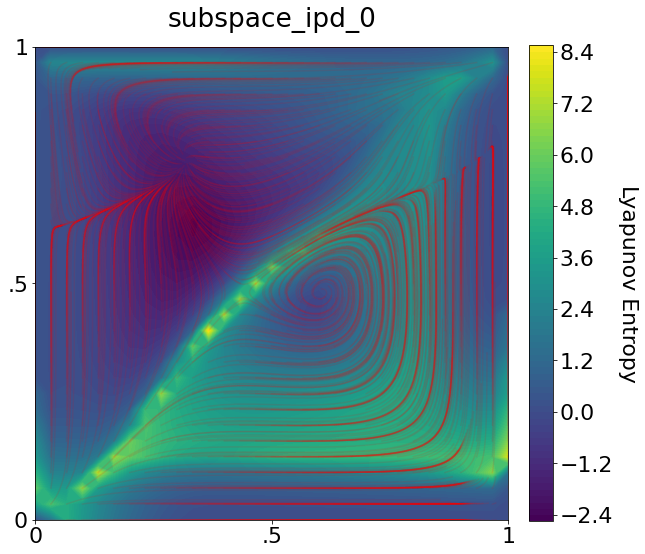}};
                        \node[left=of img11, node distance=0cm, rotate=90, xshift=1.5cm, yshift=-.9cm, font=\color{black}] {Player 1 Strategy};
                        \node[above=of img11, node distance=0cm, xshift=-.0cm, yshift=-1.25cm,font=\color{black}] {Subspace seed 0};
                    
                        \node[below=of img11, node distance=0cm, xshift=6.5cm,yshift=1.0cm,font=\color{black}] {Player 2 Strategy};
                        \node[above=of img11, node distance=0cm, xshift=6.5cm, yshift=-.75cm,font=\color{black}] {Random subspace IPD};
                    
                    \node [right=of img11, xshift=-1cm](img12){\includegraphics[trim={.25cm .25cm 4.5cm 1.25cm},clip, width=.22\linewidth]{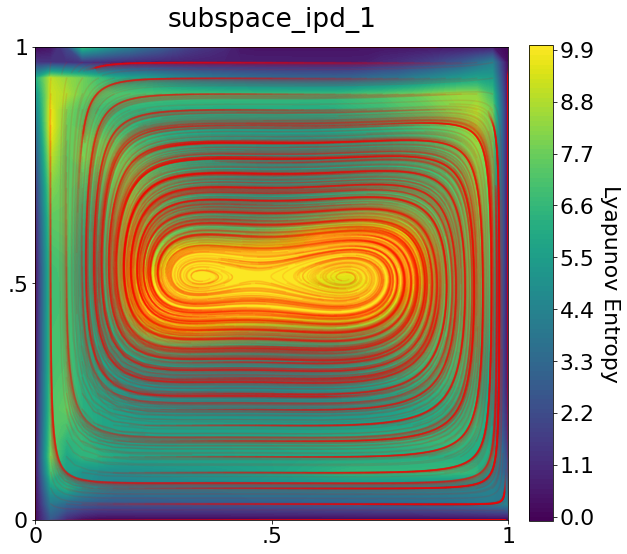}};
                        \node[above=of img12, node distance=0cm, xshift=-.0cm, yshift=-1.25cm,font=\color{black}] {Subspace seed 1};
                    
                    \node [right=of img12, xshift=-1cm](img13){\includegraphics[trim={.25cm .25cm 4.5cm 1.25cm},clip, width=.22\linewidth]{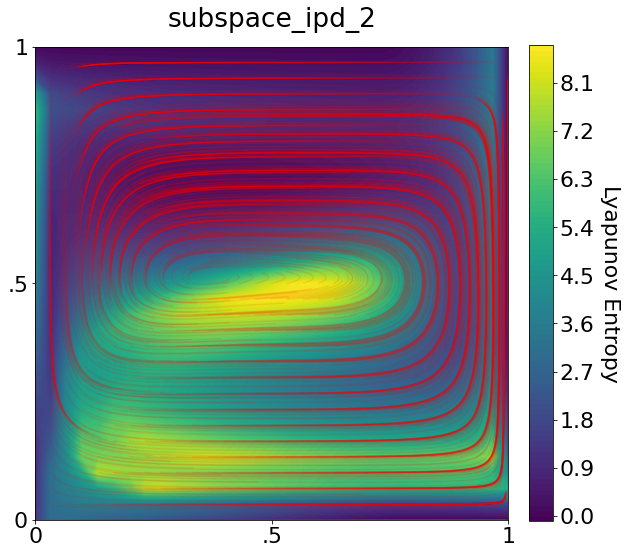}};
                        \node[above=of img13, node distance=0cm, xshift=-.0cm, yshift=-1.25cm,font=\color{black}] {Subspace seed 2};
                    
                    \node [right=of img13, xshift=-1cm](img14){\includegraphics[trim={.25cm .25cm 1.05cm 1.25cm},clip, width=.26\linewidth]{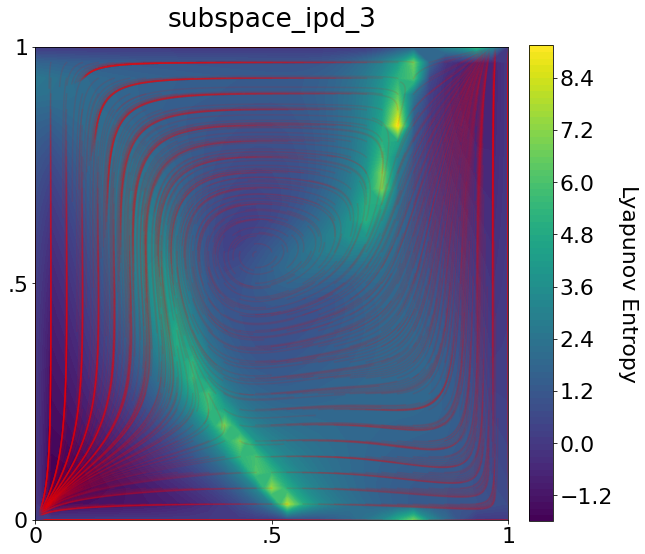}};
                        \node[above=of img14, node distance=0cm, xshift=-.0cm, yshift=-1.25cm,font=\color{black}] {Subspace seed 3};
                        
                    \node [below=of img11, xshift=.15cm, yshift=-.5cm] (img21){\includegraphics[trim={2.75cm 1.0cm 4.25cm 1.25cm},clip, width=.22\linewidth]{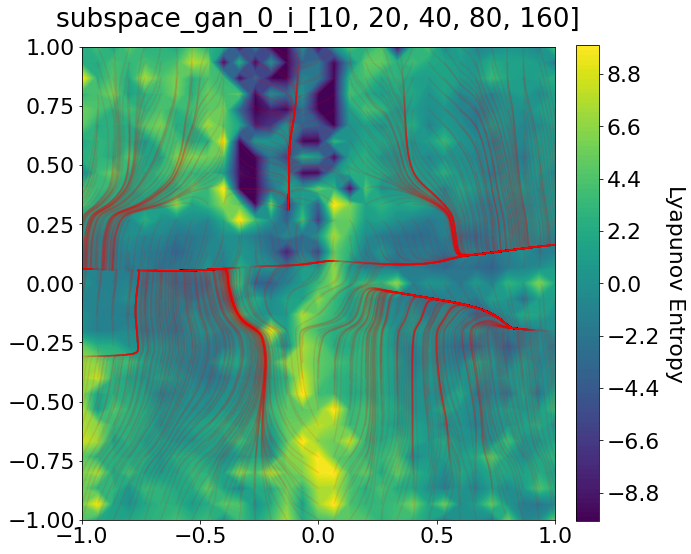}};
                        \node[left=of img21, node distance=0cm, rotate=90, xshift=1.15cm, yshift=-.75cm, font=\color{black}] {Discriminator};
                        \node[above=of img21, node distance=0cm, xshift=-.0cm, yshift=-1.25cm,font=\color{black}] {Subspace seed 0};
                        \node[below=of img21, node distance=0cm, xshift=6.5cm,yshift=1.0cm,font=\color{black}] {Generator};
                        \node[above=of img21, node distance=0cm, xshift=6.5cm, yshift=-.75cm,font=\color{black}] {Random subspace GAN};
                    
                    \node [right=of img21, xshift=-1cm](img22){\includegraphics[trim={2.75cm 1.0cm 4.25cm 1.25cm},clip, width=.22\linewidth]{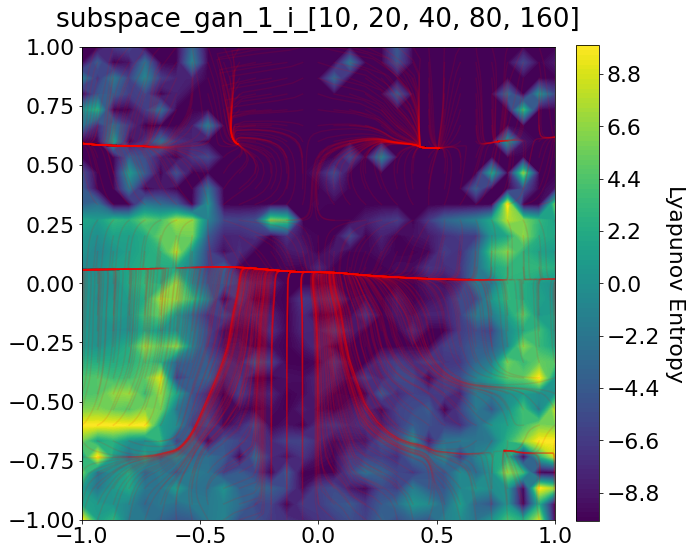}};
                        \node[above=of img22, node distance=0cm, xshift=-.0cm, yshift=-1.25cm,font=\color{black}] {Subspace seed 1};
                    
                    \node [right=of img22, xshift=-1cm](img23){\includegraphics[trim={2.75cm 1.0cm 4.25cm 1.25cm},clip, width=.22\linewidth]{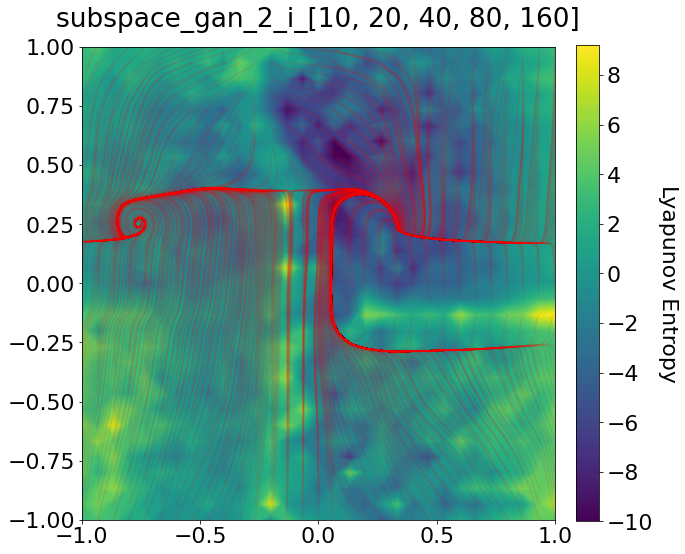}};
                        \node[above=of img23, node distance=0cm, xshift=-.0cm, yshift=-1.25cm,font=\color{black}] {Subspace seed 2};
                    
                    \node [right=of img23, xshift=-1cm](img24){\includegraphics[trim={2.75cm 1.0cm 1.05cm 1.25cm},clip, width=.26\linewidth]{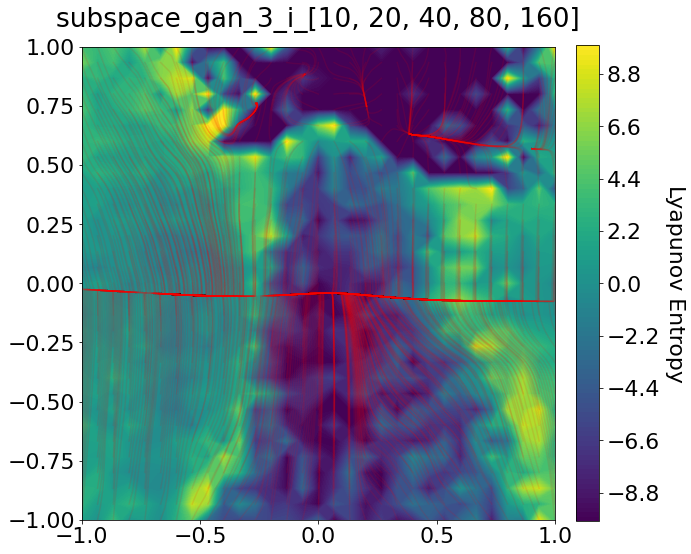}};
                        \node[above=of img24, node distance=0cm, xshift=-.0cm, yshift=-1.25cm,font=\color{black}] {Subspace seed 3};
                    
                        \node[right=of img14, node distance=0cm, rotate=270, xshift=0cm, yshift=-.9cm, font=\color{black}] {Max $10$-step Lyapunov Exponent};
                \end{tikzpicture}
                \vspace{-.01\textheight}
                \caption{
                    This reproduces Figure~\ref{fig:complicated_toy} with more random subspaces.
                    The optimization trajectories used for the exponent calculation are shown in {\color{red}red}, allowing us to see the trajectories the associated exponent measures separation over.
                    \textbf{\emph{Takeaway:}} The exponent is peaked near where trajectories separate for each subspace, showing that these strategies can find various bifurcations.
                    We display the Lyapunov exponent calculation -- as in Fig.~\ref{fig:fig_mix} --  on more complicated toy problems, to see how robustly we can find different bifurcations.
                    Section~\ref{subsec:new_problems} describes how we construct these examples by taking higher-dimensional problems and optimizing in a random subspace.
                    \emph{Top:} Different IPD subspace: we effectively highlight bifurcations -- i.e., regions where trajectory behavior qualitatively changes.
                    \emph{Bottom:} Different GAN subspaces: we are able to find bifurcations, but the highlighted structure is less crisp in this more complex toy example.
                }\label{fig:complicated_toy_moreSeeds}
            \end{figure*}
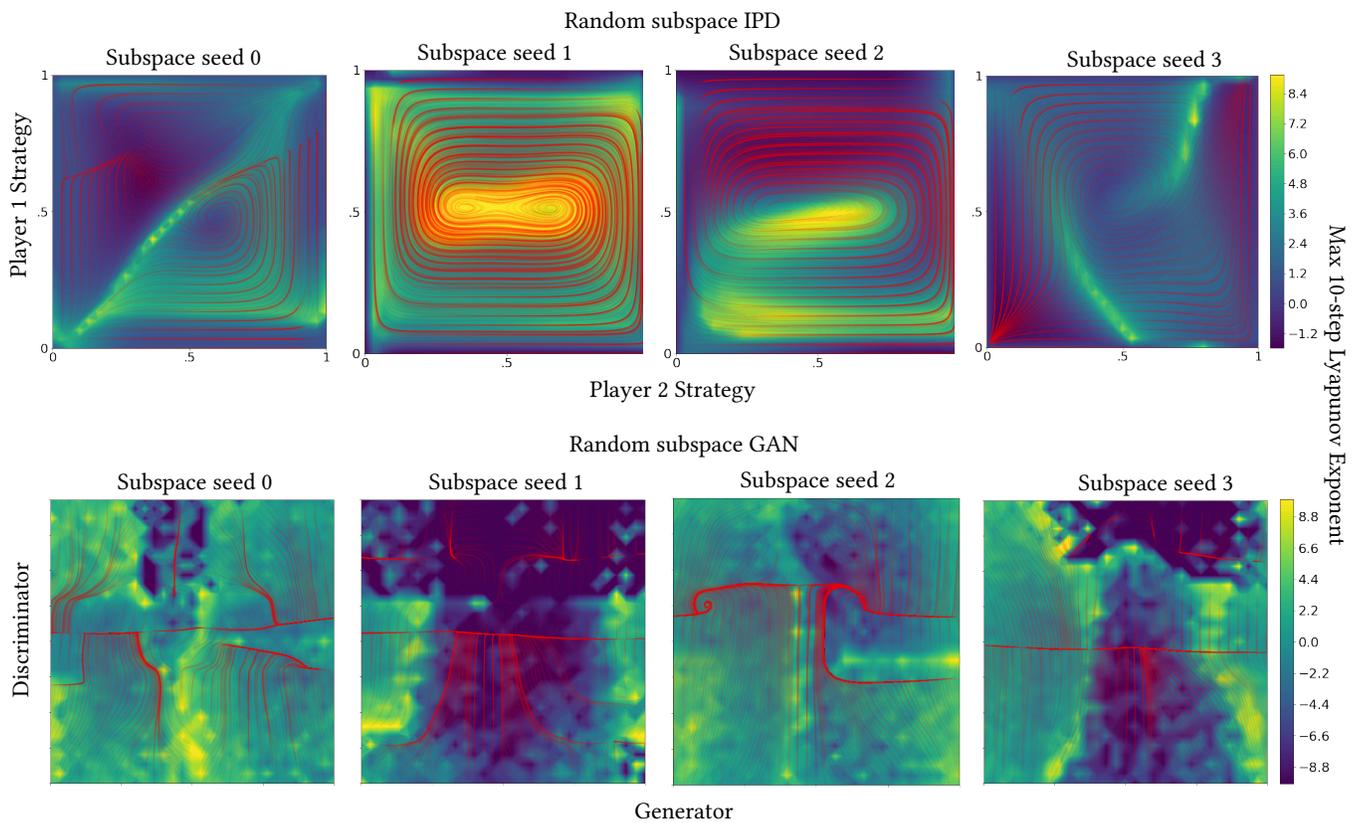
        
            \begin{figure*}
                \vspace{-0.0575\textheight}
                \centering
                \begin{tikzpicture}
                    \centering
                    \node (img11){\includegraphics[trim={1.0cm 1.85cm 55.0cm 1.15cm},clip, width=.45\linewidth]{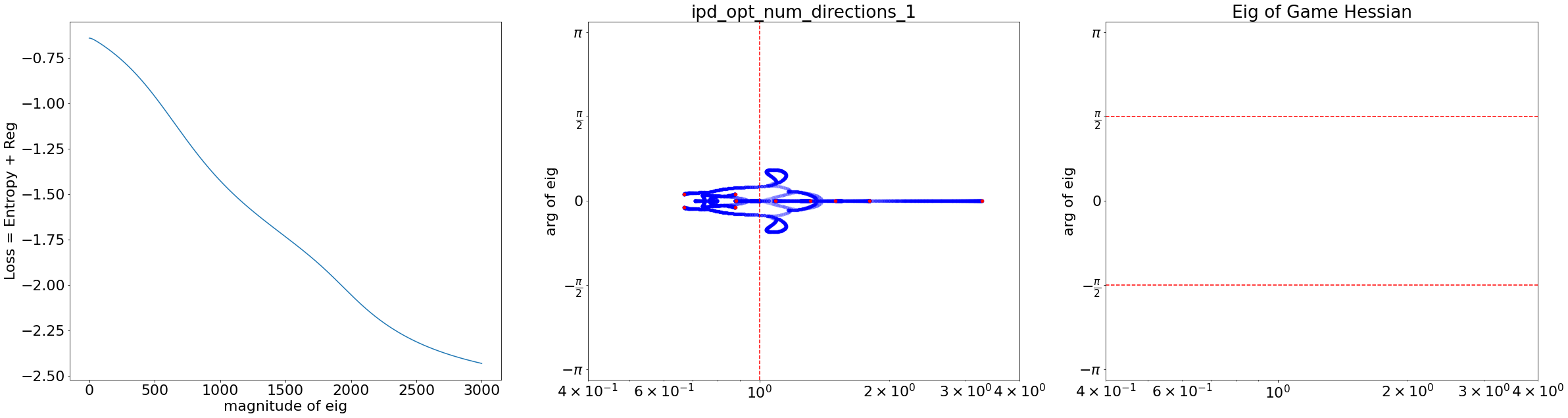}};
                        \node[left=of img11, node distance=0cm, rotate=90, xshift=2.0cm, yshift=-.75cm, font=\color{black}] {loss $\loss_{{\color{red}1}}^{\textnormal{min}}(\bothParam_0)$= -{\color{red}Max} exp loss};
                        
                        \node [right=of img11, xshift=-.5cm](img12){\includegraphics[trim={30.0cm 1.85cm 27.5cm 1.15cm},clip, width=.44\linewidth]{images/experiments/ipd/tune_lyap_ipd_1dir.png}};
                        \node[left=of img12, node distance=0cm, rotate=90, xshift=1.75cm, yshift=-.75cm, font=\color{black}] {argument of EVal $\arg(\eigval)$};
                        \node[above=of img12, node distance=0cm, xshift=-.0cm, yshift=-1.15cm,font=\color{black}] {EVals of Jac. of fixed point operator $\spectrum(\jacFixedPointOp)$};
                    
                    \node [below=of img11, yshift=1cm](img21){\includegraphics[trim={1.0cm 1.85cm 55.0cm 1.15cm},clip, width=.45\linewidth]{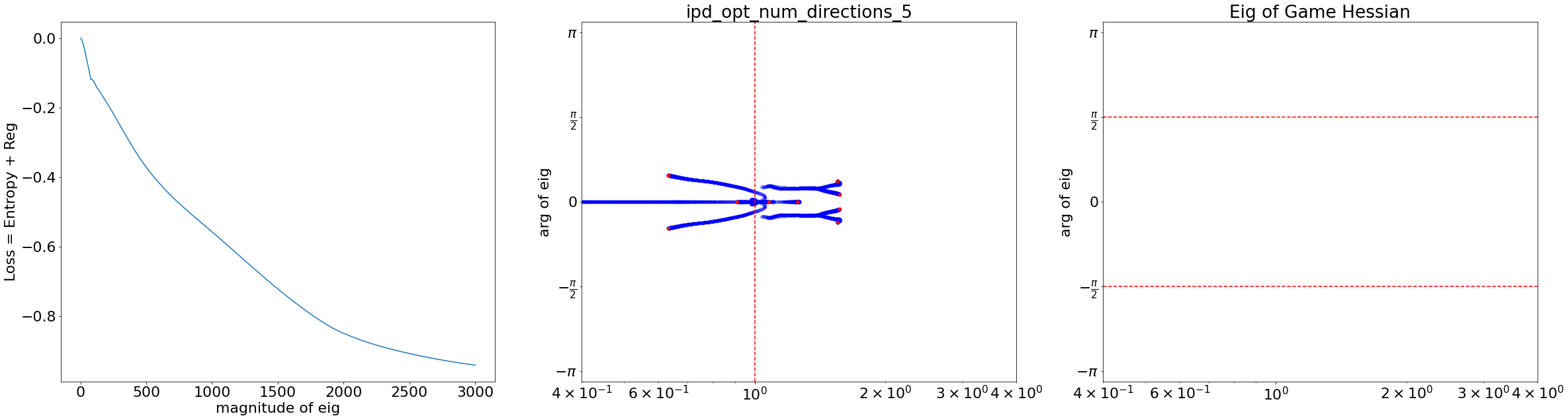}};
                        \node[left=of img21, node distance=0cm, rotate=90, xshift=2.0cm, yshift=-.75cm, font=\color{black}] {loss $\loss_{{\color{red}5}}^{\textnormal{min}}(\bothParam_0)$= -{\color{red}Min top 5} exps};
                        
                        \node [right=of img21, xshift=-.5cm](img22){\includegraphics[trim={29.5cm 1.85cm 27.5cm 1.15cm},clip, width=.44\linewidth]{images/experiments/ipd/tune_lyap_ipd_5dir.png}};
                        \node[left=of img22, node distance=0cm, rotate=90, xshift=1.5cm, yshift=-.75cm, font=\color{black}] {argument of EVal $\arg(\eigval)$};
                    
                    \node [below=of img21, yshift=1cm](img31){\includegraphics[trim={1.0cm 1.0cm 55.0cm 1.15cm},clip, width=.45\linewidth]{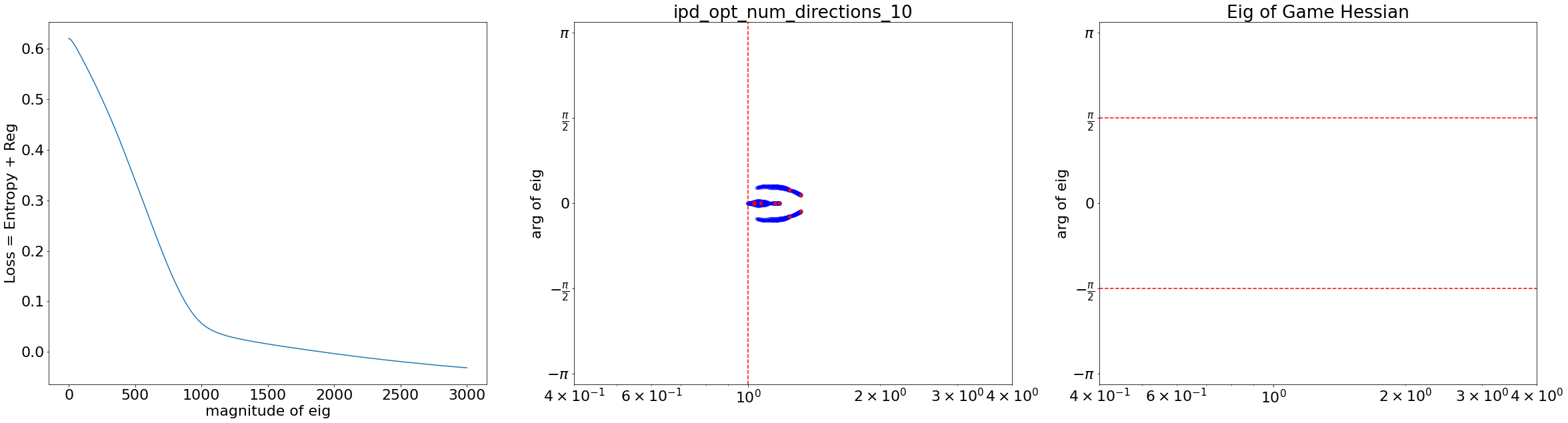}};
                        \node[left=of img31, node distance=0cm, rotate=90, xshift=2.5cm, yshift=-.75cm, font=\color{black}] {loss $\loss_{{\color{red}10}}^{\textnormal{min}}(\bothParam_0)$ = -{\color{red}Min top 10} exps};
                        \node[below=of img31, node distance=0cm, xshift=-.0cm, yshift=1.0cm,font=\color{black}] {optimization iteration};
                        
                        \node [right=of img31, xshift=-.5cm](img32){\includegraphics[trim={28.85cm 1.0cm 27.5cm 1.15cm},clip, width=.44\linewidth]{images/experiments/ipd/tune_lyap_ipd_10dir.png}};
                        \node[left=of img32, node distance=0cm, rotate=90, xshift=1.5cm, yshift=-.75cm, font=\color{black}] {argument of EVal $\arg(\eigval)$};
                        \node[below=of img32, node distance=0cm, xshift=0.0cm, yshift=1.0cm,font=\color{black}] {log-norm of EVal $\log(|\eigval|)$};
                \end{tikzpicture}
                \vspace{-0.01\textheight}
                \caption{
                    We compare the optimization for different losses using the local Lyapunov exponent with different numbers of exponents.
                    \textbf{\emph{Takeaway:}} We can effectively optimize multiple exponents, which give trajectory separation in multiple directions.
                    Our objective is the minimum of the top $n$ local Lyapunov exponents, with a different $n$'s optimization trajectory shown in each row.
                    To review this visualization also in Figure~\ref{fig:tune_lyap_ipd}:
                    The spectrum is shown with a scatter-plot in {\color{blue}blue}, with a progressively larger alpha at each iteration.
                    The final spectrum is shown in {\color{red}red}.
                    For the Jacobian of the fixed-point operator $\jacFixedPointOp$, a vertical {\color{red}red} line is shown where the EVal norm equals 1, signifying the cutoff between (locally) convergent and divergent eigenspaces.
                    \emph{Top}: Optimizing the max local Lyapunov exponent from Figure~\ref{fig:tune_lyap_ipd}, where the largest EVal is effectively maximized.
                    \emph{Middle}: Optimizing the top $5$ local exponents, where we find $5$ moderately divergent directions instead of 1 extremely divergent direction as in the top.
                    \emph{Bottom}:  Optimizing the top $10$ local exponents, which results in trajectory separation for every direction.
                    We study different objectives of the top $n$ exponents in Appendix Figure~\ref{fig:tune_lyap_ipd_difObjectives}, and study non-local $k$-step exponents in Appendix Figure~\ref{fig:tune_lyap_ipd_difNumLyap}.
                }\label{fig:tune_lyap_ipd_moreDirections}
            \end{figure*}
            
            \begin{figure*}
                \centering
                \begin{tikzpicture}
                    \centering
                    \node (img11){\includegraphics[trim={1.0cm 1.85cm 55.0cm 1.15cm},clip, width=.46\linewidth]{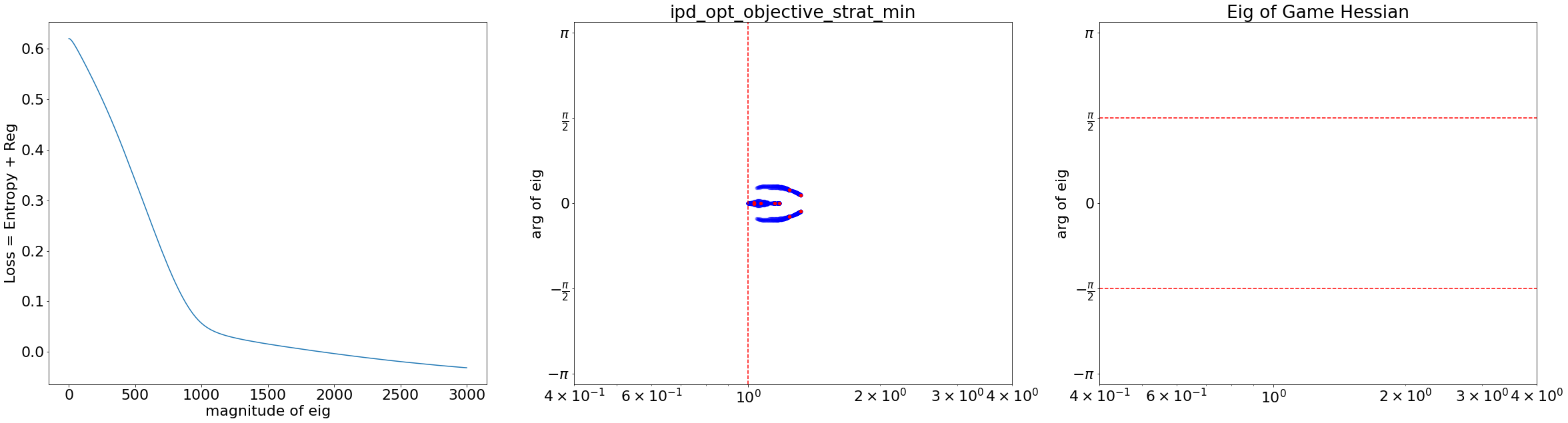}};
                        \node[left=of img11, node distance=0cm, rotate=90, xshift=2.25cm, yshift=-.9cm, font=\color{black}] {loss $\loss_{10}^{{\color{red}\textnormal{min}}}(\bothParam_0)$= -{\color{red}Min} top 10 exps};
                        
                        \node [right=of img11, xshift=-.5cm](img12){\includegraphics[trim={28.85cm 1.85cm 27.5cm 1.15cm},clip, width=.45\linewidth]{images/experiments/ipd/ipd_objective_strat/ipd_opt_objective_strat_min.png}};
                        \node[left=of img12, node distance=0cm, rotate=90, xshift=1.5cm, yshift=-.75cm, font=\color{black}] {argument of EVal $\arg(\eigval)$};
                        \node[above=of img12, node distance=0cm, xshift=-.0cm, yshift=-1.15cm,font=\color{black}] {EVals of Jac. of fixed point operator $\spectrum(\jacFixedPointOp)$};

                    \node [below=of img11, yshift=1cm](img21){\includegraphics[trim={1.0cm 1.0cm 55.0cm 1.15cm},clip, width=.46\linewidth]{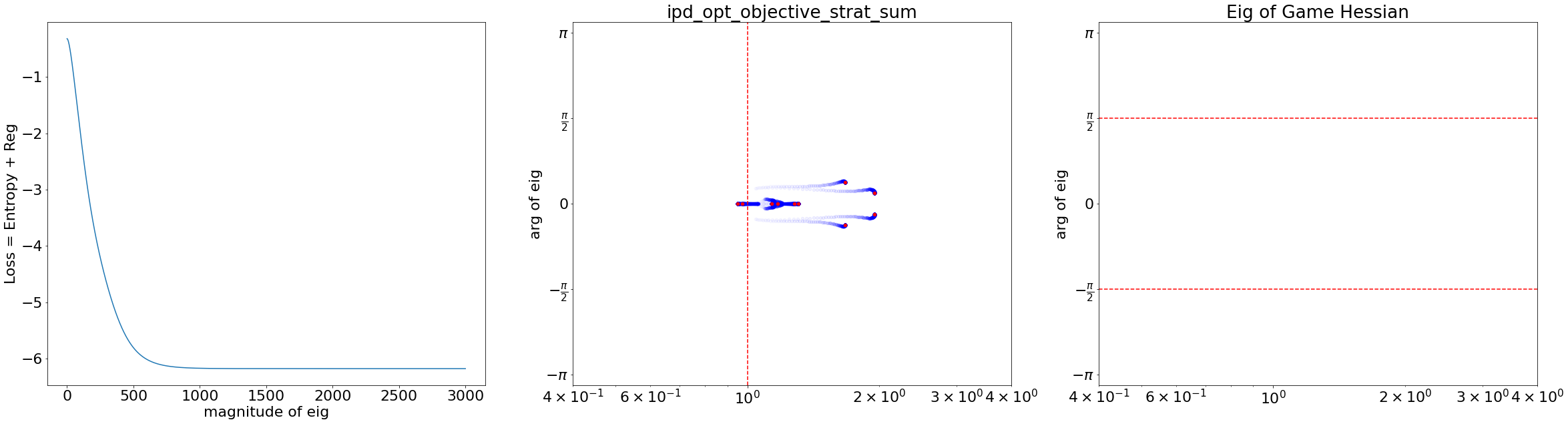}};
                        \node[left=of img21, node distance=0cm, rotate=90, xshift=3.0cm, yshift=-.9cm, font=\color{black}] {loss $\loss_{{\color{red}10}}^{\textnormal{sum}}(\bothParam_0)$= -{\color{red}Sum} top 10 exps};
                        \node[below=of img21, node distance=0cm, xshift=-.0cm, yshift=1.0cm,font=\color{black}] {optimization iteration};
                        
                        \node [right=of img21, xshift=-.5cm](img22){\includegraphics[trim={28.85cm 1.0cm 27.5cm 1.15cm},clip, width=.45\linewidth]{images/experiments/ipd/ipd_objective_strat/ipd_opt_objective_strat_sum.png}};
                        \node[left=of img22, node distance=0cm, rotate=90, xshift=1.5cm, yshift=-.75cm, font=\color{black}] {argument of EVal $\arg(\eigval)$};
                        \node[below=of img22, node distance=0cm, xshift=0.0cm, yshift=1.0cm,font=\color{black}] {log-norm of EVal $\log(|\eigval|)$};
                \end{tikzpicture}
                \caption{
                    We compare the optimization for different losses using the top $10$ local Lyapunov exponents.
                    \textbf{\emph{Takeaway:}} Using the minimum of the top exponents is a strong method to guarantee separation in many directions.
                    To review this visualization also in Figure~\ref{fig:tune_lyap_ipd}:
                    The spectrum is shown with a scatter-plot in {\color{blue}blue}, with a progressively larger alpha at each iteration.
                    The final spectrum is shown in {\color{red}red}.
                    For the Jacobian of the fixed point operator $\jacFixedPointOp$, a vertical {\color{red}red} line is shown where the EVal norm equals 1, signifying the cutoff between (locally) convergent and divergent eigenspaces.
                    \emph{Top:} We optimize the sum of the top $10$ exponents.
                    This is a simple choice, but it does not guarantee trajectory separation in every direction.
                    Our optimizer is happy with solutions that are diverging rapidly in the top \JL{TODO} directions, while converging (slowly) in the bottom directions.
                    \emph{Bottom:} To guarantee trajectory separation is every direction, we look at optimizing the minimum of the top exponents.
                    Here, at the end of training, all EVals of $\jacFixedPointOp$ are greater than $1$, signifying local trajectory separation in every direction.
                }\label{fig:tune_lyap_ipd_difObjectives}
            \end{figure*}
            
            \begin{figure*}
                \centering
                \begin{tikzpicture}
                    \centering
                    \node (img){\includegraphics[trim={1.0cm 1.0cm 55.0cm 1.15cm},clip, width=.5\linewidth]{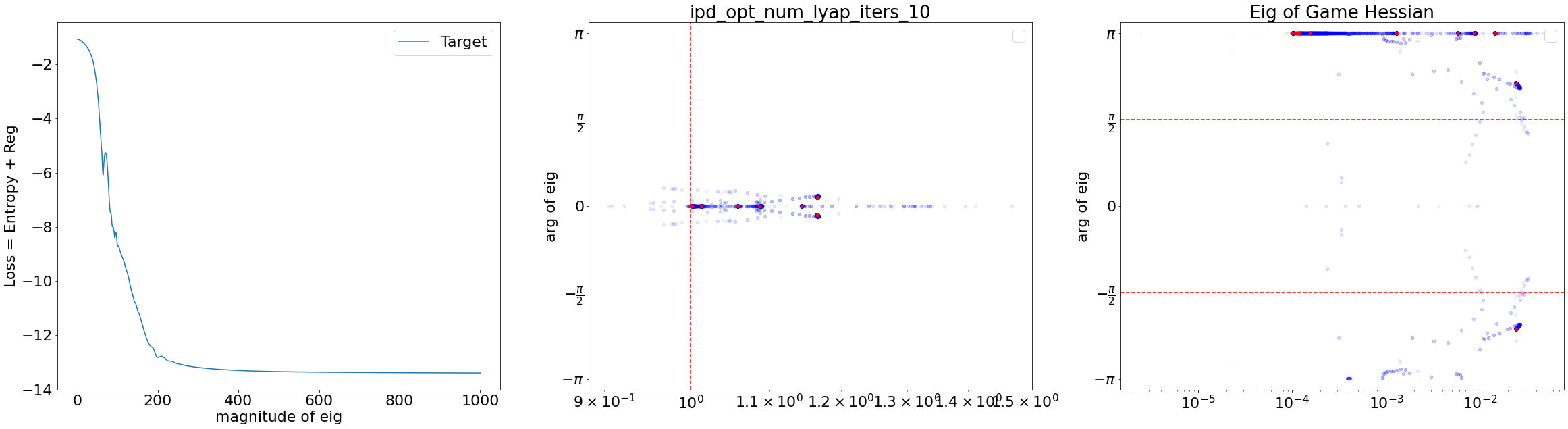}};
                    \node[left=of img, node distance=0cm, rotate=90, xshift=2.0cm, yshift=-.9cm, font=\color{black}] {loss $\loss(\bothParam_0) = -\lyap_{{\color{red}10}}^{max}(\bothParam_0)$};
                    \node[below=of img, node distance=0cm, xshift=-.0cm, yshift=1.0cm,font=\color{black}] {optimization iteration};
                \end{tikzpicture}
                \caption{
                    We investigate optimizing a loss of the max $k$-step Lyapunov exponent for $k > 1$.
                    \textbf{\emph{Takeaway:}} We are able to effectively minimize multi-step exponents in higher-dimensional problems if required.
                    To review this visualization also in Figure~\ref{fig:tune_lyap_ipd}:
                    The spectrum is shown with a scatter-plot in {\color{blue}blue}, with a progressively larger alpha at each iteration.
                    The final spectrum is shown in {\color{red}red}.
                    For the Jacobian of the fixed point operator $\jacFixedPointOp$, a vertical {\color{red}red} line is shown where the EVal norm equals 1, signifying the cutoff between (locally) convergent and divergent eigenspaces.
                }\label{fig:tune_lyap_ipd_difNumLyap}
            \end{figure*}
\end{document}